\documentclass[a4paper]{article}[11pts]

\usepackage[english]{babel}
\usepackage[utf8x]{inputenc}
\usepackage[T1]{fontenc}

\normalsize

\usepackage[a4paper,top=1in,bottom=1in,left=1in,right=1in,marginparwidth=1in]{geometry}

\usepackage{setspace}
\usepackage{amsmath}
\usepackage{amssymb} 
\usepackage{amsthm}
\usepackage{amsfonts, mathtools}
\usepackage{algorithm,algpseudocode}
\usepackage{bm}
\usepackage{smartdiagram}
\usepackage{comment}
\usepackage{graphicx}
\usepackage{multirow, booktabs}
\usepackage{caption}
\usepackage{subcaption}
\usepackage[colorinlistoftodos]{todonotes}
\usepackage[colorlinks=true, allcolors=blue]{hyperref}

\newcommand{\B}{\bm{B}}
\newcommand{\E}{\mathbb{E}}
\newcommand{\R}{\mathbb{R}}
\newcommand{\prob}{\mathbb{P}}

\newtheorem{proposition}{Proposition}
\newtheorem{lemma}{Lemma}

\newtheorem{assumption}{Assumption}
\newtheorem{theorem}{Theorem}
\newtheorem{corollary}{Corollary}

\usepackage{natbib}

\title{Fairer LP-based Online Allocation via Analytical Center}
\author{}
\date{}

\author{Guanting Chen$^\dagger$ \and Xiaocheng Li$^\Diamond$ \and  Yinyu Ye$^\ddagger$}
\date{\small 
$^\dagger$ Institute for Computational and Mathematical Engineering, Stanford University\\
$^\Diamond$ Imperial College Business School\\
$^\ddagger$Department of Management Science and Engineering, Stanford University\\
$\{$guanting, chengli1, yinyu-ye$\}$@stanford.edu\\
}

\begin{document}
\maketitle

\onehalfspacing

\begin{abstract}
In this paper, we consider an online resource allocation problem where a decision maker accepts
or rejects incoming customer requests irrevocably in order to maximize expected reward given limited resources. At each time, a new order/customer/bid is revealed with a request of some resource(s) and a reward. We consider a stochastic setting where all the orders are i.i.d. sampled from an unknown distribution. Such formulation arises from many classic applications such as the canonical (quantity-based) network revenue management problem and the Adwords problem. While the literature on the topic mainly focuses on regret minimization, our paper considers the \textit{fairness} aspect of the problem. On a high level, we define the fairness in a way that a fair online algorithm should treat similar agents/customers similarly, and the decision made for similar agents/customers should be consistent over time. To achieve this goal, we define the fair offline solution as the analytic center of the offline optimal solution set, and introduce \textit{cumulative unfairness} as the cumulative deviation from the online solutions to the fair offline solution over time. We propose a fair algorithm based on an interior-point LP solver and a mechanism that dynamically detects unfair resource spending. Our algorithm achieves cumulative unfairness on the scale of order $O(\log(T))$, while maintains the regret to be bounded without dependency on $T$. In addition, compared to the literature, our result is produced under less restrictive assumptions on the degeneracy of the underlying linear program.
\end{abstract}

\section{Introduction}
Algorithm-based decision-making systems have become widespread
 in almost every aspect
of modern life, such as policing \citep{rudin2013predictive}, hiring \citep{miller2015can}, credit lending \citep{petrasic2017algorithms}, and criminal
justice \citep{huq2018racial}. Meanwhile, the data-driven predictions and decision-making have raised increasing concerns on discriminatory and unfair treatment. The quantification and promotion of fairness for machine learning algorithms are rapidly growing areas of research. Given the abundant research focusing on fair classification and fair regression, this paper focuses on the fairness aspect of the online resource allocation problem. 

In many practical settings such as COVID-19 vaccine rollout, hiring, and service scheduling, the decision maker has to make decision online with incomplete knowledge of the demand/population distribution. Therefore, fairness concerns arise naturally in online resource allocation problems. Decision makers have to distribute limited supply of resources across multiple groups with different needs. In this paper, we formulate the online resource allocation problem as an online Linear Programming problem, which is the underlying model for various OR/OM problems such as quantity-based network revenue management \citep{talluri2006theory}, service reservations \citep{stein2020advance}, and order fulfillment \citep{simchi2010operations}.

Our motivations can be summarized from the following example. Emergency rooms are scarce resources because of the time-sensitive feature of emergency surgeries. A hospital has to make decisions on whether to accept or reject a patient. It is impossible for the hospital to accept all patients, and the rejected patients will be diverted to other hospitals. Consider a case where there is only one type of resource: the total operating hours of the emergency room. The objective is to minimize the idle time of the emergency room over a period of $T$ days. We assume there are two types of patients. Patients of type A request surgery that takes shorter time and patients of type B request surgery that takes longer time. Those two types of patients have different arrival rates $\lambda_1$ and $\lambda_2$, which are unknown to the hospital. The hospital has to make a real-time decision based on the historical information on the arrival rates and the operating time on each type of patient. 

We assume that there are more patients than the available emergency rooms. As a result, every time the hospital needs to schedule a new surgery, it has to choose from a group of patients consisting of type A and B. Since the hospital's objective is to minimize the idle time of the emergency room, there could be allocation rules that are optimal in terms of minimizing the idle time, but are unfair in some aspects.
\begin{itemize}
    \item Allocation rule a): when the hospital has the availability to schedule a new surgery, it always accepts one patient of type A from the patient group, while rejecting all patients of type B. This is obviously unfair to patients of type B.
    \item Allocation rule b): whenever a decision needs to be made before day $T/2$, the hospital always accepts one patient of type A, and when the decision needs to be made after day $T/2$, it always accepts one patient of type B. Although the hospital spends equal operating hours on servicing patients of type A and B, this policy is unfair for patients of type $B$ arriving before day $T/2$ and patients of type $A$ arriving after day $T/2$.
\end{itemize}

If the hospital knew the underlying arrival rates $\lambda_1$ and $\lambda_2$, a natural fair allocation rule would be to randomly accept one patient of type A with probability $\frac{\lambda_1}{\lambda_1+\lambda_2}$ and to randomly accept one patient of type B otherwise. It maintains the fairness between patients of type $A$ and type $B$ such that the operating times of the emergency rooms for both types of patients are proportional to their population. Moreover, both patient types will have the same probability of getting accepted across time. For such reasons, we refer to this allocation rule as fair offline decision.

In the online environment, the hospital does not know the parameter $\lambda_1$ and $\lambda_2$. To ensure fairness, the hospital has to make decisions that are close to the fair offline decision; as time increases, the online decision should converge to the fair offline decision. We also require that the treatment of similar patients across time should be similar, and therefore we expect the deviations from the online decision to the offline decision to be small.

\subsection{Overview of Contributions}

We take the fairness notion that ``similar individuals should be treated similarly'' \citep{dwork2012fairness} and formally propose two fairness requirements for online algorithms. Firstly, we require our online solution to converge to the fair offline solution, which is defined as the analytic center (see \cite{luenberger1984linear}) of the optimal solution set of the underlying LP. Secondly, we want the decision to be consistent across time such that the online decision converges to the fair offline solution in a stable way. To quantify the consistency, we define \textit{cumulative unfairness} to be the cumulative deviation from the online solutions to the fair offline solution. A fair online algorithm should maintain a moderately small cumulative unfairness.

It is common for LP problems to have multiple solutions \citep{mittelmann2005decision}. The analytic center can be viewed as the average of the optimal solutions, because it maximizes the distance to the boundary of the optimal solution set. Consequently, our algorithm ensures certain levels of ``group fairness'' such that the algorithm will not favor one optimal solution over the other, and the final allocation can be distributed more evenly across different type of agents. Also, the analytic center ensures ``individual fairness across time'' such that throughout the whole decision horizon, as long as the agents have similar reward and resource consumption patterns, we cannot treat them too differently based on other features such as the time they enter the system, demographics, race, and gender.

Next, we analyze the relationship between fairness and regret. The common approach in fair sequential decision making literature is to formulate an optimization problem and put the fairness requirements as extra constraints. Assuming the optimization problem is to maximize the objective value (we refer to it as reward), putting extra constraints will almost always decrease the reward. As a result, it is easier to get a sub-linear upper bound for regret, because regret is defined as the difference between the online reward to the offline reward benchmark, which is lower due to the extra constraint.

Instead of putting fairness constraints into the optimization problem, we consider approaches to design our sequential decision making algorithm such that it features a sub-linear upper bound for cumulative unfairness, while keeping the regret low. Our definition of regret is stronger than the definition in previous works because without the extra fairness constraint: we are comparing the online reward with the original offline reward. This ``getting fairness guarantee without fairness constraints'' approach provides a better insight on the trade-off between regret and efficiency in the online environment.

Our solution is an adaptive online linear programming algorithm that uses an interior-point type of solver and dynamically detects unfair resource spending. This algorithm features a $O(\log(T))$ upper bound for cumulative unfairness and a problem-dependent constant upper bound for regret. It implies that from the algorithm design perspective, it is possible to achieve extra fairness without affecting regret. 

Furthermore, our analysis alleviates the unique and non-degenerate condition that is commonly used in the revenue management literature \citep{jasin2012re,jasin2015performance,chen2021linear}. By leveraging properties of analytic center, we require a condition which we name it as ``dual non-degeneracy'' (non-degeneracy condition for the binding resource constraints. See Assumption \ref{assp_dual_binding} for more details), and allow the existence of multiple optimal solutions.

To conclude, our contribution can be summarized as 

\begin{itemize}
    \item We propose a LP-based online algorithm such that its online decision converges to the fair offline optimal solution in expectation, and its corresponding cumulative unfairness has an upper bound of order $O(\log(T))$.
    \item The fair algorithm achieves a bounded regret that bears no dependency on $T$ under less restrictive non-degeneracy conditions than other works in the literature.
\end{itemize}
\subsection{Related Work}
The concept of fairness has been extensively studied in machine learning, operations research and economics. We will discuss the related literature from different research streams.

\textbf{Literature on fair machine learning}

Generally speaking, the definitions of fairness in the machine learning literature can be summarized into two categories.
\begin{itemize}
    \item Group Fairness \citep{kamishima2011fairness, hardt2016equality}: also known as statistical parity or \textit{disparate impact}. It requires that the demographics of certain statistical measures (for example, positive classifiation rate) are similar to the demographics of the population as a whole.
    \item Individual Fairness \citep{dwork2012fairness}: also known as \textit{disparate treatment}. It advocates fairness through the goal that similar people should be treated similarly.
\end{itemize}

There has been a proliferate literature on promoting group fairness and individual fairness for supervised learning problems, where the algorithm takes the training data and apply the same predictive model to every new instance. Generally speaking, the existing fairness approach can be summarized into three categories:

\begin{itemize}
    \item Pre-processing \citep{zemel2013learning}: modify the training data.
    \item In-processing \citep{zafar2017fairness}: modify the learning algorithm.
    \item Post-processing \citep{hardt2016equality}: modify the output of the algorithm.
\end{itemize}

Our works is inline with the definition of individual fairness, and the way we design our algorithm to achieve fairness is an in-processing technique. It is also worth mentioning that under the case that the reward is a linear function of the resource consumption vector (such as the example in the introduction), our algorithm also exhibits group fairness property. What differentiate our work with the ones mentioned above are that, firstly, we are interested in making fair decisions instead of generate fair classifications/regressions; Secondly, in this work we focus on the online setting, where the agent receives data over time, and makes real-time decisions. Our definition of fairness can be think of as a time-dependent variant of individual fairness. We require our real time decision to achieve individual fairness and be consistent with time. 

Less fairness results have been developed in online decision making literature, and a large portion of the work focus on bandit algorithms.  \citet{joseph2016fairness} study the so-called \textit{meritocratic fairness}, which require less qualified individuals should not be
favored over more qualified individuals. The authors apply the fairness notion into classic and contextual bandits, and propose algorithm that is fair and is of sub-linear regret. \citet{jabbari2017fairness} extend
the notion of meritocratic fairness to reinforcement learning, whereby fairness requires the algorithm to prefer one action over another with low probability, if the long-term (discounted) reward of choosing the latter action is higher. Other fairness notions have also been proposed. \citet{wang2021fairness} argue that the conventional bandit formulation can lead to an undesirable and unfair winner-takes-all allocation of exposure, and propose algorithm that enables each arm to receive an amount of exposure proportional to its merit. Sub-linear fairness regret and reward regret are also guaranteed in this setting. Similarly, \citet{liu2017calibrated} study the stochastic bandit with a smoothness constraint that if two arms have a similar quality distribution, the probability of selecting each arm should be similar. \citet{patil2020achieving} and \citet{chen2020fair} study stochastic and contextual bandit problem with fairness requirement that there is a minimum rate that each arm must been pulled. Under such condition the authors obtain sub-linear regrets. Also see \citet{blum2018preserving, gillen2019online, li2019combinatorial, bechavod2019equal} for other fairness constraints in the bandit setting.

In the fair bandit setting, the most relevant work to ours is \citet{gupta2019individual}. The authors propose two fairness notions: fairness-across-time (FT) and fairness-in-hindsight (FH). FT propose a Lipchitz metric that ensures treatment of individuals to be individually
fair relative to the past as well as future, while FH only requires a one-sided notion of individual fairness that is defined relative to only the past decisions. They provide an algorithm with sub-linear regret under FH, and show linear regret is inevitable under FT. In our work, we consider a reward maximization problem under resource constraint, and we achieve fairness by controlling the cumulative deviation of online decision to the fair decision. We are not putting our unfairness metric as a constraint as in \citet{gupta2019individual}.

Another paper that shares the similar spirit is \citet{heidari2018}. The authors adopt the individual fairness notion, and require fairness across time; If two individuals are similar in the feature
space and arrive during the same time epoch, the algorithm must assign them to similar outcomes. The authors show that imposing such a time-dependent consistency constraint does not significantly affect the speed of learning. Our work differs from \citet{heidari2018} in several ways. First, our cumulative unfairness can be interpreted as a global fairness requirement, while \citet{heidari2018} require consistency to hold among recent observations only. Second, we provide theoretical result for regret, while \citet{heidari2018} analyze the convergence behavior of learning. Third, we use an in-processing technique for a specific algorithm, while \citet{heidari2018} adopt a general post-processing technique that directly modifies outcomes.

\textbf{Literature on fair operations research}

Fairness in decision
making has also attracted attentions from researchers in the field of operations research. Under a static environment, \citet{bertsimas2011price} considers a general setting where a decision maker allocates $m$ divisible resources to $n$ agents, each with a different utility function. The authors propose two measures: proportional
fairness and max-min fairness, and characterize the efficiency loss due to maximizing fairness. \citet{donahue2020fairness} analyze the fairness notion that individuals from different groups should have (approximately) equal
probabilities of receiving the resource, and characterize the trade-off between fairness and utilization in static resource allocation problems. The authors provide upper bounds for the gap between the maximum possible
utilization of the resource and the maximum possible utilization
subject to this fairness condition.

In the online decision making setting, one model that is close to ours is \citet{elzayn2019fair}. The authors consider a centralized agent allocating a scarce resource amongst several groups. The agent have to maximize objective value under the ``equality of opportunity'' constraint that the probability of an individual receiving the scarce resource is approximately independent of the individual’s group. The authors propose a learning algorithm that
converges to an optimal fair allocation without distribution knowledge of the population. In contrast, our definition of fairness has an extra requirement that the decision must be consistent with respect to time. Our model is more general such that it allows multiple resource constraints. In addition to providing sub-linear bound for cumulative unfairness, we also prove the extra property on bounded regret.

Similarly, \citet{sinclair2020sequential} consider a fairness objective which is the maximum difference between the algorithms allocation and the fair offline solution. The authors prove that an approximation for such objective gives an allocation rule that achieves approximate Pareto-efficiency and envy-freeness. \citet{ma2020group} formulate the online resource allocation problem into the online bipartite matching model. The authors study the fairness objective of maximizing the minimum service rate across all groups, and analyze the competitive ratio of online algorithms. \citet{manshadi2021fair} consider fair online rationing such that each arriving agent receives a fair share of resources proportional to its demand, and provide upper bounds on both the expected minimum fill rate and minimum expected fill rate.

In the pricing and revenue management literature, \citet{li2016behavior} study the impact of consumers fairness concerns
on firm's pricing strategy, profits, consumer surplus, and social welfare. \citet{cohen2019price} consider a single-period model with a known linear demand, and study the impact of imposing various fairness criterion on the seller's profit, consumer surplus, and social
welfare. 

In a similar motivation to ours, \citet{cohen2021dynamic} study the problem of dynamic pricing with
unknown demand and with two fairness constraints. The fairness constraint for price requires similar prices for different customer groups, and
ensures that the prices over time for each customer group are relatively stable. Under price constraint the authors propose an UCB type of algorithm that yields a near-optimal regret. The fairness constraint for demand requires that by setting price accordingly, the resulting demand from different
customer groups is relatively similar. Under such constraint the authors design an algorithm that also achieves a near-optimal performance. 

An important distinction that differentiates our work from the works mentioned previously is that, in the problem formulation, we are not imposing extra fairness constraint to the offline problem, therefore our online algorithm is comparing against the unconstrained benchmark.

\textbf{Literature on LP-based revenue management problem}

There has been a proliferate literature on algorithm design and analysis for the online revenue management problem. We provide a short summary of our result on regret and the related literature in Table \ref{tab:result}.

\begin{table}[]
    \centering
    \begin{tabular}{c|c|c}
    \toprule
          & Regret Bound &  Key Assumption(s)\\
         \midrule
        \cite{jasin2012re} & Bounded & Unique, nondegenerate, and distributional knowledge\\ 
        \cite{jasin2015performance}  &$\tilde{O}(\log T)$ & Unique and nondegenerate \\ 
        \cite{chen2021linear}  & Bounded & Unique,  nondegenerate, and finite support \\
        This paper  & Bounded & Dual nondegenate and finite support\\
         \bottomrule
    \end{tabular}
    \caption{Result comparison against literature: We note that all the regret bounds here are problem-dependent bound that mainly focuses on the dependence on horizon length $T$ but will inevitably involve certain parameters related to the underlying distribution/optimization problem.}
    \label{tab:result}
\end{table}

\begin{figure}[t]
\centering
    \includegraphics[height = 3cm, width = 5.25cm]{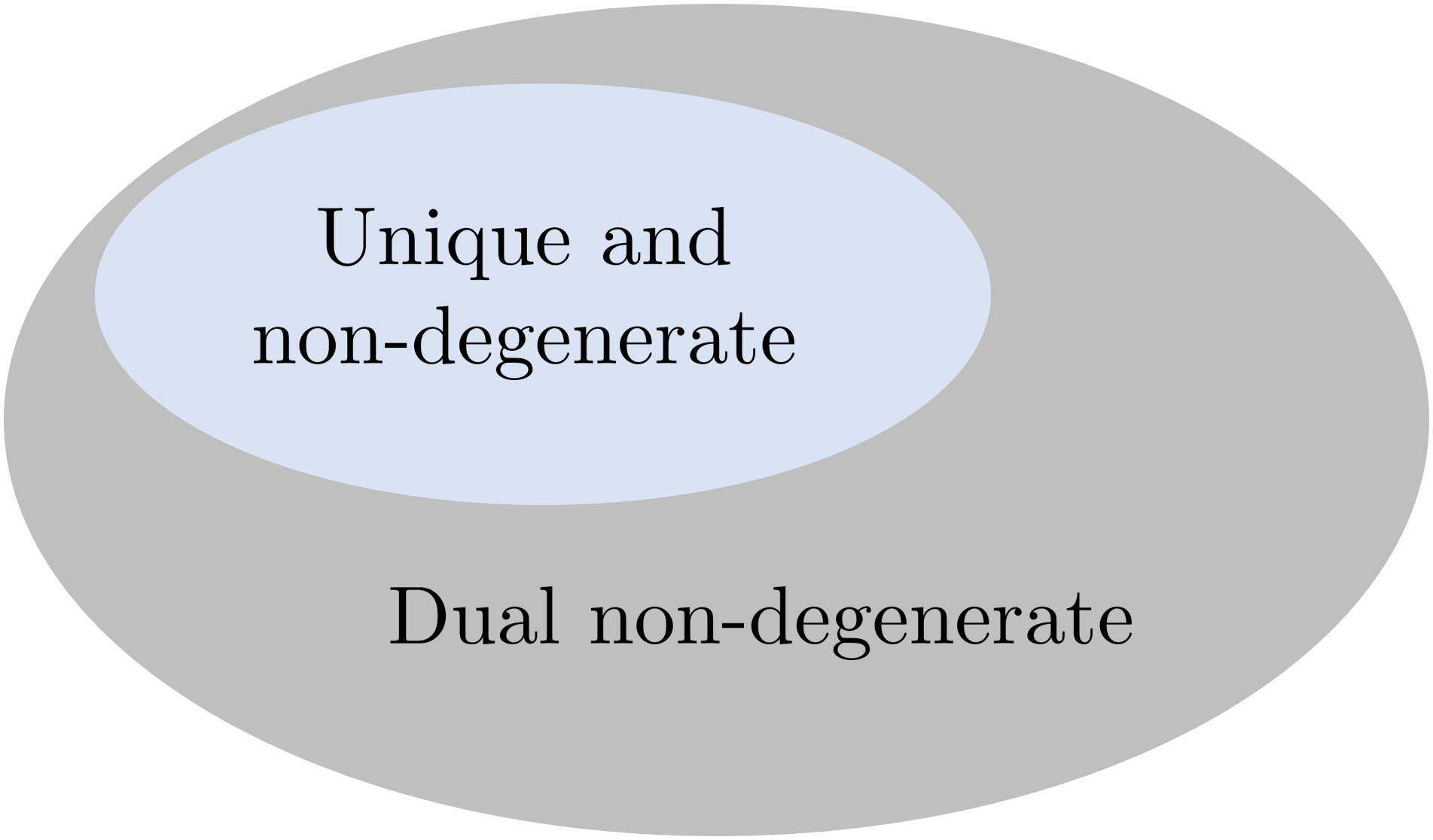}
    \caption{Notice that the dual nondegeneracy condition is a weaker condition than the widely assumed unique and non-degenerate condition in the literature}\label{fig:assump}
\end{figure}

One stream of literature investigates the existence of an online algorithm that achieves bounded regret (not dependent on the number of decision variables or time horizon). Specifically, under the knowledge of the customer arrival distribution, a line of works \citep{jasin2012re, wu2015algorithms, bumpensanti2020re} study the canonical quantity-based network revenue management problem and design algorithms that achieve bounded regret. \citet{jasin2012re} show that the LP-based re-solving algorithm achieves bounded regret where the algorithm computes a control policy by periodically solving a linear program specified by the known arrival distribution. A critical condition is that the underlying LP should be unique and non-degenerate. A subsequent work \citep{bumpensanti2020re} fully re-solves the problem through an infrequent re-solving scheme which updates the control policy only at a few selected
points. Another line of works \citep{vera2019online, vera2019bayesian, banerjee2020uniform} devise algorithms that achieve bounded regret for various problems including dynamic pricing, knapsack problem, and bin packing problem. The authors develop a novel and intuitive approach called ``compensated coupling'' to derive regret upper bound for an online algorithm. The idea is to bound the cumulative expected loss induced by the different decisions made by the online algorithm) against the hindsight optimal (which has all the future information). As in the aforementioned works on network revenue management, this line of works are also built upon the knowledge of the underlying distribution. 

There are two types of benchmark for the online revenue management literature: fluid approximation and expected hindsight optimal. Fluid approximation benchmark is obtained by taking expectation on the reward and resource distribution and then solving the optimization problem. Expected hindsight optimal can be calculated from expectation of the maximization problem, where under the expectation, we observe the realizations of arrivals and solve for the optimal value in hindsight. As a result, fluid benchmark is stronger than hindsight benchmark, and from \citet{bumpensanti2020re} the difference can be of order $\Omega(\sqrt{T})$ under degenerate environment.

In our paper, we consider the fluid approximation LP (also known as Deterministic LP, which we refer to as DLP) as the benchmark for regret analysis. Under a similar setting, \citet{chen2021linear} show that bounded regret against the fluid benchmark is achievable for the online revenue management problem if the underlying LP is unique and non-degenerate, and the number of customer/order types are finite. We improve this result by extending the condition to that the DLP is dual non-degenerate. When the DLP is degenerate, both \cite{bumpensanti2020re} and \cite{vera2019online} show that the gap between the fluid benchmark and the hindsight benchmark is $\Omega(\sqrt{T})$, thereby arguing that the fluid benchmark can be too strong to compete against. Our result provides a positive view of the fluid benchmark under a weaker non-degenerate condition. In other words, our work shows that the large gap between the benchmarks and the deterioration of the algorithm performance is fully caused by the degeneracy of the \textit{binding} constraints in the DLP. When the binding constraints are non-degenerate, bounded regret is still achievable against the fluid benchmark even when the distribution is unknown.

The rest of the paper is organized as follows. In section \ref{sec_model}, we introduce the model, performance metrics and briefly discuss the fair algorithm. In section \ref{sec_fair}, we give a more detailed analysis on how we derived the algorithm with fair decisions and low regret. In section \ref{sec_reg_general}, we discuss the technical framework for our theoretical analysis. In section \ref{sec_numerical}, we verify our findings in several simulated environments. Extra proofs are left in the appendix.  

\section{Model, Performance Metric, and Algorithm}
\label{sec_model}

\subsection{Problem Formulation}

Consider a resource allocation problem over a horizon of $T$. The objective is to maximize the cumulative reward over the horizon subject to the budget constraints on $m$ types of resources. At each time period, a customer order/bid/request arrives and it asks for certain amount of each resource. The decision maker needs to decide whether to accept or reject the customer order. Upon the acceptance of the order, we need to satisfy the resource request, and we will collect the reward associated with the order. From the resource viewpoint, we act as a market maker who allocates the resources among all the customer orders. The allocation decisions are made in an online manner and all the past decisions are irrevocable. In this paper, we study a stochastic model where the request-reward pair made by each customer is assumed to be i.i.d. and follows an unknown distribution. In addition, we assume the distribution has a finite support, i.e., there is a finite number of customer types and customers of each type place identical orders.

More specifically, our paper analyzes a class of resource allocation problems which take the following LP as its underlying (offline) form:
\begin{align}
   \max \ \ & \sum_{t=1}^T \bm{r}^\top_t \bm{x}_t \label{eqn:multiILP}  \\
    \text{s.t. }\ & \sum_{t=1}^T \bm{A}_t \bm{x}_t \le \bm{B} \nonumber  \\ 
    & \bm{1}^\top \bm{x}_t \le 1, \ \ \bm{x}_t \ge \bm{0}, \ \ t=1,...,T \nonumber
\end{align}
where $\bm{r}_t = (r_{t1},...,r_{tk})^\top \in \R^k$, $\bm{A}_t = (\bm{a}_{t1},...,\bm{a}_{tk}) \in \R^{m\times k}$, and $\bm{a}_{ts} = (a_{1ts},...,a_{mts})^\top\in\mathbb{R}^m,$ for $t=1,...,T$ and $s=1,...,k.$ The right-hand-side vector $\bm{B}=(B_1,...,B_m)^\top$ encapsulates the capacity for each resource. The decision variables are $\bm{x}=\left(\bm{x}_1,...,\bm{x}_T\right)$ where $\bm{x}_t=(x_{t1},...,x_{tk})^\top$ for $t=1,...,T$. In an online setting, the parameters of the optimization problem (\ref{eqn:multiILP}) are revealed in an online fashion and one needs to determine the value of decision variables sequentially. At each time $t,$ the coefficients $(\bm{r}_t, \bm{A}_t)$ are revealed, and we need to decide the value of $\bm{x}_t$ instantly. Different from the offline setting, at time $t$, we do not have the information of the subsequent coefficients to be revealed, i.e., $\{(\bm{r}_{t'}, \bm{A}_{t'})\}_{t'=t+1}^T$. The problem \eqref{eqn:multiILP} in an online setting is often referred to as \textit{online linear programming} \citep{agrawal2014dynamic,kesselheim2014primal}. With different specification of the input of the LP, the problem encompasses a wide range of applications, including secretary problem \citep{ferguson1989solved}, knapsack problem \citep{kellerer2003knapsack},  network routing problem \citep{buchbinder2009online}, matching and Adwords problem \citep{mehta2005adwords}, combinatorial auction problem \citep{agrawal2009}, etc.

In the rest of the paper, for notational simplicity we focus on the one-dimensional case where $k=1$. We note that the results and proof of the general case follow the same idea of the one-dimensional case. For $k=1$, the online formulation of (\ref{eqn:multiILP}) reduces to a one-dimensional online LP problem,
\begin{align}
   \max \ \ & \sum_{t=1}^T r_t x_t \label{eqn:OLP}  \\
    \text{s.t. }\ & \sum_{t=1}^T \bm{a}_t x_t \le \bm{B} \nonumber  \\ 
    & 0 \le x_t \le 1,  \ \ t=1,...,T \nonumber
\end{align}
where $\bm{a}_t = (a_{1t},...,a_{mt})^\top \in \R^{m}$ and the decision variables are $\bm{x}=\left({x}_1,...,{x}_T\right)^\top\in \R^{T}.$ There are $m$ constraints and $T$ decision variables. Throughout the paper, we use $i$ to index constraints and $t$ to index decision variables.

\subsection{Performance measure} 
In this section we introduce the performance measure. Before doing so, we note that it is necessary to state the assumption on the distribution that governs the arrival of $(r_t, \bm{a}_t)$'s.
\begin{assumption}[Distribution] We assume
\begin{itemize}
    \item[(a)] Stochastic:  The column-coefficient pair $(r_t,\bm{a}_t)$'s are i.i.d. sampled from a distribution $\mathcal{P}.$ The distribution $\mathcal{P}$ takes a finite support $\{(\mu_j, \bm{c}_j)\}_{j=1}^n$ where $\mu_j \in \mathbb{R}$ and $\bm{c}_j \in \mathbb{R}^m$. Specifically, $$\prob((r_t, \bm{a}_t) = (\mu_j, \bm{c}_j)) = p_j$$ for $j=1,...,n$. The probability vector $\bm{p} = (p_1,...,p_n)^\top$ is unknown. 
    \item[(b)] Positiveness and boundedness: $0\le\mu_j\le1$, $\bm{c}_j\ge \bm{0}$ and $\|\bm{c}_j\|_\infty\le 1$ for $j=1,...,n.$
    \item[(c)] Linear growth: The right-hand-side $\bm{B}=T\bm{b}$ for some $\bm{b}=(b_1,...,b_m)^\top > \bm{0}$.
\end{itemize}
\label{assp}
\end{assumption}

Assumption \ref{assp} (a) imposes a stochastic assumption for the customer orders. In addition, it states that the support of the distribution is finite and known, but the parameters of the distribution are unknown. In other words, it means that there are $n$ known customer types, and the customer type at each time $t$ follows a multinomial distribution with unknown parameters. Assumption \ref{assp} (b) requires all the entries of $(\mu_j,\bm{c}_j)$ are between $0$ and $1$. We remark that all the results in this paper still hold (up to a constant) when this part is violated, and the positiveness and boundedness are introduced only for simplicity of notation. Lastly, the linear growth condition in Assumption \ref{assp} (c) is commonly assumed in online resource allocation problems \citep{asadpour2019online, li2019online, bumpensanti2020re}. In our context, the condition is mild in that if $\bm{B} = o(T)$, we can always adjust the time horizon with $T'\ll T$ such that $\bm{B} = T'\bm{b}$, and consequently the linear growth condition holds for $T'$.

Under Assumption \ref{assp} a), a commonly considered performance benchmark is the \textit{fluid approximation} version of the ``offline'' LP \eqref{eqn:OLP},
\begin{equation}
    \begin{aligned}
   \max \ \ & \sum_{j=1}^n p_j \mu_j y_j \label{eqn:DLP}  \\
    \text{s.t. }\ & \sum_{j=1}^n p_j \bm{c}_j \cdot y_j \le \bm{b} \\ 
    & 0 \le y_j \le 1,  \ \ j=1,...,n.
    \end{aligned}
\end{equation}
Recall from Assumption \ref{assp} that $\mu_j$ and $\bm{c}_j$ represent the reward and requested resource consumption of the $j$-th customer type. The right-hand-side $\bm{b}=\bm{B}/T$ represents the average resource capacity per time period, and $p_j$ is the probability of the $j$-th customer type. The decision variables $y_j$'s prescribe a ``probabilistic'' decision rule for the orders, and $y_j$ can be interpreted as the proportion of accepted orders (or the probability of accepting orders) for the $j$-th customer type. The LP \eqref{eqn:DLP} can be viewed as a deterministic version obtained by taking expectation of the objective and the left-hand-side of LP \eqref{eqn:OLP}. For such reason, we refer to \eqref{eqn:DLP} as the deterministic LP (DLP).

We define $\bm{y}^*$ to be the analytic center (see \cite{luenberger1984linear}) of the optimal solution set of the DLP \eqref{eqn:DLP}. Generally speaking, analytic center can be viewed as the average of optimal solutions and can be treated as the fair solution. We define $\bm{y}_t = (y_{1,t}, \cdots, y_{n,t})$ to be the probability of accepting corresponding order types at time $t$. More specifically, if at time $t$ the arriving order is of type $j$, then the algorithm will accept the arriving order with probability $y_{j,t}$. For fairness reason, we want the online allocation policy to stay close to $\bm{y}^*$. Therefore, we define the \textit{cumulative unfairness} of the online algorithm $\pi$ as 

\begin{align}\label{eqn:cu}
    \text{UF}_T(\pi) = \E \left[\sum_{t=1}^T \|\bm{y}_t-\bm{y}^*\|_2^2\right].
\end{align}

After defining the fairness metric, we begin to define the regret. For the online problem, at each time $t$ we decide the value of $x_t$. $x_t=1$ means that we accept the order and allocate $\bm{a}_t$ amount of resources to this order accordingly; $x_t=0$ means that we reject the order. Like the offline problem, we need to conform to the constraints throughout the procedure, i.e., no shorting of the resources is allowed. In this paper, we consider \textit{regret} as the performance measure, formally defined as
$$\text{Reg}_{T}^\pi \coloneqq \E\left[T\cdot \text{OPT}_{\text{D}} - \sum_{t=1}^T r_tx_t\right],$$
where the quantity $\text{OPT}_{\text{D}}$ represents the optimal objective value of the DLP \eqref{eqn:DLP}, and $(x_1,...,x_T)$ represent the online solution. The superscript $\pi$ denotes the online algorithm/policy according to which the online decisions are made. The expectation is taken with respect to $(r_t,\bm{a}_t)$'s and the randomness introduced by the algorithm (for example, $\bm{y}_t$).

\subsection{LP-based Adaptive Algorithms}

In this section we introduce the commonly used adaptive algorithms (also known as re-solving technique \citep{reiman2008asymptotically, jasin2012re, jasin2015performance, bumpensanti2020re} in the revenue management literature), and introduce modifications to make it a fairer algorithm.

We firstly introduce a few additional notations to characterize the constraint consumption process. Define $\bm{B}_1=\bm{B}$ and $\B_t=(B_{1t},...,B_{mt})^\top$ as the remaining resource capacity at the beginning of time $t$, i.e.
$$\bm{B}_t = \bm{B}_{t-1} - \bm{a}_{t-1} x_{t-1}.$$
Accordingly, we define $\bm{b}_t = \bm{B}_t/(T-t+1)$ as the average resource capacity at time $t$.
In addition, we use $\bm{B}_{T+1}$ to denote the remaining constraint at the end of horizon, and use $\bm{b}_1 = (b_{1,1},...,b_{1,m})^\top = \bm{B}_1/T= \bm{B}/T=\bm{b}$ to denote the initial average resource.

Let $N_{j}(t)$ denote the counting process of the $j$-th customer type, i.e., the number of observations $(\mu_j,\bm{c}_j)$ up to time $t$ (inclusively) for $j=1,...,n$, and denote $[n]$ the set $\{1,...,n\}$. Because no shorting is allowed, the remaining constraint vector $\bm{B}_{t}$ must be element-wise non-negative for all $t=1,...,T$. Since the true probability distribution $\bm{p}=(p_1,...,p_n)$ is unknown, the counts $N_{j}(t)$'s will be used by the online algorithm to construct empirical estimates for the corresponding probabilities. More specifically, denote 
$$\hat{\bm{p}}_t = (\hat{p}_{1,t}, \cdots, \hat{p}_{n,t}) = \left(\frac{N_1(t-1)}{t-1}, \cdots,\frac{N_n(t-1)}{t-1}\right)$$ to be the empirical estimation at time $t$.

Now we present the LP-based adaptive algorithm. At each time $t$, the algorithm solves a sampled linear program
\begin{align}
   \max \ \ & \sum_{j=1}^n \hat{p}_{j,t} \mu_j y_{j,t} \label{eqn:adpt_lp}  \\
    \text{s.t. }\ & \sum_{j=1}^n  \hat{p}_{j,t}\bm{c}_j\cdot y_{j,t} \le \bm{b}_t \nonumber  \\
    & 0 \le y_{j,t} \le 1,  \ \ j=1,...,n \nonumber
\end{align} to compute $\bm{y}_t = (y_{1,t},\cdots,y_{n,t})$, the probability of acceptance for each customer type $(\mu_j, \bm{c}_j).$ The sampled LP \eqref{eqn:adpt_lp} takes a similar form as the DLP \eqref{eqn:DLP} but differs in two aspects: (i) the probabilities $p_j$'s in \eqref{eqn:DLP} are replaced by their empirical estimates since the underlying distribution is assumed unknown; (ii) the right-hand-side $\bm{b}$ in \eqref{eqn:DLP} is replaced with its adaptive counterpart $\bm{b}_t$. It then uses the sampled LP's optimal solution $\bm{y}^*_t$ to determine the online solution $x_t$ at time $t$. Given that there is enough inventory remaining, this probabilistic decision rule aims to follow the prescription of the optimal solution by accepting the $j$-th customer type with probability $y^*_{j,t}$. We informally summarize this approach in Algorithm \ref{alg:DAA}.
\begin{algorithm}[ht!]
\caption{(Informal) A typical adaptive algorithm without fairness concern}
\label{alg:DAA}
\begin{center}
\smartdiagramset{border color=none, uniform color list=gray!60 for 4 items,  module x sep=3.5, back arrow distance=0.75, module minimum width = 2.8cm, text width = 2.5cm} 
\smartdiagram[flow diagram:horizontal]{Update $\hat{\bm{p}}_t$ and $\bm{b}_t$,Solve $\bm{y}_t^*$ for the sampled LP \eqref{eqn:adpt_lp}, Allocate resource based on $\bm{B}_t$ and $\bm{y}_t^*$, Observe new data and proceed as $t \leftarrow t+1$}
\end{center}
\end{algorithm}

The performance of this type of algorithm has been analyzed under different conditions. When the distribution is known, we can replace the sample estimation $\hat{\bm{p}}_t$ by $\bm{p}$ when solving \eqref{eqn:adpt_lp}. \cite{jasin2012re} derive a bounded regret under a nondegeneracy assumption for the deterministic LP; \cite{bumpensanti2020re} show that an infrequent re-solving scheme with carefully designed re-solving time will also achieve bounded regret for the degenerate case. Without distributional knowledge, \citet{jasin2015performance} derive a $O(\log(T))$ upper bound for the regret, and \citet{chen2021linear} established that the bounded regret can also be achieved if the distribution is unknown but have finite support. 

However, this type of algorithm can be unfair if not implemented correctly. In the following, we will analyze how to maintain a sub-linear upper bound at the order of $O(\log(T))$ for cumulative unfairness defined in \eqref{eqn:cu}, while maintaining the regret to be bounded. 

We make two improvements to the algorithm introduced above.
\begin{itemize}
    \item When solving the sampled LP \eqref{eqn:adpt_lp}, we need a solver that outputs the analytic center of the sampled LP. For example, an interior-point type algorithm will output the analytic center.
    \item For the sampled LP \eqref{eqn:adpt_lp}, at each time $t$ we have to determine the constraint $\bm{b}_t$ for two categories of resources: For resource $i$ that belongs to the ``binding'' category, we use the adaptive process $b_{i,t} = B_{i,t}/(T-t+1)$; For resource $i$ that belongs to the ``non-binding'' category, we use initial average process $b_{i,t} = B_i/T$ instead.
\end{itemize}
We choose the analytic center as the solution because the analytic center is more tractable than other types of solutions. We can expect the analytic center of the sampled LP \eqref{eqn:adpt_lp} to converge to $\bm{y}^*$, the analytic center of the DLP \eqref{eqn:DLP}. As for dynamically adjusting the resource constraint for ``binding'' and ``non-binding'' categories, the intuition is that the existence of ``non-binding'' resource suggests that for this specific type of resource, on average the demand is less than the constraint. As time goes on, the average remaining resource for this type will increase. Adjusting it to the initial average resource prevent us from allocating resources inconsistently across time due to the increase of the average remaining resources. 
Algorithm \ref{alg:DAA_fair} details our fair allocation scheme and the next section contains more detailed analysis on this fair algorithm.

\begin{algorithm}[ht!]
\caption{Fair Adaptive Allocation Algorithm}
\label{alg:DAA_fair}
\begin{algorithmic}[1]
\State Input: $\bm{B}, T, \{(\mu_j, \bm{c}_j)\}_{j=1}^n$
\State Initialize $\bm{B}_1 = \bm{b}$, $\bm{b}_1=\bm{B}_1/T$
\State Set $x_1=1$
\For {$t=2,..., n$}
\State Compute $\bm{B}_t=\bm{B}_{t-1}-\bm{a}_{t-1}x_{t-1}$
\State Compute $\bm{b}_t=\bm{B}_{t}/(T-t+1)$
\State Solve the following linear program where the decision variables are $(y_1,...,y_n)$:
\begin{align}\label{eqn:LP_algorithm_beforefair}
   \max \sum_{j=1}^n \hat{p}_{j,t} \mu_j y_j \,\,\,\,\,\,
    \text{s.t. }\ \sum_{j=1}^n  \hat{p}_{j,t}\bm{c}_j\cdot y_j \le \bm{b}_t,\,\,\,\,\,\,
    & 0 \le y_j \le 1,  \ \ j=1,...,n
\end{align}
\State Denote the optimal interior primal solution for \eqref{eqn:LP_algorithm_beforefair} as $\bm{y}_t'=(y_{1t}',...,y_{nt}')$, and find the binding set $\mathcal{B}_t = \left\{i\left|b_{i,t} = \sum_{j=1}^n\hat{p}_{j,t}\bm{c}_jy_{jt}'\right.\right\}$ and non-binding set $\mathcal{N}_t = [m]/\mathcal{B}_t$.
\State Set $\bm{b}'_t$ such that $\bm{b}'_{\mathcal{B}_t, t} = \bm{b}_{\mathcal{B}_t, t}$ and $\bm{b}'_{\mathcal{N}_t, t} = \bm{b}_{\mathcal{N}_t}$
\State Solve the following linear program where the decision variables are $(y_1,...,y_n)$:
\begin{align}\label{eqn:LP_algorithm_afterfair}
   \max \sum_{j=1}^n \hat{p}_{j,t} \mu_j y_j \,\,\,\,\,\,
    \text{s.t. }\ \sum_{j=1}^n  \hat{p}_{j,t}\bm{c}_j\cdot y_j \le \bm{b}_t',\,\,\,\,\,\,
    & 0 \le y_j \le 1,  \ \ j=1,...,n
\end{align}
\State Denote the optimal interior primal solution for \eqref{eqn:LP_algorithm_afterfair} as $\bm{y}_t^*=(y_{1t}^*,...,y_{nt}^*)$. Observe $(r_t,\bm{a}_t)$ and identify $(r_t,\bm{a}_t) = (\mu_j,\bm{c}_j)$ for some $j$, and let
\begin{align*}
    x_t = \begin{cases}
    1, & \text{ with probability } y_{jt}^*\\
    0, & \text{ with probability } 1-y_{jt}^*
    \end{cases}
\end{align*}
when the constraint permits; otherwise set $x_t=0.$ 
\State Update the counts for $\hat{\bm{p}}_{t+1}$ such that
\begin{align*}
    N_{j}(t) = \begin{cases}
    N_j(t-1)+1, & \text{ if } (r_t,\bm{a}_t) = (\mu_j,\bm{c}_j)\\
    N_j(t-1), & \text{ otherwise }
    \end{cases}
\end{align*}
\EndFor
\State Output: $\bm{x} = (x_1,...,x_T)$
\end{algorithmic}
\end{algorithm}

\section{Results and Analysis}\label{sec_fair}
In this section, we present the theoretical results on the cumulative unfairness and regret for Algorithm \ref{alg:DAA_fair}. Moreover, we provide a more detailed analysis on the properties of Algorithm \ref{alg:DAA_fair}, especially on the effects of the two modifications we summarized in the previous section.

\subsection{Main Results}
We firstly introduce the additional assumption needed for our theoretical results. The assumption is on the non-degeneracy for the binding resource, which we refer to as dual non-degeneracy of the underlying LP. We define the set of \textit{binding} resource to be 
\begin{equation*}
    \begin{aligned}
   \mathcal{B} := \left\{i\left|\sum_{j=1}^np_j\bm{c}_{ij}y_j^* = b_i, \bm{y}^* \text{ is the analytic center for \eqref{eqn:DLP}} \right.\right\}.
    \end{aligned}
\end{equation*}
The standard form of the dual program for the DLP \eqref{eqn:DLP} is
\begin{equation}\label{eqn:DDLP}
    \begin{aligned}
   \min \ \ & \bm{b}^\top{\bm{\lambda}} + \sum_{j=1}^n \gamma_j \\
    s.t. \ & p_j\bm{c}_j^\top{\bm{\lambda}} + \gamma_j \ge p_j\mu_j, \ \
    j=1,...,n \\ 
    & \bm{\lambda}\ge \bm{0}, \gamma_j \ge 0, \,\,j = 1,...,n,
    \end{aligned}
\end{equation}
where $\bm{\lambda}$ and $\bm{\gamma}$ are decision variables. The dual program can also be simplified as a program with decision variable $\bm{\lambda}$.
\begin{equation}\label{eqn:ADDLP}
    \begin{aligned}
   \min \ \ & \bm{b}^\top\bm{\lambda} + \sum_{j=1}^np_j(\mu_j - \bm{c}_j^\top\bm{\lambda})_+  \\
    \text{s.t. }\ & \lambda_i \ge 0, \,\,i = 1,\cdots,m.
    \end{aligned}
\end{equation}
Next, we introduce our second assumption.
\begin{assumption}[Dual Nondegereacy]\label{assp_dual_binding}
    Any optimal solution $\bm{\lambda}$ of \eqref{eqn:ADDLP} must satisfy $\lambda_i > 0$ for $i \in \mathcal{B}$.
\end{assumption}

Assumption \ref{assp_dual_binding} is a weaker assumption than the widely used unique and non-degenerate assumption in the literature \citep{jasin2012re, jasin2015performance, chen2021linear}. In practice, many common LP problems are degenerate and have multiple optimal solutions \citep{mittelmann2005decision}. Assumption \ref{assp_dual_binding} only requires the nondegenreate condition for the binding resource constraints, and allows degeneracy in other non-binding constraints. Moreover, it allows the primal and dual program to have multiple optimal solutions.

Now, we state the main results.

\begin{theorem}\label{thm_fair_regret}
Under Assumption \ref{assp} and \ref{assp_dual_binding}, Algorithm \ref{alg:DAA_fair} has the following upper bound on regret
\begin{align*}
    \text{Reg}_{T}^\pi &\leq O(m+n) +  O(T^{3/2}\exp(-T/2)).
\end{align*}
\end{theorem}
This theorem implies that with an interior-point type of solver and modifications on the adaptive process $\bm{b}_t$, we can still achieve bounded regret under a weaker assumption on the DLP. In terms of the regret analysis, it is different from that in \cite{chen2021linear} in two ways. First, without the unique and non-degenerate assumption, the stability analysis of the DLP is different. Second, we have to show that the introduction of the binding/non-binding detection mechanism in Algorithm \ref{alg:DAA_fair} will not affect the regret. We also note that the regret upper bound has extra dependence on $n$ that is hidden in some problem-dependent constants characterizing the continuity/Lipchitz property for the dual program. For more details see the proof in \ref{sec_ap_thm1}.

Next, we present the bound for the cumulative unfairness $\E \left[\sum_{t=1}^T \|\bm{y}_t-\bm{y}^*\|_2^2\right]$.
\begin{proposition}\label{prop_cu}
    Under Assumption \ref{assp} and \ref{assp_dual_binding}, Algorithm \ref{alg:DAA_fair} has the following bound on cumulative unfairness
    $$\E \left[\sum_{t=1}^T \|\bm{y}_t^*-\bm{y}^*\|_2^2\right] \leq O(mn^9\log(T)).$$
\end{proposition}

The proposition states that the expected cumulative distance of our online decision to the offline fair decision has a logarithmic upper bound. The choice of $\bm{y}_t^*$ and $\bm{y}^*$ being analytic center plays an essential role for obtaining this result. To establish this bound, we leverage the fact that the change of the analytic center is continuous with respect to the amount of perturbation on the DLP \eqref{eqn:DLP}. If our average inventory process $\bm{b}_t$ is ``stable'' with respect to $\bm{b}$, and $\hat{\bm{p}}_t$ is close to $\bm{p}$, we can expect that the analytic center of the sampled LP \eqref{eqn:adpt_lp} to be close to $\bm{y}^*$. Therefore, in order to bound the cumulative unfairness, it suffices to characterize the distance of $\bm{b}_t$ to $\bm{b}$ and $\hat{\bm{p}}_t$ to $\bm{p}$. Lastly, we remark that the $n^9$ dependence in this bound is due to the concentration inequality we used and we think there is room for improvement by adopting different techniques or parameter choices. The proof of Proposition \ref{prop_cu} is deferred in Section \ref{sec:pf_prop1}.

For the rest of the section, we will give a more comprehensive discussion on the two modifications we made when deriving Algorithm \ref{alg:DAA_fair}.

\subsection{Fairness Property of Analytic Center}
Since Assumption \ref{assp_dual_binding} does not exclude the situation of the DLP \eqref{eqn:DLP} having multiple and degenerate solutions, one would naturally raise the question ``what would be the fair solution for the DLP''? When solving LPs, the basic optimal solution can be viewed as the corner (simplex) solution in the optimal solution set, and can often favor one group over the other. 

For example, consider a dynamic resource allocation problem with three different order types and two limited resources. The orders follow a multinomial distribution with $\bm{p} = [0.3, 0.3, 0.4]$, and each type of order has a reward of $1, 1$, and $2$. The resource consumption vectors for each type of order are $[1,0], [0,1]$, and $[1,1]$, and we have average resources $\bm{b} = [0.2, 0.2]$. This sequential decision making problem has the DLP in the form of 
\begin{equation*}
    \begin{aligned}
   \max \ \ & 0.3y_1 + 0.3y_2 +0.8y_3 \\
    \text{s.t. }\ & 0.3y_1 + 0.4y_3 \leq 0.2 \\ 
    & 0.3y_2 + 0.4y_3  \le 0.2,  \\
    &0\leq y_j \leq 1, \ \ j=1,...,3.
    \end{aligned}
\end{equation*}
The optimal corner (simplex) solutions for the DLP are $\bm{y}_1 = [2/3, 2/3, 0]$ and $\bm{y}_2 = [0,0,1/2]$. None of these solutions are fair because $\bm{y}_1$ allocates all resources to order $1$ and $2$, and $\bm{y}_2$ allocates all resources to order $3$. The analytic center $\bm{y}^* = [0.37, 0.37, 0.22]$ would be a fairer solution because it is not only optimal but also distributes resources across different orders more evenly. The analytic center is uniquely defined and it maximizes the distance to the boundary of the optimal solution set. Intuitively, it could be understood as the average of all optimal solutions.

\begin{figure}[t]
\centering
    \includegraphics[width=16cm, height=4.3cm]{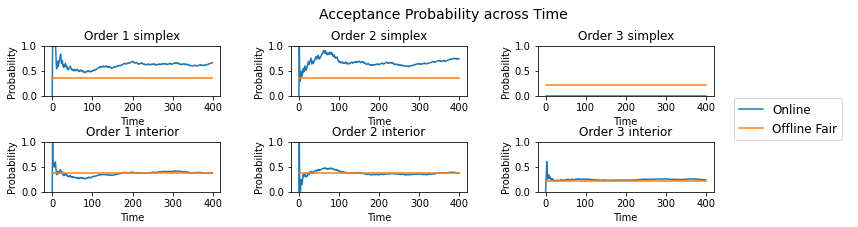}
    \caption{On a simulated environment, the plots show the acceptance probability of each order type across time. For the plots in the first row, the probability is obtained by solving the LP with a simplex type of algorithm. For the plot on the second row, the probability is obtained by solving the LP with an interior-point type of algorithm. }\label{fig:simplex}
\end{figure}

Since the interior-point type solver will output analytic center as solution, we call $\bm{y}^*$ the interior primal solution. To compare the difference, we test Algorithm \ref{alg:DAA_fair} based on a simplex solver and an interior-point type of solver, and plot the acceptance probability of three types of orders across time. From Figure \ref{fig:simplex} we can see the interior-point type of algorithm generate a more evenly distributed allocation rule compared to the simplex solver.
Moreover, we can see the online decision generated from an interior-point solver is converging to the analytic center of DLP.

It seems that we can also define a ``fair'' solution by taking weighted average of all the optimal basic solution. We argue that the analytic center is better than weighted averages, because it is continuous (and Lipchitz) with respect to perturbations of the DLP. It has been shown that if the perturbation on the DLP is small, then the analytic center of the perturbed LP is close to the analytic center of the original LP \citep{holder2001marginal}. Therefore, ensuring the sampled LP being close to the DLP implies that the analytic center of the sampled LP being close to that of the DLP.
Therefore, this property makes analytic center more suitable for being the fair solution.

To conclude, choosing the interior primal solution at time $t$ provides extra consistency and convergence result. As will be shown later, the Lipchitz continuous property of the analytic center is essential to bound the cumulative unfairness. Notice that such continuity result is exclusive for analytic center of primal solutions, and the analytical center of dual solutions might be discontinuous with respect to the same type of perturbation \citep{holder2001marginal}.

\subsection{Addressing Unfairness from Degeneracy}\label{subsec_resource non degen}
Another source of unfairness comes from the degeneracy of non-binding resource. For non-binding resource $i \in \mathcal{N}:=[m]/\mathcal{B}$, $\sum_{j=1}^np_jc_{i,j}y_j^* < b_i$ implies that in expectation, the DLP will not use all the resource $i$, and as the algorithm proceeds the average leftovers will accumulate. Therefore, if we are running Algorithm \ref{alg:DAA}, where at each time we are solving sampled LP \eqref{eqn:adpt_lp} as a proxy of DLP \eqref{eqn:DLP}, as time goes on the average remaining resource $b_{i,t}$ should be larger than the initial average resource $b_i$. As a result, as $t$ increases, with high probability we will solve a sampled LP with more resource budget for those non-binding resource. If the constraint on resource $i$ is degenerate, those agents arriving late in the system will be allocated under more resource budget, and that will make the allocation rule different from that of the agents entering the system early. To eliminate such unfairness across time, we need to modify the algorithm such that every time we are solving a linear program with an inventory constraint process that is more stable.

For example, consider the following linear program.
\begin{equation}\label{eqn:lp_example_b}
    \begin{aligned}
   \max \ \ & 3.6y_A +  0.12y_B \\
    \text{s.t. }\ & 1.2y_A + 0.04y_B \le b_1\\ 
    &  0.6y_A + 4y_B \le b_2  \\ 
    & 0 \le y_A \le 1,  \ \ 0 \le y_B \le 1,
    \end{aligned}
\end{equation}
where we have $\bm{b} = [b_1, b_2] = [1, 2]$. There are two types of order $A$ and $B$, and two types of resource $1$ and $2$ (corresponding to $b_1$ and $b_2$). From the formulation of the LP, type $A$ order will generate more revenue than type $B$ order. In terms of resource consumption, type $A$ consumes more resource $1$ than resource $2$, and type $B$ consumes more resource $2$ than resource $1$.

\begin{figure}[h]
\centering
\begin{subfigure}{.45\textwidth}
  \centering
  \includegraphics[width=.9\linewidth]{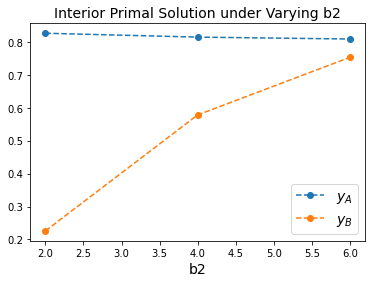}
  \caption{$y_B^*$ as $b_2$ changes}
  \label{fig:solution}
\end{subfigure}
\begin{subfigure}{.45\textwidth}
  \centering
  \includegraphics[width=.9\linewidth]{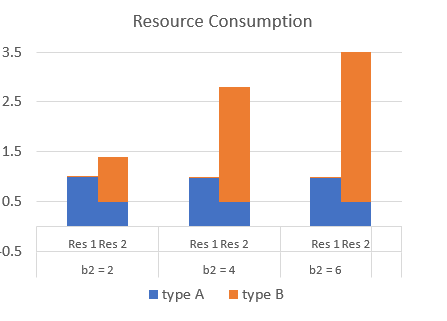}
  \caption{resource consumption as $b_2$ changes}
  \label{fig:resourece}
\end{subfigure}%
\caption{
    Although the second resource in \eqref{eqn:lp_example_b} is non-binding, the nondegenereacy condition still create inconsistent allocation results as $b_2$ increase. Notice that the type $B$ order still gets allocated with a tiny amount of resource $1$, but is getting more resource $2$ as $b_2$ increases. 
}\label{fig:b_nonbinding}
\end{figure}

Observing the above, we can determine that resource $1$ is more preferred than resource $2$. Actually, the interior primal solution $\bm{y}^*$ confirms this finding. Under the setup $\bm{b} = [1, 2]$, we have $\bm{y}^* = [0.826, 0.226]$, and the resource consumption condition is
\begin{equation*}
    \begin{aligned}
     1.2y_A^* + 0.04y_B^* &= b_1\\ 
     0.6y_A^* + 4y_B^* &< b_2.  \\ 
    \end{aligned}
\end{equation*}
Notice that this LP satisfy Assumption \ref{assp_dual_binding}, however it has degeneracy on the second resource. As indicated in Figure \ref{fig:b_nonbinding}, if we fix $b_1$ to be a constant and increase $b_2$, we will find $y_B^*$ increases as $b_2$ goes up. It is important to notice that the total revenue will always be the same with the increase of $b_2$, and the problem is that the optimal solution set will be different, thereby changing the analytic center.
 
\begin{figure}[h]
\centering
\begin{subfigure}{0.45\textwidth}
  \centering
    \includegraphics[width=.9\linewidth]{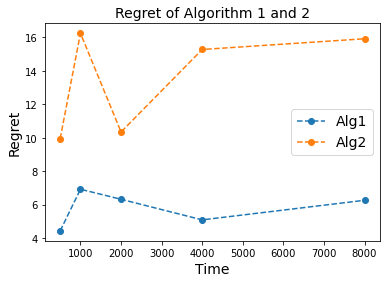}
    \label{fig:eps_test}
\end{subfigure}%
\begin{subfigure}{0.45\textwidth}
  \centering
    \includegraphics[width=1\linewidth]{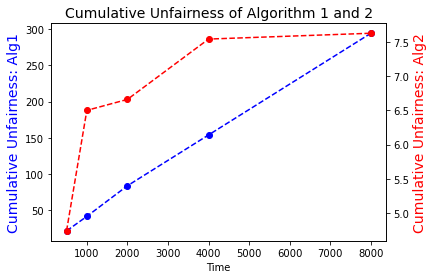}
    \label{fig:efficient_test_S}
\end{subfigure}
\caption{Performance of Algorithm \ref{alg:DAA} and Algorithm \ref{alg:DAA_fair} for the online linear programming problem that has fluid approximation LP of the form \eqref{eqn:lp_example_b}. The number of trails is $30$ and we record the sample mean. We can see that both algorithms have bounded regret. However there is a dramatic difference in the cumulative unfairness, due to the existence of degenerate non-binding resource. 
}
\label{fig:sec4_alg}
\end{figure}

\subsection{Details on Fair Algorithms}
In this section we elaborate on the intuition of choosing the ``right'' inventory process for the fair algorithm. We will also compare the performance of Algorithm \ref{alg:DAA} to our modified algorithm in figure \ref{fig:sec4_alg} based on example \eqref{eqn:lp_example_b}.
As mentioned earlier, with high probability the average remaining non-binding resource will increase as the algorithm proceed. This gives inconsistent allocation rule across time for those orders that consume more non-binding resources than binding resource.

Our solution is to modify the inventory process for the right-hand-side of \eqref{eqn:adpt_lp}. An intuitive way is to replace those non-binding inventory process to be the same as the initial inventory process. In this way, we can always have a consistent allocation rule for the non-binding resources. However, the obstacle is that we do not know the categorization of binding and non-binding resources, and have to learn it from the distribution as the algorithm proceeds.

More specifically, at every time $t$, when we obtain $\bm{y}_t^*$, the interior primal solution for sample LP $\eqref{eqn:adpt_lp}$, we should check the binding resource and the non-binding resource, and denote the sets $$\mathcal{B}_t = \left\{i\left|b_{i,t} = \sum_{j=1}^n\hat{p}_{j,t}\bm{c}_jy_{j,t}^*\right.\right\}$$ and  $\mathcal{N}_t = [m]/\mathcal{B}_t$. Next, we pretend all the resource in set $\mathcal{B}_t$ to be binding and all the resource in set $\mathcal{N}_t$ to be non-binding. Define $\bm{b}'_t$ such that
\begin{align*}
    \bm{b}'_{\mathcal{B}_t, t} = \bm{b}_{\mathcal{B}_t, t},\,\,\,\,\bm{b}'_{\mathcal{N}_t, t} = \bm{b}_{\mathcal{N}_t},
\end{align*}
where $\bm{b}_{t}$ is the regular average inventory process and $\bm{b}$ is the initial average inventory process. We re-solve the LP again using $\bm{b}'_t$ instead of $\bm{b}_t$ for the right-hand-side of \eqref{eqn:adpt_lp}. Under the sampled LP with different constraint $\bm{b}'_t$, we calculate the interior primal solution $\bm{y}_t^*$, and use it to make the probabilistic allocation at time $t$.

We remark that although we use $\bm{b}_t'$ for the decision $\bm{y}^*_t$, we still have to maintain the regular inventory process $\bm{b}_t$. The reasons are twofolds. First, we have to keep track of the real inventory process to determine the end time for the algorithm. Second, the process $\bm{b}_t$ is essential for proving that the modified algorithm still have bounded regret.

Our approach is justified by Corollary \ref{cor_bindingness}. We know that once we gain enough sample from the distribution, the sample estimates $\hat{\bm{p}}_t$ will concentrate well to $\bm{p}$. With good concentration, Corollary \ref{cor_bindingness} ensures that we can correctly learn the binding/non-binding structure with high probability.

We compare Algorithm $\ref{alg:DAA_fair}$ to Algorithm $\ref{alg:DAA}$ in the previous example \eqref{eqn:lp_example_b}. For the online environment, by setting $\bm{p} = [0.6, 0.4]$, $\bm{\mu} = [6, 0.3]$, $\bm{c}_1 = [2, 1]$, and $\bm{c}_2 = [0.1, 10]$ we know that \eqref{eqn:lp_example_b} is the underlying fluid approximation LP. From Figure \ref{fig:sec4_alg}, we find that both regrets are bounded, and the regret of Algorithm \ref{alg:DAA_fair} is higher than Algorithm \ref{alg:DAA}. This is expected since there are chances that Algorithm \ref{alg:DAA_fair} mistakenly takes the binding resource as non-binding and that will terminate the algorithm faster. There is a drastic difference between the cumulative unfairness of Algorithm \ref{alg:DAA} and Algorithm \ref{alg:DAA_fair}. We find that the cumulative unfairness of Algorithm \ref{alg:DAA} is linear while its counterpart in Algorithm \ref{alg:DAA_fair} is logarithmic.

\section{Technical Analysis}
\label{sec_reg_general}
In this section we discuss the the general approach for proving the fairness result and the bounded regret. Because the proof of fairness result (Proposition \ref{prop_cu}) partially relies on the proof of bounded regret (Theorem \ref{thm_fair_regret}), we discuss the proof of Theorem \ref{thm_fair_regret} first.

We briefly discuss the main idea of the proof. Since there are two random variables $\hat{\bm{p}}_t$ and $\bm{b}_t$ for our algorithm, we want to find a ``stable region'' for those two processes such that as long as the processes are within the stable region, our regret will remain bounded. Under the Assumption \ref{assp} and \ref{assp_dual_binding}, we determine the stable region in Lemma \ref{lem_region_subset}. Next, based on the stable region, we analyze the exit time of $\hat{\bm{b}}_t$ out of the region and the concentration behavior of $\hat{\bm{p}}_t$ to $\bm{p}$. More specifically, Corollary \ref{cor_general_regret_form} (based on Proposition \ref{prop_adaptive_regret_form}) characterizes the regret as a summation of two components. In Lemma \ref{lem_bounding_acceptance}, we prove that the first component is bounded based on the notion of stable region and analytic center; The second component is analyzed in Proposition \ref{prop_stoppingtime}. It involves characterizing the exit time of $\bm{b}_t$ out of the stable region. Notice that $\bm{b}_t$ is not a martingale and we approximate it by a martingale-like process and analyze the distance. For the result on cumulative unfairness, we use continuity argument (see Lemma \ref{lem_lipchitz}) of the analytic center. It turns out that the analytic center of the primal solution is Lipchitz continuous with respect to random perturbation of the sampled LP. Therefore, we can bound the cumulative unfairness by bounding the cumulative deviation of the random process $\bm{b}_t$ and $\hat{\bm{p}}_t$.
\subsection{Setup}
To facilitate the analysis, we introduce the standard form of the DLP $\eqref{eqn:DLP}$. With a slight abuse of notation (omitting the probability vector $\bm{p}$), we use $\bm{\mu}$ to denote the vector $(p_1\mu_1, ..., p_n\mu_n)^\top$ and $\bm{C}$ to denote the matrix $(p_1\bm{c}_1, ..., p_n\bm{c}_n)$. The standard form of the DLP $\eqref{eqn:DLP}$ can be written compactly as

\begin{equation}
    \begin{aligned}
   \max \ \ & \bm{\mu}^\top \bm{y} \label{eqn:PSDDLP}  \\
    \text{s.t. }\ & \bm{C}\bm{y} + \bm{s} = \bm{b}   \\
    & \bm{y} + \bm{z} = \bm{1}\\
    & \bm{y}, \bm{s}, \bm{z} \geq \bm{0},
    \end{aligned}
\end{equation}
where $\bm{s} \in \mathbb{R}^m$ and $\bm{z} \in \mathbb{R}^n$ are slack variables.
Notice that if we denote $I_m$ and $I_n$ the corresponding identity matrix of dimension $m$ and $n$, and define $\bar{\bm{C}}\in \mathbb{R}^{(m+n)\times(m+2n)}, \bar{\bm{y}}, \bar{\bm{\mu}}\in\mathbb{R}^{(m+2n)}, \bar{\bm{b}}\in\mathbb{R}^{(m+n)}$ as

\[\bar{\bm{C}} :=
\begin{pmatrix}
   \bm{C} & I_m & 0\\
   I_n & 0 & I_n
    \end{pmatrix},\,\,\,\,\,
    \bar{\bm{y}}
    =    \begin{pmatrix}
   \bm{y}\\
   \bm{s}\\
   \bm{z}
    \end{pmatrix} ,\,\,\,\,\,
    \bar{\bm{\mu}}
    =    \begin{pmatrix}
   \bm{\bm{\mu}}\\
   \bm{0}\\
   \bm{0}
    \end{pmatrix},\,\,\,\,\,
    \bar{\bm{b}}
    =   \begin{pmatrix}
   \bm{b}\\
   \bm{1}\\
    \end{pmatrix},\,\,\,\,\,
\]
then $\eqref{eqn:PSDDLP}$ could be reformulated as
\begin{equation}
    \begin{aligned}
   \max \ \ & \bar{\bm{\mu}}^\top\bar{\bm{y}} \label{eqn:SDDLP}  \\
    \text{s.t. }\ & \bar{\bm{C}}\bar{\bm{y}} = \bar{\bm{b}}  \\
    & \bar{\bm{y}} \geq \bm{0}.
    \end{aligned}
\end{equation}
This standard form of LP will be useful when we analyze the cumulative unfairness. Then, we note that the dual of \eqref{eqn:SDDLP}  is the same as the dual \eqref{eqn:DDLP} of the original DLP. From its alternative form \eqref{eqn:ADDLP},
we know that the dual program will always have feasible region and the dual optimal solution exist, which implies the existence of optimal primal solution. 

Next, we focus on setting up the notations for analyzing the stable region of $\bm{b}_t$. Recall that $\bm{y}^*$ is the analytic center of the solution in \eqref{eqn:DLP}. Similarly, we denote $\bm{y}^*(\bm{b}')$ as the analytic center of the solution in \eqref{eqn:DLP} with right-hand-side replaced by $\bm{b}'$,
\begin{equation*}
    \begin{aligned}
    \mathcal{B} &:= \{i: b_i - (\bm{C}\bm{y}^*)_i= 0\}\\
    \mathcal{B}(\bm{b}') &:= \{i: b_i' - (\bm{C}\bm{y}(\bm{b}')^*)_i= 0\}\\
    \mathcal{N} &:= [m]/\mathcal{B}, \,\,\,\mathcal{N}(\bm{b}') = [m]/\mathcal{B}(\bm{b}'),
    \end{aligned}
\end{equation*}
where $\mathcal{B}$ stands for binding and $\mathcal{N}$ stands for non-binding. The intuition of the definition is that we define the resource $i$ to be binding if and only if for any optimal solution $\bm{y}$, we always have the $i$-th inventory constraint being zero, i.e. $b_i - (\bm{C}\bm{y})_i = 0$.
Because the analytic center of the optimal phase can be interpreted as the ``average'' of optimal solutions, it suffices to define the notions in the way above.

Moreover, we denote the basic optimal solution set for $\eqref{eqn:ADDLP}$ to be $\mathcal{I}^*$ such that 
\begin{equation*}
    \begin{aligned}
    \mathcal{I}^* &:= \{\bm{\lambda} | \bm{\lambda} \text{ is the basic optimal solution for } \eqref{eqn:ADDLP} \}\\
    \mathcal{I}^*(\bm{b}') &:= \{\bm{\lambda} | \bm{\lambda} \text{ is the basic optimal solution for } \eqref{eqn:ADDLP} \text{ with input } \bm{b}' \},\\
    \end{aligned}
\end{equation*}
and most importantly, 
\begin{equation*}
    \begin{aligned}
    \mathcal{B}_{\text{strict}} &:= \cap_{\bm{\lambda\in\mathcal{I}^*}}\{i:\lambda_i > 0\} \\
    \mathcal{B}_{\text{strict}}(\bm{b}') &:= \cap_{\bm{\lambda\in\mathcal{I}^*(\bm{b}')}}\{i:\lambda_i > 0\}. \\
    \end{aligned}
\end{equation*}
We have to define the notion of ``strict binding'' in this way because for the analytic center of the dual \eqref{eqn:DDLP}, $\lambda^*_i > 0$ does not necessarily mean $\lambda_i > 0$ for any optimal $\bm{\lambda}$, and this notion is quite essential in defining the condition for a stable region.
We then state an lemma as an implication of Assumption \ref{assp_dual_binding}.
\begin{lemma}\label{lem_binding}
    Under Assumption \ref{assp} and \ref{assp_dual_binding}, there exists a constant $\Delta > 0$ and a region defined as $\left(\bigotimes_{i\in\mathcal{B}} \left[b_{i}-\Delta, b_{i}+\Delta\right]\right)\bigotimes \left(\bigotimes_{i\in\mathcal{N}} \left[b_{i}-\Delta, +\infty\right)\right)$
    such that for any $\bm{b'}$ in this region, $\mathcal{B} = \mathcal{B}(\bm{b}') = \mathcal{B}_{\text{strict}} = \mathcal{B}_{\text{strict}}(\bm{b}')$.
\end{lemma}

The property $\mathcal{B} = \mathcal{B}_{\text{strict}}$ is very important because if not, we can still have $\lambda_i = 0$ while $b_i - (\bm{C}\bm{y}^*)_i = 0$. Imposing such requirement for all $\bm{b}' \in \Omega$ ensure the non-degeneracy condition for all $i \in \mathcal{B}$.

\subsection{Decomposing the Regret}
To see the relationship between the dual formulation and the regret, we define $N_i^a(T)$ to be the number of accepted order $i$ during time horizon $T$. Moreover, we denote the optimal solution set for $\eqref{eqn:ADDLP}$ to be $\mathcal{D}^*$ such that 
$$\mathcal{D}^* = \{\bm{\lambda} | \bm{\lambda} \text{ is the optimal solution for } \eqref{eqn:ADDLP} \}.$$
Next, we state a proposition for the regret decomposition formula. This proposition is not directly applicable to our conditions, but the main idea is relevant.

\begin{proposition}[\cite{chen2021linear}]\label{prop_general_regret_form}

Under Assumption \ref{assp}, for any $\bm{\lambda} \in \mathbb{R}^m$ such that $\bm{\lambda} \in \mathcal{D}^*$, we have
\begin{equation}\label{eqn_general_regret_form}
    \begin{aligned}
    \text{Reg}_{T}^\pi &\leq \bm{\lambda}^{\top} \E\left[\bm{B}_{\tau}\right] + \E\left[\sum_{j=1}^n (Tp_j
    - N_j^a(T))(\mu_j - \bm{c}_j^\top\bm{\lambda})_+\right] + \E\left[ \sum_{j=1}^nN_j^a(T)(\bm{c}_j^\top\bm{\lambda} - \mu_j)_+ \right].\\
    \end{aligned}
\end{equation}
\end{proposition}
Here, $\tau$ is the stopping time that some resource is depleted, and $\bm{B}_\tau$ denotes the remaining resource vector when the algorithm terminates. Thus the first part on the right-hand-side of \eqref{eqn_general_regret_form} represents a penalization for wasting resources. In particular, the first term suggests that only wasting the resources such that $\lambda_i > 0$ will be penalized.
As to the second and third terms on the right-hand-side of \eqref{eqn_general_regret_form}, they capture the costs induced by order acceptance. More specifically, the last two terms in  \eqref{eqn_general_regret_form} helps to identify the orders into three different categories. For any $\bm{\lambda} \in \mathcal{D}^*$, we can have the following categories of order

\begin{itemize}
    \item All-accepted orders: $\bm{\mu}_j-\bm{c}_j^\top \bm{\lambda}>0$. For these orders, the optimal decision should be to accept all of them. On expectation, we will observe $Tp_j$ such orders throughout the horizon and aim to have the number of acceptance $N_{j}^a(T)$ to be close to that. 
    \item All-rejected orders:  $\bm{\mu}_j-\bm{c}_j^\top \bm{\lambda}<0$. On the opposite of the previous case, the optimal decision should be to reject all of these orders. Each acceptance of such order will induce a cost of $\bm{c}_j^\top \bm{\lambda}-\mu_j$: resources of value $\bm{c}_j^\top \bm{\lambda}$ are spent, but only reward of value $\mu_j$ is received.
    \item Partially-accepted orders: $\bm{\mu}_j-\bm{c}_j^\top \bm{\lambda}=0$. The condition may lead to a partial acceptance of the orders, i.e., $0\le y_j\le 1$. In terms of regret analysis, there is no need to worry about these orders because they do not contribute to the right-hand-side of \eqref{eqn_general_regret_form}.
\end{itemize}

Notice that the optimal primal solution $\bm{y}$ is responsible for making the decisions, while the optimal dual solution $\bm{\lambda}$ is used for prescribing the accept/reject rule to bound the regret. If the solution for \eqref{eqn:SDDLP} is unique and non-degenerate, from complementary slackness condition we know that we must have $y_j^* = 1$ (accept with probability one) corresponding to $\mu_j > \bm{c}_j^\top \bm{\lambda}^*$ and $y_j^* = 0$ (reject with probability one) corresponding to $\mu_j < \bm{c}_j^\top \bm{\lambda}^*$. This relationship is consistent with our characterization of the categories above, and the last two terms of $\eqref{eqn_general_regret_form}$ can be bounded. However, once the solution is non-unique and degenerate, if we choose arbitrary optimal primal solution $\bm{y}$ and optimal dual solution $\bm{\lambda}$, and we will not always get the property that $y_j = 1$ corresponding to $\mu_j > \bm{c}_j^\top \bm{\lambda}$.

For this reason, we present another regret decomposition formula that has the freedom of choosing $\bm{\lambda}$ adaptively. We define $\bm{\lambda}_{max} \in \mathbb{R}^m$ such that 
\begin{equation*}
    \begin{aligned}
    \lambda_{max, i} = 
    \begin{cases}
    \max\{||\bm{\lambda}||_{\infty}|\bm{\lambda} \in \mathcal{D}^*\} \,\,&\text{if } \exists \bm{\lambda}\in\mathcal{D}^* \,\,s.t.\,\, \lambda_i > 0,\\
    0 &\text{otherwise,}
    \end{cases}
    \end{aligned}
\end{equation*}
define $\mathcal{H}_{t}$ to be the information generated by $\{(r_s, \bm{a}_s)\}_{s=1}^{t}$ up to $t$, and state the following regret decomposition formula.
\begin{proposition}\label{prop_adaptive_regret_form}
Under Assumption \ref{assp}, for $\{\bm{\lambda}_t\}_{t=1}^T$ such that $\bm{\lambda}_t$ is $\mathcal{H}_{t-1}$-measurable and $\bm{\lambda}_t \in \mathcal{D}^*$, we have
\begin{equation*}
    \begin{aligned}
    \text{Reg}_{T}^\pi &\leq \bm{\lambda}_{max}^{\top} \E\left[\bm{B}_{\tau}\right] + \sum_{t=1}^T\E\left[(r_t - \bm{a}_t^\top\bm{\lambda}_t)_+(1-x_t)\right] + \sum_{t=1}^T\E\left[(\bm{a}_t^\top\bm{\lambda}_t - r_t )_+x_t\right].\\
    \end{aligned}
\end{equation*}
\end{proposition}

We defer the proof of the proposition to \ref{sec:pf_prop_regretdecom}. The difference of the above proposition to Proposition \ref{prop_general_regret_form} is that, here we are using $\bm{\lambda}_t$ that is adaptive and optimal, instead of fixing an accept/reject rule in Proposition \ref{prop_general_regret_form} and accumulate the mistakes for each order. By choosing $\bm{\lambda}_t$ adaptively we can have more flexibility when we need to bound the regret.

Our adaptive procedure can be summarized as follows. At time $t$, we choose an optimal dual solution $\bm{\lambda}_t$ that gives us an accept/reject rule, and then choose optimal primal solution $\bm{y}_t$ (which generates $x_t$) that solves the perturbed LP $\eqref{eqn:adpt_lp}$ and is consistent with the accept/reject rule. Proposition \ref{prop_adaptive_regret_form} implies that an appropriate optimal dual price $\bm{\lambda}_t$ has to satisfy the following properties:
\begin{itemize}
    \item $\bm{\lambda}_t \in \mathcal{D}^*$: This is because by setting $\bm{\lambda}_t \in \mathcal{D}^*$, the duality gap between the primal and dual of the DLP is zero. Any other dual solution will result in a linear regret hence useless for our analysis.
    \item Consistent with $\bm{y}_t$: Once $\bm{\lambda}_t$ is determined, an algorithm features low regret should have $\bm{y}_{j,t} = 1$ for the order such that $\mu_j > \bm{c}_j^\top\bm{\lambda}_t$, and $\bm{y}_{j,t} = 0$ for the order such that $\mu_j < \bm{c}_j^\top\bm{\lambda}_t$.
\end{itemize}

Next, we introduce the following corollary that separates the upper bound of the regret into two components. For now take $\tau_S$ to be a stopping time such that $\tau_S \leq \tau < T$ a.s., and we have 

\begin{corollary}
\label{cor_general_regret_form}
Under Assumption \ref{assp}, for $\{\bm{\lambda}_t\}_{t=1}^T$ such that $\bm{\lambda}_t$ is $\mathcal{H}_{t-1}$-measurable and $\bm{\lambda}_t \in \mathcal{D}^*$, we have the following inequality holds 
\begin{align}
    \text{Reg}_{T}^\pi  & \leq \bm{\lambda}_{max}^{\top} \E\left[\bm{B}_{\tau_S}\right]
   +\left(T -\E[\tau_S]\right)\cdot \max_{j\in[n], \bm{\lambda}\in\mathcal{D}^*}|\mu_j-\bm{c}_j^\top \bm{\lambda}| \label{eqn:cor_regret_line1} \\ & \ \ \ \  \ \ \ + \sum_{t=1}^{\tau_S}\E\left[(r_t - \bm{a}_t^\top\bm{\lambda}_t)_+(1-x_t)\right] + \sum_{t=1}^{\tau_S}\E\left[(\bm{a}_t^\top\bm{\lambda}_t - r_t )_+x_t\right]\label{eqn:cor_regret_line2}.
\end{align}
\end{corollary}

From Corollary \ref{cor_general_regret_form}, we can see that the first component \eqref{eqn:cor_regret_line1} involves bounding the expected stopping time $\mathbb{E}[T-\tau_S]$, and if we can bound the stopping time, $\mathbb{E}[\bm{B}_{\tau_S}]$ can be bounded easily. For the second component \eqref{eqn:cor_regret_line2}, we have to bound $\sum_{t=1}^{\tau_S}\E\left[(r_t - \bm{a}_t^\top\bm{\lambda}_t)_+(1-x_t)\right] + \sum_{t=1}^{\tau_S}\E\left[(\bm{a}_t^\top\bm{\lambda}_t - r_t )_+x_t\right]$, which is to accept/reject the right order following the accept/reject rule prescribed by $\bm{\lambda}_t$.

In the following section, we are going to find a ``stable region'' that helps to define $\tau_S$ exactly and $\mathbb{E}[T -\tau_S] = O(1)$. We will also characterize $\{\bm{\lambda}_t\}_{t=1}^T$ such that $\bm{\lambda}_t \in \mathcal{H}_{t-1}$ and
$$\sum_{j=1}^T\E\left[(r_t - \bm{a}_t^\top\bm{\lambda}_t)_+(1-x_t)\right] + \sum_{j=1}^T\E\left[(\bm{a}_t^\top\bm{\lambda}_t - r_t )_+x_t\right] = O(1)$$
for $x_t$ generated by Algorithm \ref{alg:DAA_fair}.

\subsection{Determine the stable region}
In this section we characterize the stable region $\mathcal{K}$. This region can be viewed as a neighbourhood of $\bm{p}$ and $\bm{b}$, and characterizes ranges for the processes $\hat{\bm{p}}_t$ and $\bm{b}_t$ such that as long as $\hat{\bm{p}}_t$ and $\bm{b}_t$ belongs to the region, the sampled LP \eqref{eqn:adpt_lp} is ``close'' to the DLP \eqref{eqn:DLP}. With this condition we are able to provide a proof of the bounded regret based on the decomposition from Corollary \ref{cor_general_regret_form}.

\begin{lemma}\label{lem_region_subset}
There exists problem dependent constant $L > 0$ depending on the DLP (\eqref{eqn:DLP} or \eqref{eqn:SDDLP}) such that the region $\mathcal{K}$, as a domain of $(\hat{\bm{b}}, \hat{\bm{p}})$ defined by
\begin{equation}\label{eqn_region_K}
    \begin{aligned}
     \hat{b}_i &\in
        \begin{cases}
        [b_i - L, +\infty),\,\,\, i\in\mathcal{N}\\
        [b_i - L, b_i + L],\,\,\, i\in\mathcal{B}
        \end{cases}\\
     \hat{p}_{j} &\in [p_j - L, p_j + L],\,\,\, j = [n],
    \end{aligned}
\end{equation}
has the property that for every $(\bm{b}_t,\hat{\bm{p}}_t) \in \mathcal{K}$, the dual optimal solution set of the sampled LP \eqref{eqn:adpt_lp} is a subset of $\mathcal{D}^*$.
\end{lemma}

We leave the details of the proof in \ref{sec:pf_stableregion}. Lemma \ref{lem_region_subset} characterizes the notion of stable region for $\bm{b}_t$ and $\hat{\bm{p}}_t$. The result that the optimal dual solution set of \eqref{eqn:adpt_lp} is a subset of $\mathcal{D}^*$ is equivalent to the statement that the current accept/reject decision prescribed by $\bm{\lambda}_t$ based on the sampled LP is a correct choice corresponding to the DLP. Notice that the stability analysis is different from that of \citet{chen2021linear}. We do not have the unique and non-degenerate condition so we cannot use the similar matrix inequality to get the result. We circumvent this difficulty by analyzing the structure of the dual program, and the fact that the distribution has finite support is essential in the proof. For convenience, we set the constant $L$ to be strictly smaller than $\Delta$ in Lemma \ref{lem_binding}. 

Another implication of the stability result is the following corollary.
\begin{corollary}\label{cor_bindingness}
For the sampled LP \eqref{eqn:adpt_lp}, given $(\bm{b}_t, \hat{\bm{p}}_t) \in \mathcal{K}$, we have that 
\begin{equation*}
    \begin{aligned}
    \left(\sum_{j=1}^n  \hat{p}_{j,t}\bm{c}_j\cdot y_{j,t}-\bm{b}_t\right)_i &> 0 \text{ for $i\in\mathcal{N}$}\\
    \left(\sum_{j=1}^n  \hat{p}_{j,t}\bm{c}_j\cdot y_{j,t}-\bm{b}_t\right)_i &= 0 \text{ for $i\in\mathcal{B}$}.\\
    \end{aligned}
\end{equation*}
Moreover, $L$ can also meet the condition that if we are solving the sampled LP in \eqref{eqn:LP_algorithm_afterfair} with $(\bm{b}_t', \hat{\bm{p}}_t)$ in the ``$\mathcal{K}$-projected'' region defined as
\begin{equation*}
    \begin{aligned}
     \hat{b}_i &\in
        \begin{cases}
        \{b_i\},\,\,\, i\in\mathcal{N}\\
        [b_i - L, b_i + L],\,\,\, i\in\mathcal{B}
        \end{cases}\\
     \hat{p}_{j} &\in [p_j - L, p_j + L],\,\,\, j = [n],
    \end{aligned}
\end{equation*}
we have that for $i \in \mathcal{N}$, $\left(\sum_{j=1}^np_j\bm{c}_{j}\bm{y}_{t}^*\right)_i < b_i - L$.
\end{corollary}

Corollary \ref{cor_bindingness} states that under the stable region, the sampled LP's solution features the same binding/nonbiding structure for the resource constraints as in the DLP \eqref{eqn:DLP}. Moreover, the second statement concerns $\mathbb{E}[\bm{a}_tx_t] = \sum_{j=1}^np_j\bm{c}_{j}\bm{y}_{t}^*$, where $x_t$ is based on $\bm{y}_t^*$, the solution of the sampled LP. It ensures that the expected resource consumption for the non-binding resource $i\in\mathcal{N}$ is lower than $b_i - L$, given that we set the non-binding resources $\bm{b}_{\mathcal{N},t}'$ to be the initial average resource $\bm{b}_{\mathcal{N}}$. This condition will be useful in Section \ref{sec:pf_maintext} where the dynamics of the average process $\bm{b}_t$ is studied. The proof can be found in \ref{sec_ap_cor2}.

With the results on the stable region, we can work on bounding the regret in the next section. 

\subsection{Bound for Regret}\label{sec:pf_maintext}
In this section, we first discuss bounding the second component \eqref{eqn:cor_regret_line2} in Corollary \ref{cor_general_regret_form}, and then derive the bound for the first component \eqref{eqn:cor_regret_line1}. 
From Lemma \ref{lem_binding} we know that there exists region depending on $\Delta$ such that the bindingness structure of \eqref{eqn:DLP} remains the same when $\bm{b}' \in \left(\bigotimes_{i\in\mathcal{B}} \left[b_{i}-\Delta, b_{i}+\Delta\right]\right)\bigotimes \left(\bigotimes_{i\in\mathcal{N}} \left[b_{i}-\Delta, +\infty\right)\right)$. Combined with Lemma \ref{lem_region_subset}, without loss of generality we assume that $L < \Delta$ and take
\begin{align*}
\Omega \coloneqq \left(\bigotimes_{i\in\mathcal{B}} \left[b_{i}-L, b_{i}+L\right]\right)\bigotimes \left(\bigotimes_{i\in\mathcal{N}} \left[b_{i}-L, +\infty\right)\right).
\end{align*}
We also precisely define the stopping time $\tau_S$ mentioned in Corollary \ref{cor_general_regret_form} as the exit time of $\bm{b}_t$ out of $\Omega$
$$\tau_S \coloneqq \min \left\{t\le T: \left|b_{it}-b_i\right| > L  \text{ for some }i\in\mathcal{B} \right\} \cup \left\{t\le T: b_{it}-b_i < - L \text{ for some }i\in\mathcal{N} \right\} \cup \{T+1\}.$$

Next, we present the result on the bound of \eqref{eqn:cor_regret_line2}.

\begin{lemma}\label{lem_bounding_acceptance}
    Under Assumption \ref{assp} and \ref{assp_dual_binding}, Algorithm \ref{alg:DAA_fair} gives the following bound
    \begin{equation*}
        \begin{aligned}
        \sum_{t=1}^{\tau_S}\E\left[(r_t - \bm{a}_t^\top\bm{\lambda}_t)_+(1-x_t)\right] + \sum_{t=1}^{\tau_S}\E\left[(\bm{a}_t^\top\bm{\lambda}_t - r_t )_+x_t\right] \leq \frac{2n\cdot\max_{j\in[n],\bm{\lambda}\in\mathcal{D}^*}|\mu_j - \bm{c}_j^\top\bm{\lambda}|}{1-\exp(-2L^2)}.
        \end{aligned}
    \end{equation*}
\end{lemma}
Notice that in Algorithm \ref{alg:DAA_fair}, the choice of interior primal ($\bm{y}_t^*$) and dual ($\bm{\lambda}_t^*$) solution of the sampled LP \eqref{eqn:adpt_lp} plays an essential role for the result. With Lemma \ref{lem_region_subset}, we can show that the event $\{(\bm{b}_t, \hat{\bm{p}}_t) \in \mathcal{K}\}$ will happen with high probability. Under the stable region $\mathcal{K}$, our interior dual solution $\bm{\lambda}_t^*$ will characterize the accept/reject type of order according to the sampled LP \eqref{eqn:adpt_lp} at time $t$, and the interior solution $\bm{y}_t^*$ will make sure that we accept the order $j$ such that $\bm{\mu}_j - \bm{c}_j^\top\bm{\lambda}_t^* > 0$ and reject the order $j$ such that $\bm{\mu}_j - \bm{c}_j^\top\bm{\lambda}_t^* < 0$. Moreover, Lemma \ref{lem_region_subset} makes sure that when $(\bm{b}_t, \hat{\bm{p}}_t) \in \mathcal{K}$, the characterization of the accept/reject type based on the sampled LP \eqref{eqn:adpt_lp} is also a ``correct'' characterization based on the DLP. Therefore, with high probability, we will make the right accept/reject decision so that \eqref{eqn:cor_regret_line2} will be bounded. We defer the details for the proof of the lemma to Appendix \ref{ap:pf_acceptance}.

With Lemma \ref{lem_bounding_acceptance} and Corollary \ref{cor_general_regret_form}, to prove the bounded regret it remains to show that 
$$\bm{\lambda}_{max}^{\top} \E\left[\bm{B}_{\tau_S}\right]
   +\left(T -\E[\tau_S]\right)\cdot \max_{j\in[n], \bm{\lambda}\in\mathcal{D}^*}|\mu_j-\bm{c}_j^\top \bm{\lambda}| = O(1),$$
which involves analyzing the exit time of $\bm{b}_t$ out of $\Omega$. The main approach is similar to Appendix C of \cite{chen2021linear}, and the difference is that in this paper we are using a fairer inventory constraint $\bm{b}_t'$ for the allocation policy (see \eqref{eqn:LP_algorithm_afterfair}). Generally speaking, we define another process $\tilde{\bm{b}}_t$ which is similar to a martingale, and is close to the process $\bm{b}_t$ with high probability. To ensure the closeness, we characterize high probability events under which the process $\bm{b}_t$ is well behaved. Then, the exit time of $\bm{b}_t$ is related to the exit time of $\tilde{\bm{b}}_t$ and some high probability events that we will define below.

Now we formally define event $\mathcal{E}_t$ as follows. Let
$$\epsilon_t \coloneqq \begin{cases}
1 & t \leq \kappa T, \\
\frac{1}{t^{1/4}} & t >  \kappa T,
\end{cases}$$
where $\kappa$ is a positive constant that will be specified in Proposition \ref{prop_stoppingtime}. Next, denote the event 
\begin{equation*}
    \begin{aligned}
    \mathcal{E}_t &\coloneqq \left\{\mathcal{H}_{t-1}\Big \vert \sup_{\bm{b}'\in \Omega }\left\|\mathbb{E}[\bm{a}_{\mathcal{B}, t}x_{t}(\bm{b}')|\mathcal{H}_{t-1}]-\bm{b}'_{\mathcal{B}}\right\|_\infty \le \epsilon_{t-1} \right\} \cap \left\{\mathcal{H}_{t-1}\Big \vert||\hat{\bm{p}}_t - \bm{p}||_{\infty}\leq L \text{ for } t>\kappa T\right\}\\
    \end{aligned}
\end{equation*}
where the subscript $\mathcal{B}$ denotes the corresponding dimensions of the vectors.

Now, we provide some intuitions for the definition of $\mathcal{E}_t.$ Ideally, for the binding resouces, we hope that the expected resource consumption at each time $t$ stays close to $\bm{b}_t.$ Because this will prevent the resources from an early depletion or a left-over at the end of the horizon. As for the condition $||\hat{\bm{p}}_t - \bm{p}||_{\infty} \leq L$, combined with the condition $\bm{b}_t\in\Omega$ introduced later, we can ensure the stable region so that the binding/nonbinding structure of the sampled LP stays the same as the DLP (as ensured by Corollary \ref{cor_bindingness}). This helps to deal with the effect of step $8$ and $9$ in Algorithm \ref{alg:DAA_fair} to the stability of the process $\bm{b}_t$.

Here we only define tolerance level $\epsilon$ for binding dimensions, because for non-binding dimensions with high probability the average inventory process will be increasing, and we can tolerant larger deviation for the resource consumption as long as it does not sabotage the non-bindingness. In fact, the condition $||\hat{\bm{p}}_t - \bm{p}||_{\infty} \leq L$ will ensure such requirement. The event $\mathcal{E}_t$  characterizes the algorithm behavior regarding to this aspect, and it defines a ``good'' event: At time $t$, if $\bm{b}_t' \in \Omega$, the binding structure for resources is the same as the DLP \eqref{eqn:DLP}, and $\mathbb{E}[\bm{a}_{\mathcal{B},t}x_{t}(\bm{b}')|\mathcal{H}_{t-1}]$, the expected binding resource consumption at time $t$, stays close to $\bm{b}_{\mathcal{B}, t}'$ with a range of $\epsilon_{t-1}$. 

To justify our choice of $\epsilon_t$, in the first $\kappa T$ time periods, since we still have a lot of resources, we can tolerate a relatively large deviation, and the event $\mathcal{E}_t$
will happen almost surely for the first $\kappa T$ time periods. For $t>\kappa T$, we define $\epsilon_t = \frac{1}{t^{1/4}}$. Intuitively, at time $t$, the estimation error for the binding resource is on the order of $\frac{1}{t^{1/2}}$, which is higher. We remark two points on the choice of $\epsilon_t$: (i) it guarantees that the event $\mathcal{E}_t$ will happen with high probability; (ii) its value is not too large that the adaptive (re-solving) mechanism can still ensure the stability of the process $\bm{b}_t.$ For completeness, we define $\mathcal{E}_0$ as the whole space.

Now we define our approximation process $\tilde{\bm{b}}_t$. According to the event $\mathcal{E}_t$, we can define the stopping time 
$$\tilde{\tau} \coloneqq \min \{t\le T:\bm{b}_t\notin \Omega \text{ or } \mathcal{H}_{t-1} \notin \mathcal{E}_t\} \cup \{T+1\}.$$
and the approximation process
$$\tilde{\bm{b}}_t = \begin{cases} 
\bm{b}_t, & t<\tilde{\tau},\\
\bm{b}_{\tilde{\tau}}, & t\ge \tilde{\tau}.
\end{cases}$$
These two definitions lead to the following decomposition for the exit time.
\begin{equation}
    \begin{aligned}\label{decompose}
    \prob\left(\tau_S \leq t\right) &= \prob\left(\bm{b}_s\notin \Omega \text{ for some }s\le t \right)\\ 
    &\le \prob\left(\bm{b}_s\notin\Omega \text{ for some }s\le t, \cap_{s=1}^t \mathcal{E}_s \right) + \prob\left(\bm{b}_s\notin\Omega \text{ for some }s\le t, \cup_{s=1}^t \bar{\mathcal{E}}_s \right)\\
    &\le \prob\left(\tilde{\bm{b}}_s\notin\Omega \text{ for some }s\le t \right) + \sum_{s=1}^t \prob(\bar{\mathcal{E}}_s).
    \end{aligned}
\end{equation}

The objective of our analysis is to bound the probability of $\bm{b}_t\notin \Omega$. The inequality (\ref{decompose}) separates the probability into two components. The first component involves the process $\tilde{\bm{b}}_t$, which is a very ``regular'' process in that if $t<\tilde{\tau}$, the process $\tilde{\bm{b}}_t$'s fluctuation is subject to the event $\mathcal{E}_t$ and if $t\ge \tilde{\tau}$, the process freezes. The event $\mathcal{E}_t$ further regulates the fluctuation for the process $\tilde{\bm{b}}_t$, and it makes the process behaves roughly like a martingale. This creates much convenience in analyzing the first component of (\ref{decompose}). Next, the second component is about the probability of $\bar{\mathcal{E}}_t$'s, which can be analyzed individually for each $t$. Overall, the inequality \eqref{decompose} disentangles the stability of the process $\bm{b}_t$ from the estimation error. Its first component concerns the stability of the process given a good estimate, while its second component concerns the probability of obtaining the good estimate for the model parameters. By combining those two estimates, we have the following proposition. Note that $\underline{b} \coloneqq \min_{1 \leq i \leq m} b_i$ is the smallest value of the initial average constraint level $\bm{b}$.

\begin{proposition}\label{prop_stoppingtime}
If we execute Algorithm \ref{alg:DAA_fair} under Assumption \ref{assp} and \ref{assp_dual_binding}, we have
\begin{equation*}
    \begin{aligned}
    \prob(\tau_S \leq t)
    &\leq 2m\exp\left(-\frac{L^2(T-t)}{8}\right) + \frac{n}{L^2}\exp\left(-2L^2\kappa T\right) + 4n^3T^{1/2}\exp\left(-\frac{\kappa^{1/2}}{n^2}T^{1/2}\right)\\
    \end{aligned}
\end{equation*}
where $\kappa$ is set by $\kappa = \left(1-\exp(-\frac{L}{8})\right) \wedge \left(1-\exp(-\frac{L}{8(1+L-\underline{b})})\right)$. Moreover
\begin{equation*}
    \begin{aligned}
    \mathbb{E}[\tau_S]
    &\ge T - 2 - \frac{16m}{L^2} - \cfrac{nT}{L^2} \exp\left(-2L^2 \kappa T\right) - 4n^3T^{3/2}\exp\left({-\frac{\kappa ^{1/2}}{n^2}\cdot T^{1/2}}\right).
    \end{aligned}
\end{equation*}
\end{proposition}
We leave the proof in \ref{sec_ap_stptime}. From Proposition \ref{prop_stoppingtime} we know that $\E[T-\tau_S] = O(1)$, hence \eqref{eqn:cor_regret_line1} has an $O(1)$ bound and Theorem \ref{thm_fair_regret} is true. For more details of the proof of Theorem \ref{thm_fair_regret} see \ref{sec_ap_thm1}.

\subsection{Bound for Cumulative Unfairness} \label{sec_cu}
In this section we begin to bound the cumulative unfairness. With the general DLP form \eqref{eqn:SDDLP}, we can transfer the sampled LP \eqref{eqn:adpt_lp} to another LP that is in similar form of \eqref{eqn:SDDLP} and the perturbation is on the right-hand-side constraint ($\bar{\bm{b}}$) only. In contrast, for the sampled LP \eqref{eqn:adpt_lp} the perturbations are both in the left-hand-side ($\hat{\bm{p}}_t$) and right-hand-side ($\bm{b}_t$). The reason we are formulating the sampled LP in this alternative form is that for the LP \eqref{eqn:SDDLP},  $\bar{\bm{y}}^*$, the analytic center of the optimal solution, is continuous with the right-hand-side constraint $\bar{\bm{b}}$. Therefore, we can use the cumulative perturbation on $\bar{\bm{b}}$ to give an upper bound for the cumulative perturbation on $\bar{\bm{y}}^*$ over time.

The next lemma states the upper bound for the cumulative distance of $\bm{b}_t'$ (the resource constraint we put on the sampled LP \eqref{eqn:LP_algorithm_afterfair}) to $\bm{b}$ (the initial average resource).
\begin{lemma}\label{lem_bt}
If we execute Algorithm \ref{alg:DAA_fair} under Assumption \ref{assp} and \ref{assp_dual_binding}, we have $\mathbb{E}\left[\sum_{t=1}^T||\bm{b}_{t}' - \bm{b}||_2^2\right] = O(mn^9\log(T))$.
\end{lemma}
The proof can be found in \ref{sec_ap_lem_bt}. Such result cannot be attained if we replace $\bm{b}_t'$ with $\bm{b}_t$ under the existence of non-binding resources. However, for the case that every resource is binding, we are able to get the same result for $\mathbb{E}\left[\sum_{t=1}^T||\bm{b}_{t} - \bm{b}||_2^2\right]$. Also, the $n^9$ dependence comes from our choice of characterizing the good event (see \eqref{eqn:event_AB}) as a subset of $\mathcal{E}_t$ and the concentration inequality we used. We note that the $n^9$ dependence is not optimal and can be reduced by adopting different characterization of the event or different concentration inequalities.

In order to leverage the continuity property, we introduce the LP in standard form. Denote the perturbed reward vector to be $\bm{\mu}_t := [\hat{p}_{1,t}\mu_1, \cdots \hat{p}_{n,t}\mu_n]$, resource matrix $\bm{C}_t := [\hat{p}_{1,t}\bm{c}_1, \cdots \hat{p}_{n,t}\bm{c}_n]$, and the general matrix form
\begin{equation}\label{eqn:Ct}
    \begin{aligned}
    \bar{\bm{C}}_t :=
\begin{pmatrix}
   \bm{C}_t & I_m & 0\\
   I_n & 0 & I_n
    \end{pmatrix},\,\,\,\,\,
    \bar{\bm{y}}
    =    \begin{pmatrix}
   \bm{y}\\
   \bm{s}\\
   \bm{z}
    \end{pmatrix} ,\,\,\,\,\,
    \bar{\bm{\mu}}_t
    =    \begin{pmatrix}
   \bm{\bm{\mu}}_t\\
   \bm{0}\\
   \bm{0}
    \end{pmatrix},\,\,\,\,\,
    \bar{\bm{b}}_t
    =   \begin{pmatrix}
   \bm{b}_t\\
   \bm{\xi}_t\\
    \end{pmatrix},\,\,\,\,\,
    \bm{\xi}_t
    =   \begin{pmatrix}
   \frac{\hat{p}_{1,t}}{p_1}\\
   \vdots\\
   \frac{\hat{p}_{n,t}}{p_n}\\
    \end{pmatrix}.\,\,\,\,\,
    \end{aligned}
\end{equation}
From Algorithm \ref{alg:DAA_fair}, we can rewrite the sample LP $\eqref{eqn:LP_algorithm_afterfair}$ in the following form 
\begin{equation*}
    \begin{aligned}
   \max \ \ & \bm{\mu}_t^\top \bm{y}   \\
    \text{s.t. }\ & \bm{C}_t\bm{y} \leq \bm{b}_t' \\
    & 0 \leq y_j \leq 1,\,\,\,\, j \in [n].
    \end{aligned}
\end{equation*}
If we denote $y_j' = y_j\frac{\hat{p}_{j,t}}{p_j}$, the above LP is equivalent to 
\begin{equation*}
    \begin{aligned} 
   \max \ \ & \bm{\mu}^\top \bm{y}'  \\
    \text{s.t. }\ & \bm{C}\bm{y}' \leq \bm{b}_t'   \\
    & 0 \leq y_j' \leq \frac{\hat{p}_{j,t}}{p_j},\,\,\,\, j \in [n].
    \end{aligned}
\end{equation*}
By denoting $\bar{\bm{b}}'_t = (\bm{b}_t', \bm{\xi}_t)$ and $\bar{\bm{y}}'$ the corresponding decision variable in the standard form for $\bm{y}'$, the final standard form is 
\begin{equation}\label{eqn:SDSLP}
    \begin{aligned}
   \max \ \ & \bar{\bm{\mu}}^\top\bar{\bm{y}}'  \\
    \text{s.t. }\ & \bar{\bm{C}}\bar{\bm{y}}' = \bar{\bm{b}}'_t\\
    & \bar{\bm{y}}' \geq \bm{0}.
    \end{aligned}
\end{equation}
To formalize the continuity argument, we consider the function $\bar{\bm{b}}'(\theta, \Delta\bar{\bm{b}})$ to be the right-hand-side constraint for the LP
\begin{equation}\label{eqn:DLP_lem8_std}
    \begin{aligned}
   \max \ \ & \bar{\bm{\mu}}^\top\bar{\bm{y}}  \\
    \text{s.t. }\ & \bar{\bm{C}}\bar{\bm{y}} = \bar{\bm{b}}(\theta, \Delta\bar{\bm{b}})\\
    & \bar{\bm{y}} \geq \bm{0},
    \end{aligned}
\end{equation}
where $\bar{\bm{b}}(\theta, \Delta\bar{\bm{b}}) = \bar{\bm{b}} + \theta \Delta\bar{\bm{b}}$ is a function of two variables: a scalar $\theta \in \mathbb{R}$, and an arbitrary directional vector $\Delta\bar{\bm{b}} \in \mathbb{R}^{m+n}$. We denote the optimal solution for \eqref{eqn:DLP_lem8_std} to be $\bar{\bm{y}}^*(\theta, \Delta\bar{\bm{b}})$, which is again a function of $\theta$ and $\Delta\bar{\bm{b}}$. From \cite{holder2001marginal} we know that $\bar{\bm{y}}^*(\theta, \Delta\bar{\bm{b}})$ is continuous and differentiable with respect to $\theta$ for every fixed $\Delta\bar{\bm{b}}$. We present the next lemma that shows the Lipchitz property.

\begin{lemma}\label{lem_lipchitz}
    \begin{align*}
        ||\bar{\bm{y}}^*(\theta, \Delta\bar{\bm{b}}) - \bar{\bm{y}}^*(0, \Delta\bar{\bm{b}})||_2 \leq \chi_{\bm{\bar{C}}}||\Delta\bar{\bm{b}}||_2|\theta|,
    \end{align*}
    where $\chi_{\bm{C}} := \sup\{||(\bm{C}\bm{D}\bm{C})^{-1}\bm{C}\bm{D}||_2: \bm{D}\text{ is positive diagonal}\}$.
\end{lemma}

We leave its proof in \ref{sec_ap_lem_lipchitz}. It is worth noticing that $\bar{\bm{y}}^*(0, \Delta\bar{\bm{b}})$ is the interior solution of the standard form DLP \eqref{eqn:SDDLP}. Lemma \ref{lem_lipchitz} states that the interior primal solution is Lipchitz with respect to the perturbation on the right-hand-side. We note that such Lipchitz property holds for the interior primal solution, but does not hold for the interior dual solution. Therefore, to prove the convergence of the dual solution, one have to assume that the dual solution is unique. See \cite{li2019online} for the convergence of the dual solution.

Lastly, Proposition \ref{prop_cu} can be derived from Lemma \ref{lem_bt} and \ref{lem_lipchitz}. We already know $\bm{b}_t'$ is ``close'' to $\bm{b}$ from Lemma \ref{lem_bt}, and it is also easy to show $\xi_{j,t} = \frac{\hat{p}_{j,t}}{p_j}$  being close to $1$ with high probability. Therefore, with high probability $\bar{\bm{b}}_t' = (\bm{b}_t', \bm{\xi}_t)$ stays close to $\bar{\bm{b}}$, hence the online decision will be close to the fair solution. The rest of the proof for Proposition \ref{prop_cu} is left in \ref{sec:pf_prop1}.

\section{Numerical Experiments}\label{sec_numerical}
In this section, we evaluate the performance of different heuristics in the online resource allocation problem under different environments. We find that the numerical results are inline with our theoretical findings. It provides insights on how to choose algorithms under different problem instances. Based on the discussions above, we test the following three algorithms:
\begin{itemize}
    \item Adaptive Simplex: this is Algorithm \ref{alg:DAA} with a simplex solver.
    \item Adaptive Interior: this is Algorithm \ref{alg:DAA} with an interior-point solver.
    \item Adaptive Fair: this is Algorithm \ref{alg:DAA_fair}.
\end{itemize}
We test the three algorithms above under simulated environments. 
\begin{itemize}
    \item In the first environment, there exists both binding and non-binding resources, and one of the non-binding resource is degenerate.
    \item In the second environment, there exists both binding and non-binding resources, but no degeneracy condition on non-binding resources exists.
    \item In the third environment, we have less supply than demand such that all resources are binding.
\end{itemize}
Intuitively, from the theoretical results and the design of Algorithm \ref{alg:DAA} and \ref{alg:DAA_fair}, one should expect that the adaptive fair algorithm has lower cumulative unfairness when the non-binding resource has degeneracy, and it should have higher regret than the adaptive simplex and adaptive interior for all environments. Our tests on those environments confirm this finding. Moreover, we find Algorithm \ref{alg:DAA_fair} features a lower cumulative unfairness in all environments mentioned above.

\subsection{An Example with Non-binding Degeneracy} \label{sec_41}
We choose the model such that 
\begin{equation*}
    \begin{aligned}
    \bm{p} &= [0.15,0.15,0.15,0.15,0.15, 0.15, 0.1]\\
    \bm{\mu} &= [7 , 3.5, 4, 3.5, 6.5, 0.4, 0.7]\\
    \bm{C} &= \begin{pmatrix}
   2 & 1 & 1 & 1 & 1 & 0.1 & 0.2\\
   1 & 0.5 & 1 & 0.5 & 2 & 0.1 & 0.1\\
   1 & 2 & 0.5 & 0.2 & 0.5 & 10 & 7\\
    \end{pmatrix}\\
    \bm{b} &= [0.5, 1, 2],
    \end{aligned}
\end{equation*}
and the corresponding fluid approximation LP is 
\begin{equation*}
    \begin{aligned}
    \max\hspace{5mm} &0.3y_1 + 0.15y_2 + 0.15y_3 + 0.15y_4 + 0.225y_5 + 0.015y_6 + 0.02y_7\\
    s.t.\hspace{5mm} &1.05y_1 + 0.525y_2 + 0.6y_3 + 0.525y_4 + 0.975y_5 + 0.06y_6 + 0.07y_7 \leq 0.5\\
    &0.15y_1 + 0.075y_2 + 0.15y_3 + 0.075y_4 + 0.3y_5 + 0.015y_6 + 0.01y_7 \leq 1\\
    &0.15y_1 + 0.3y_2 + 0.075y_3 + 0.03y_4 + 0.075y_5 + 1.5y_6 + 0.7y_7 \leq 2\\
    &0 \leq y_j \leq 1, j \in [7].
    \end{aligned}
\end{equation*}
We choose this model because the first dimension of resource is binding, and the third dimension of resource is degenerate. Therefore this model could be interpreted as a more complicated model for Section \ref{subsec_resource non degen}.

\begin{figure}[t]
\centering
\begin{subfigure}{0.33\textwidth}
  \centering
    \includegraphics[width=.98\linewidth]{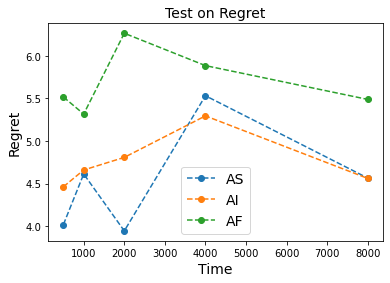}
    \label{fig:s5p1}
\end{subfigure}%
\begin{subfigure}{0.33\textwidth}
  \centering
    \includegraphics[width=.98\linewidth]{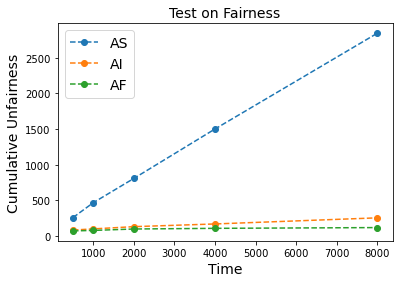}
    \label{fig:s5p2}
\end{subfigure}
\begin{subfigure}{0.33\textwidth}
  \centering
    \includegraphics[width=.98\linewidth]{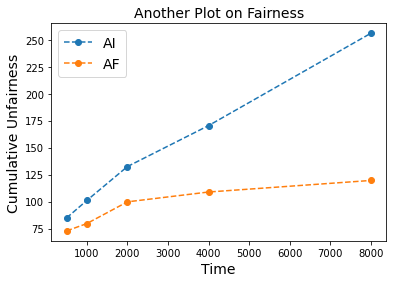}
    \label{fig:s5p3}
\end{subfigure}
\caption{Performance evaluation for Section \ref{sec_41}. AS stands for adaptive simplex, AI stands for adaptive interior,  and AF stands for adaptive fair. The plots above suggests that the adaptive interior can balance the regret and cumulative unfairness well.
}
\label{fig:sec5_1}
\end{figure}

As expected, from Figure \ref{fig:sec5_1} we find that all the adaptive algorithms have $O(1)$ regret. We also notice similar behavior as in Section \ref{subsec_resource non degen} that the Adaptive Interior (Algorithm \ref{alg:DAA}) has a linear growth of cumulative unfairness while the Adaptive Fair (Algorithm \ref{alg:DAA_fair}) has an logarithmic order.

\subsection{An Example with Non-binding Nondegeneracy}\label{sec_42}
To modify the environment such that there is no non-degenereacy in non-binding resource, we change the reward vector, resource consumption matrix, and resource vector such that
\begin{equation*}
    \begin{aligned}
    \bm{p} &= [0.15,0.15,0.15,0.15,0.15, 0.15, 0.1]\\
    \bm{\mu} &= [7, 7, 6.5, 3.2, 5.5, 5, 3.5]\\
    \bm{C} &= \begin{pmatrix}
   2 & 1 & 1 & 1 & 1.5 & 1 & 1\\
   1 & 2 & 2 & 0.5 & 1 & 1 & 0.5\\
   1 & 1 & 0.5 & 0.2 & 0.5 & 1 & 0.5\\
    \end{pmatrix}\\
    \bm{b} &= [0.5, 0.5, 1.5],
    \end{aligned}
\end{equation*}
and the corresponding fluid approximation LP is 
\begin{equation*}
    \begin{aligned}
    \max\hspace{5mm} &1.05y_1 + 1.05y_2 + 0.975y_3 + 0.48y_4 + 0.825y_5 + 0.75y_6 + 0.35y_7\\
    s.t.\hspace{5mm} &0.3y_1 + 0.15y_2 + 0.15y_3 + 0.15y_4 + 0.225y_5 + 0.15y_6 + 0.1y_7 \leq 0.5\\
    &0.15y_1 + 0.3y_2 + 0.3y_3 + 0.075y_4 + 0.15y_5 + 0.15y_6 + 0.05y_7 \leq 0.5\\
    &0.15y_1 + 0.15y_2 + 0.075y_3 + 0.03y_4 + 0.075y_5 + 0.15y_6 + 0.05y_7 \leq 1.5\\
    &0 \leq y_j \leq 1, j \in [7],
    \end{aligned}
\end{equation*}
where the first and second dimension of resource are binding, and the third dimension of resource is non-binding and non-degenerate.

\begin{figure}[t]
\centering
\begin{subfigure}{0.33\textwidth}
  \centering
    \includegraphics[width=.98\linewidth]{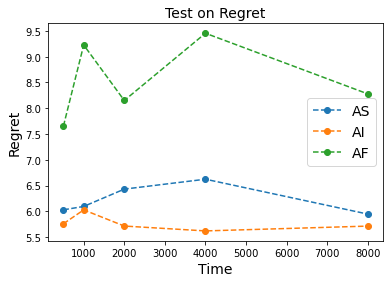}
\end{subfigure}%
\begin{subfigure}{0.33\textwidth}
  \centering
    \includegraphics[width=.98\linewidth]{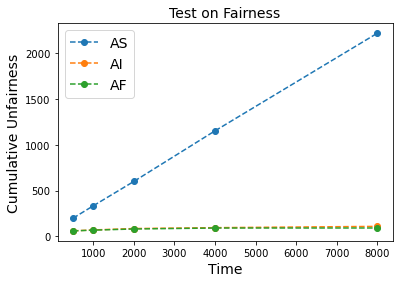}
\end{subfigure}
\begin{subfigure}{0.33\textwidth}
  \centering
    \includegraphics[width=.98\linewidth]{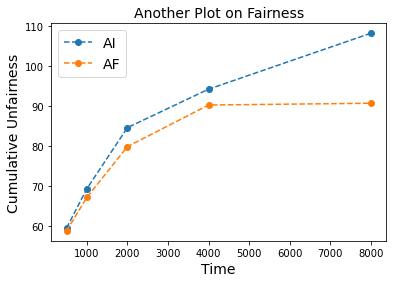}
\end{subfigure}
\caption{Performance evaluation for Section \ref{sec_42}. AS stands for adaptive simplex, AI stands for adaptive interior,  and AF stands for adaptive fair.
}
\label{fig:sec5_2}
\end{figure}

From Figure \ref{fig:sec5_2} we find that all the adaptive algorithms have $O(1)$ regret. Since we do not have degeneracy in non-binding resource, the Adaptive Interior (Algorithm \ref{alg:DAA}) has a sub-linear growth of cumulative unfairness, and it is still higher than the Adaptive Fair algorithm.

\subsection{An Example with All-binding Resources}\label{sec_43}
Lastly we consider an environment that all resources have a higher demand than supply. We change the reward vector, resource consumption matrix, and resource vector such that
\begin{equation*}
    \begin{aligned}
    \bm{p} &= [0.15,0.15,0.15,0.15,0.15, 0.15, 0.1]\\
    \bm{\mu} &= [4, 4, 4.5, 2.7, 3, 3, 2.8]\\
    \bm{C} &= \begin{pmatrix}
   2 & 1 & 1 & 1 & 1.5 & 1 & 1\\
   1 & 2 & 2 & 0.5 & 1 & 1 & 0.5\\
   1 & 1 & 1.5 & 1.2 & 0.5 & 1 & 1.3\\
    \end{pmatrix}\\
    \bm{b} &= [0.5, 0.5, 0.5],
    \end{aligned}
\end{equation*}
and the corresponding fluid approximation LP is 
\begin{equation*}
    \begin{aligned}
    \max\hspace{5mm} &0.6y_1 + 0.6y_2 + 0.675y_3 + 0.405y_4 + 0.45y_5 + 0.45y_6 + 0.28y_7\\
    s.t.\hspace{5mm} &0.3y_1 + 0.15y_2 + 0.15y_3 + 0.15y_4 + 0.225y_5 + 0.15y_6 + 0.1y_7 \leq 0.5\\
    &0.15y_1 + 0.3y_2 + 0.3y_3 + 0.075y_4 + 0.15y_5 + 0.15y_6 + 0.05y_7 \leq 0.5\\
    &0.15y_1 + 0.15y_2 + 0.225y_3 + 0.18y_4 + 0.075y_5 + 0.15y_6 + 0.13y_7 \leq 0.5\\
    &0 \leq y_j \leq 1, j \in [7],
    \end{aligned}
\end{equation*}
where all resources are binding.

\begin{figure}[h]
\centering
\begin{subfigure}{0.33\textwidth}
  \centering
    \includegraphics[width=.98\linewidth]{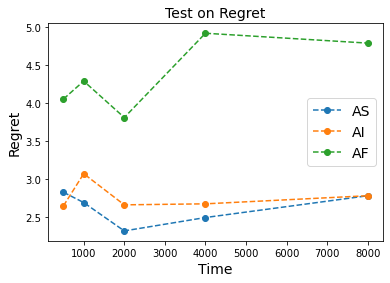}
\end{subfigure}%
\begin{subfigure}{0.33\textwidth}
  \centering
    \includegraphics[width=.98\linewidth]{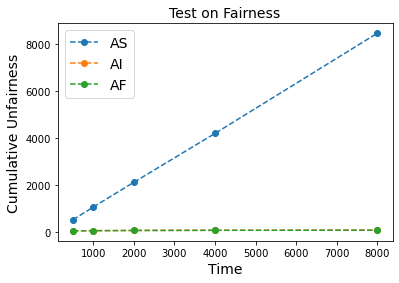}
\end{subfigure}
\begin{subfigure}{0.33\textwidth}
  \centering
    \includegraphics[width=.98\linewidth]{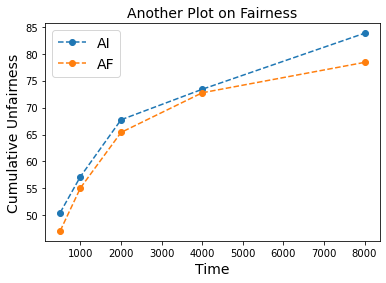}
\end{subfigure}
\caption{Performance evaluation for Section \ref{sec_43}. AS stands for adaptive simplex, AI stands for adaptive interior,  and AF stands for adaptive fair.
}
\label{fig:sec5_3}
\end{figure}

From Figure \ref{fig:sec5_3} we find that all the adaptive algorithms have $O(1)$ regret, and the performance is similar to previous example where there is no non-binding degeneracy.

\bibliographystyle{informs2014} 
\bibliography{main} 

\newpage

\section{Proof of Section \ref{sec_reg_general}}
\section{Proof of Section \ref{sec_reg_general}}

\renewcommand{\thesubsection}{A\arabic{subsection}}

\subsection{Proof of Theorem \ref{thm_fair_regret}}\label{sec_ap_thm1}
Recall that from Corollary \ref{cor_general_regret_form} and Lemma \ref{lem_bounding_acceptance}, we have that 
\begin{align*}
    \text{Reg}_{T}^\pi  & \leq \bm{\lambda}_{max}^{\top} \E\left[\bm{B}_{\tau_S}\right]
   +\left(T -\E[\tau_S]\right)\cdot \max_{j\in[n], \bm{\lambda}\in\mathcal{D}^*}|\mu_j-\bm{c}_j^\top \bm{\lambda}|  \\ & \ \ \ \  \ \ \ + \sum_{t=1}^{\tau_S}\E\left[(r_t - \bm{a}_t^\top\bm{\lambda}_t)_+(1-x_t)\right] + \sum_{t=1}^{\tau_S}\E\left[(\bm{a}_t^\top\bm{\lambda}_t - r_t )_+x_t\right]\\
   &\leq \bm{\lambda}_{max}^{\top} \E\left[\bm{B}_{\tau_S}\right]
   +\left(T -\E[\tau_S]\right)\cdot \max_{j\in[n], \bm{\lambda}\in\mathcal{D}^*}|\mu_j-\bm{c}_j^\top \bm{\lambda}| + \frac{2n\cdot\max_{j\in[n],\bm{\lambda}\in\mathcal{D}^*}|\mu_j - \bm{c}_j^\top\bm{\lambda}|}{1-\exp(-2L^2)}.
\end{align*}
Next, from Proposition \ref{prop_stoppingtime} we know that 
\begin{equation*}
    \begin{aligned}
     T - \mathbb{E}[\tau_S]
    &\le  2 + \frac{16m}{L^2} + \cfrac{nT}{L^2} \exp\left(-2L^2 \kappa T\right) + 4n^3T^{3/2}\exp\left({-\frac{\kappa ^{1/2}}{n^2}\cdot T^{1/2}}\right).
    \end{aligned}
\end{equation*}
Lastly, to finish the proof of bounded regret, we notice that
\begin{equation*}
    \begin{aligned}
    \bm{\lambda}_{max}^{\top}\E\left[\bm{B}_{\tau_S}\right] \leq \bm{\lambda}_{max}^{\top}\E\left[(T-\tau_S+2)\bm{b}_{\tau_S-1}\right] \leq \bm{\lambda}_{max}^{\top}\left(\bm{b} + \frac{L}{2}\right)\E\left[(T-\tau_S+2)\right],
    \end{aligned}
\end{equation*}
where the first inequality is because $\bm{B}_{t} \leq \bm{B}_{t-1}$, and the second inequality is becasue of the definition of $\tau_S$.

\subsection{Proof of Proposition \ref{prop_cu}}\label{sec:pf_prop1}
Before we prove Proposition \ref{prop_cu}, we need to review some notation. At time $t$ the sampled LP we are solving is 
\begin{equation*}
    \begin{aligned}
   \max \ \ & \bar{\bm{\mu}}^\top\bar{\bm{y}}'  \\
    \text{s.t. }\ & \bar{\bm{C}}\bar{\bm{y}}' = \bar{\bm{b}}'_t\\
    & \bar{\bm{y}}' \geq \bm{0},
    \end{aligned}
\end{equation*}
where $\bm{y}' = \bm{\xi}_t\circ\bm{y}$ ($\circ$ stands for entry-wise multiplication) and $\bar{\bm{b}}_t'  = (\bm{b}_t', \bm{\xi}_t)$. To apply Lemma \ref{lem_lipchitz},
we set $\theta = 1$ and $\Delta\bar{\bm{b}} = (\bar{\bm{b}}_t' - \bar{\bm{b}})$. By recalling the formulation in Section \ref{sec_cu} we have $\bar{\bm{y}}'^*(\theta, \Delta\bar{\bm{b}})_{1:n} = \bm{\xi}_t\circ\bm{y}_t^*$ and $\bar{\bm{y}}'^*(0, \Delta\bar{\bm{b}})_{1:n} = \bm{y}^*$. From Lemma \ref{lem_lipchitz}, we know that
\begin{equation*}
    \begin{aligned}
     ||\bm{\xi}_t\circ\bm{y}_t^* - \bm{y}^*||_2^2 &= ||\bar{\bm{y}}'^*(1, \Delta\bar{\bm{b}})_{1:n} - \bar{\bm{y}}'^*(0, \Delta\bar{\bm{b}})_{1:n}||_2^2\\
    &\leq \chi_{\bm{\bar{C}}}^2||\Delta\bar{\bm{b}}||_2^2 = \chi_{\bm{\bar{C}}}^2||\bar{\bm{b}}_t' - \bar{\bm{b}}||_2^2\\
    &= \chi_{\bm{\bar{C}}}^2\left(||\bm{b}_t' - \bm{b}||_2^2 + ||\bm{\xi}_t - \bm{1}||_2^2  \right).
    \end{aligned}
\end{equation*}
Therefore, to calculate $||\bm{y}_t^* - \bm{y}^*||_2^2$, we decompose it into
\begin{equation*}
    \begin{aligned}
     ||\bm{y}_t^* - \bm{y}^*||_2^2 &\leq 2||\bm{y}_t^* - \bm{\xi}_t\circ\bm{y}_t^*||_2^2 + 2||\bm{\xi}_t\circ\bm{y}_t^* - \bm{y}^*||_2^2\\
    &\leq 2||\bm{\xi}_t - \bm{1}||_2^2 + 2\chi_{\bm{\bar{C}}}^2\left(||\bm{b}_t' - \bm{b}||_2^2 + ||\bm{\xi}_t - \bm{1}||_2^2  \right).
    \end{aligned}
\end{equation*}
Then, it is easy to observe that for $t > 1$, from Hoeffding's inequality we have
\begin{equation*}
    \begin{aligned}
     \mathbb{E}\left[||\bm{\xi}_t - \bm{1}||_2^2\right] &= \sum_{j=1}^n\mathbb{E}\left[\left(\frac{\hat{p}_{t,j}}{p_j} - 1\right)^2\right]\\
     &= \sum_{j=1}^n\int_{0}^{+\infty}\frac{x}{2}P\left(\left|\frac{\hat{p}_{t,j}}{p_j} - 1\right| > x\right)dx\\
     &\leq \sum_{j=1}^n\int_{0}^{+\infty}x\exp\left(-2(t-1)x^2p_j^2\right)dx\\
     &= \frac{n}{4(t-1)\underline{p}^2},\\
    \end{aligned}
\end{equation*}
where $\underline{p} = \min_{j\in[n]}p_j$. Therefore, we conclude the proof by applying Lemma \ref{lem_bt} and the fact that $\sum_{t=1}^T 1/t = O(\log (T))$ to show
\begin{equation*}
    \begin{aligned}
     \mathbb{E}\left[\sum_{t=1}^T||\bm{y}_t^* - \bm{y}^*||_2^2\right]
    &\leq (2+2\chi_{\bar{\bm{C}}}^2)\sum_{t=1}^T \mathbb{E}\left[||\bm{\xi}_t - \bm{1}||_2^2\right] + 2\chi_{\bar{\bm{C}}}^2\sum_{t=1}^T \mathbb{E}\left[||\bm{b}_t' - \bm{b}||_2^2\right]\\
    &= O(mn^9\log(T)).
    \end{aligned}
\end{equation*}

\subsection{Proof of Lemma \ref{lem_binding}}
We denote $OPT(\bm{b}')$ to be the optimal value of the dual LP \eqref{eqn:ADDLP} with the input being $\bm{b}'$, and define
$$\mathcal{I}(\bm{b}') = \{\bm{\lambda}|\bm{\lambda} \text{ is the basic optimal solution for \eqref{eqn:ADDLP} with input } \bm{b}' \} .$$
We can define the gap between the optimal and the best suboptimal objective value as
$$ \varepsilon := \min_{\bm{\lambda} \in \bar{\mathcal{I}}(\bm{b})} \left\{   \bm{b}^\top\bm{\lambda} + \sum_{j=1}^n p_j(\mu_j - \bm{c}_j^\top\bm{\lambda})_+   \right\} - OPT(\bm{b}),$$
where $\bar{\mathcal{I}}(\bm{b})$ is the complement set, i.e., the non-optimal basic solution. For the dual program, the change in $\bm{b}$ will not change the set of basic solutions. Consequently, for every fixed basic solution $\bm{\lambda}$,  $\bm{b}^\top\bm{\lambda} + \sum_{j=1}^n p_j(\mu_j - \bm{c}_j^\top\bm{\lambda})_+$ can be viewed as a continuous function with variable $\bm{b}$. The function is continuous and it implies the existence of $\Delta > 0$ such that for $||\bm{b}' - \bm{b}||_{\infty} \leq \Delta$, we have
$$\max_{\bm{\lambda} \in \mathcal{I}(\bm{b})} \left\{ \bm{b}'^\top\bm{\lambda} + \sum_{j=1}^n p_j(\mu_j - \bm{c}_j^\top\bm{\lambda})_+ \right\} < OPT(\bm{b}) + \frac{\varepsilon}{2}$$
$$\min_{\bm{\lambda} \in \bar{\mathcal{I}}(\bm{b})} \left\{ \bm{b}'^\top\bm{\lambda} + \sum_{j=1}^n p_j(\mu_j - \bm{c}_j^\top\bm{\lambda})_+ \right\} > OPT(\bm{b}) + \frac{\varepsilon}{2}.$$
The above inequality implies that for a specific $\bm{b}'$ such that $||\bm{b}' - \bm{b}||_{\infty} \leq \Delta$, there cannot exists basic solution $\bm{\lambda}\in \bar{\mathcal{I}}(\bm{b})$ such that $\bm{\lambda} \in \mathcal{I}(\bm{b}')$. Therefore we can have the result that $\mathcal{I}(\bm{b}') \subseteq \mathcal{I}(\bm{b})$. However there is still one step forward, which is to show $\mathcal{I}(\bm{b}') \subseteq \mathcal{I}(\bm{b})$ for
$$\bm{b}' \in \left(\bigotimes_{i\in\mathcal{B}} \left[b_{i}-\Delta, b_{i}+\Delta\right]\right)\bigotimes \left(\bigotimes_{i\in\mathcal{N}} \left[b_{i}-\Delta, +\infty\right)\right).$$
To show this argument, we begin with the definition of projection $h$ such that
\begin{equation*}
    \begin{aligned}
     h(\bm{b}')_i &=
        \begin{cases}
        b_i-\Delta,\,\,\, i\in\mathcal{N}\\
        b_i',\,\,\, i\in\mathcal{B}
        \end{cases}\\
    \end{aligned}
\end{equation*}
From the definition, we know that $||h(\bm{b}') - \bm{b}||_{\infty} \leq \Delta$, therefore from previous argument we know $\mathcal{I}(h(\bm{b}')) \subseteq \mathcal{I}(\bm{b})$. We aim to show $\mathcal{I}(\bm{b}')  \subseteq \mathcal{I}(h(\bm{b}'))$.

For any $\bm{\lambda}_1 \in \mathcal{I}(h(\bm{b}'))$ and $\bm{\lambda}_2 \in \bar{\mathcal{I}}(h(\bm{b}'))$, we know that 
$$h(\bm{b}')^\top\bm{\lambda}_1 + \sum_{j=1}^n p_j(\mu_j - \bm{c}_j^\top\bm{\lambda}_1)_+ < h(\bm{b}')^\top\bm{\lambda}_2 + \sum_{j=1}^n p_j(\mu_j - \bm{c}_j^\top\bm{\lambda}_2)_+.$$

From Assumption \ref{assp_dual_binding}, we know that $\bm{\lambda}_1$ must have non-binding entry being zero. Since $\bm{b}' - h(\bm{b}')$ is positive only for the non-binding constraint, we have that 
$$\bm{b}'^\top\bm{\lambda}_1 + \sum_{j=1}^n p_j(\mu_j - \bm{c}_j^\top\bm{\lambda}_1)_+ = h(\bm{b}')^\top\bm{\lambda}_1 + \sum_{j=1}^n p_j(\mu_j - \bm{c}_j^\top\bm{\lambda}_1)_+.$$
Moreover, since $\bm{b}' - h(\bm{b}') \geq 0$, we have 
$$h(\bm{b}')^\top\bm{\lambda}_2 + \sum_{j=1}^n p_j(\mu_j - \bm{c}_j^\top\bm{\lambda}_2)_+ < \bm{b}'^\top\bm{\lambda}_2 + \sum_{j=1}^n p_j(\mu_j - \bm{c}_j^\top\bm{\lambda}_2)_+,$$
and arrive at the inequality that
$$\bm{b}'^\top\bm{\lambda}_1 + \sum_{j=1}^n p_j(\mu_j - \bm{c}_j^\top\bm{\lambda}_1)_+ < \bm{b}'^\top\bm{\lambda}_2 + \sum_{j=1}^n p_j(\mu_j - \bm{c}_j^\top\bm{\lambda}_2)_+.$$
Since $\bm{\lambda}_1 \in \mathcal{I}(h(\bm{b}'))$, the above inequality ensures that we must have $\mathcal{I}(\bm{b}') \subseteq \mathcal{I}(h(\bm{b}'))$.

Therefore, for $\bm{b}' \in \left(\bigotimes_{i\in\mathcal{B}} \left[b_{i}-\Delta, b_{i}+\Delta\right]\right)\bigotimes \left(\bigotimes_{i\in\mathcal{N}} \left[b_{i}-\Delta, +\infty\right)\right)$, from the result that 
$\mathcal{I}(\bm{b}') \subseteq \mathcal{I}(\bm{b})$ we know for any $\bm{\lambda} \in \mathcal{I}(\bm{b}')$, it must have property such that $\lambda_i > 0$ for $i \in \mathcal{B}$ as well. Therefore $\mathcal{B} = \mathcal{B}(\bm{b}') = \mathcal{B}_{\text{strict}} = \mathcal{B}_{\text{strict}}(\bm{b}')$.\label{sec:pf_lem_binding}

\subsection{Proof of Proposition \ref{prop_adaptive_regret_form}}\label{sec:pf_prop_regretdecom}

We begin with the following decomposition. For any $\bm{\lambda} \in \mathcal{D}^*$
\begin{equation*}
    \begin{aligned}
    \text{Reg}_{T}^\pi &= \E\left[T\cdot \text{OPT}_D\right] - \sum_{t=1}^T\E\left[r_tx_t\right]\\
    &= \E\left[\bm{\lambda}^{\top}\bm{B} + \sum_{t=1}^T (r_t - \bm{a}_t^\top\bm{\lambda})_+\right] - \sum_{t=1}^T\mathbb{E}\left[r_t x_t\right]\\
    &= \E\left[\bm{\lambda}^{\top}\bm{B} + \sum_{t=1}^T (r_t - \bm{a}_t^\top\bm{\lambda})_+\right] - \sum_{t=1}^T\mathbb{E}\left[(r_t - \bm{a}_t^\top\bm{\lambda})x_t + \bm{a}_t^\top\bm{\lambda}x_t\right]\\
    &= \E\left[\bm{\lambda}^{\top}\left(\bm{B}- \sum_{t=1}^T\bm{a}_tx_t\right)\right] + \sum_{t=1}^T\mathbb{E}\left[ (r_t - \bm{a}_t^\top\bm{\lambda})_+ - (r_t - \bm{a}_t^\top\bm{\lambda})x_t \right]\\
    &= \E\left[\bm{\lambda}^{\top}\left(\bm{B}- \sum_{t=1}^T\bm{a}_tx_t\right)\right] + \sum_{t=1}^T\mathbb{E}\left[ (r_t - \bm{a}_t^\top\bm{\lambda})_+(1-x_t) + (\bm{a}_t^\top\bm{\lambda} - r_t)_+x_t \right].\\
    \end{aligned}
\end{equation*}
For each $t$, if we choose $\bm{\lambda}_t \in \mathcal{D}^*$ and being $\mathcal{H}_{t-1}$-measurable, we can treat $\bm{\lambda}_t$ as a constant (conditioned on $\mathcal{H}_{t-1}$) that belongs to $\mathcal{D}^*$, therefore for any fixed $\bm{\lambda} \in \mathcal{D}^*$, we have
\begin{equation*}
    \begin{aligned}
    \mathbb{E}[\text{OPT}_D-r_tx_t|\mathcal{H}_{t-1}]&=\E\left[\bm{\lambda}_t^{\top}\left(\bm{b}- \bm{a}_tx_t\right) + (r_t - \bm{a}_t^\top\bm{\lambda}_t)_+(1-x_t) + (\bm{a}_t^\top\bm{\lambda_t} - r_t)_+x_t | \mathcal{H}_{t-1} \right]\\
    &= \E\left[\bm{\lambda}^{\top}\left(\bm{b}- \bm{a}_tx_t\right) + (r_t - \bm{a}_t^\top\bm{\lambda})_+(1-x_t) + (\bm{a}_t^\top\bm{\lambda} - r_t)_+x_t | \mathcal{H}_{t-1} \right].\\
    \end{aligned}
\end{equation*}
Therefore, by summing up we have
\begin{equation*}
    \begin{aligned}
    &\hspace{5mm}\E\left[\bm{\lambda}_t^{\top}\left(\bm{B}- \sum_{t=1}^T\bm{a}_tx_t\right)\right] + \sum_{t=1}^T\mathbb{E}\left[ (r_t - \bm{a}_t^\top\bm{\lambda}_t)_+(1-x_t) + (\bm{a}_t^\top\bm{\lambda}_t - r_t)_+x_t \right]\\
    &=\E\left[\sum_{t=1}^T\left(\bm{\lambda}_t^{\top}\left(\bm{b}- \bm{a}_tx_t\right) + (r_t - \bm{a}_t^\top\bm{\lambda}_t)_+(1-x_t) + (\bm{a}_t^\top\bm{\lambda_t} - r_t)_+x_t\right) \right]\\
    &=\E\left[\sum_{t=1}^T\E\left[\left.\left(\bm{\lambda}_t^{\top}\left(\bm{b}- \bm{a}_tx_t\right) + (r_t - \bm{a}_t^\top\bm{\lambda}_t)_+(1-x_t) + (\bm{a}_t^\top\bm{\lambda_t} - r_t)_+x_t\right) \right|\mathcal{H}_{t-1}\right]\right]\\
    &=\E\left[\sum_{t=1}^T\E\left[\left.\left(\bm{\lambda}^{\top}\left(\bm{b}- \bm{a}_tx_t\right) + (r_t - \bm{a}_t^\top\bm{\lambda})_+(1-x_t) + (\bm{a}_t^\top\bm{\lambda} - r_t)_+x_t\right) \right|\mathcal{H}_{t-1}\right]\right]\\
    &=\E\left[\sum_{t=1}^T\left(\bm{\lambda}^{\top}\left(\bm{b}- \bm{a}_tx_t\right) + (r_t - \bm{a}_t^\top\bm{\lambda})_+(1-x_t) + (\bm{a}_t^\top\bm{\lambda} - r_t)_+x_t\right) \right]\\
    &=\E\left[\bm{\lambda}^{\top}\left(\bm{B}- \sum_{t=1}^T\bm{a}_tx_t\right)\right] + \E\left[ \sum_{t=1}^T\left((r_t - \bm{a}_t^\top\bm{\lambda})_+(1-x_t) + (\bm{a}_t^\top\bm{\lambda} - r_t)_+x_t\right) \right],\\
    \end{aligned}
\end{equation*}
and the last line is equal to the decomposition of $\text{Reg}_{T}^\pi$. We also observe that the first term $$\E\left[\bm{\lambda}^{\top}\left(\bm{B}- \sum_{t=1}^T\bm{a}_tx_t\right)\right] \leq \bm{\lambda}_{max}^\top\mathbb{E}[\bm{B}_{\tau}],$$ and we finish the proof by showing
\begin{equation*}
    \begin{aligned}
    \text{Reg}_{T}^\pi &= \E\left[\bm{\lambda}_t^{\top}\left(\bm{B}- \sum_{t=1}^T\bm{a}_tx_t\right)\right] + \sum_{t=1}^T\mathbb{E}\left[ (r_t - \bm{a}_t^\top\bm{\lambda}_t)_+(1-x_t) + (\bm{a}_t^\top\bm{\lambda}_t - r_t)_+x_t \right]\\
    &\leq \bm{\lambda}_{max}^{\top} \E\left[\bm{B}_{\tau}\right] + \sum_{t=1}^T\E\left[(r_t - \bm{a}_t^\top\bm{\lambda}_t)_+(1-x_t)\right] + \sum_{t=1}^T\E\left[(\bm{a}_t^\top\bm{\lambda}_t - r_t )_+x_t\right].
    \end{aligned}
\end{equation*}

\subsection{Proof of Lemma \ref{lem_region_subset}}\label{sec:pf_stableregion}
The proof follows the similar idea to the proof of Lemma \ref{lem_binding}, and the difference is that there is another variable $\bm{p}'$ involved. We denote $OPT(\bm{b}', \bm{p}')$ to be the optimal value of the dual LP \eqref{eqn:ADDLP} with the input being $\bm{b}'$ and $\bm{p}'$,
and define
$$\mathcal{I}(\bm{b}',\bm{p}') = \{\bm{\lambda}|\bm{\lambda} \text{ is the basic optimal solution for \eqref{eqn:ADDLP} with input } \bm{b}',\bm{p}' \}$$
such that for $\bm{\lambda} \in \mathcal{I}(\bm{b}',\bm{p}')$ we have
$$OPT(\bm{b}', \bm{p}') = \bm{b}'^\top\bm{\lambda} + \sum_{j=1}^np'_j(\mu_j - \bm{c}_j^\top\bm{\lambda})_+.$$
Next, for the input $\bm{b}$ and $\bm{p}$, we define the gap between optimal and the best suboptimal objective value as
$$ \varepsilon := \min_{\bm{\lambda} \in \bar{\mathcal{I}}(\bm{b}, \bm{p})} \left\{   \bm{b}^\top\bm{\lambda} + \sum_{j=1}^n p_j(\mu_j - \bm{c}_j^\top\bm{\lambda})_+   \right\} - OPT(\bm{b}, \bm{p}).$$
Where $\bar{\mathcal{I}}(\bm{b}, \bm{p})$ is the complement set, i.e., the non-optimal basic solution. Notice that here the set of basic solution will not change with the change of $\bm{b}$ and $\bm{p}$. To see this one can recall the formation \eqref{eqn:Ct}, look at the dual of the standard form of the sampled LP \eqref{eqn:SDSLP}, and find out that the constraint for the dual variable is the same as the dual of \eqref{eqn:SDDLP}. To derive an uniform bound, we note that $(\mu_j - \bm{c}_j^\top\bm{\lambda})_+$ will be uniformly bounded for all basic solutions of $\bm{\lambda}$ such that $\bm{\lambda} \in \mathcal{I}(\bm{b}, \bm{p})\cup\bar{\mathcal{I}}(\bm{b}, \bm{p})$. Therefore there exists $\eta$ such that whenever $||\bm{p}' - \bm{p}||_{\infty} < \eta$, we have that for all $\bm{\lambda} \in \mathcal{I}(\bm{b}, \bm{p})\cup\bar{\mathcal{I}}(\bm{b}, \bm{p})$,
$$\sum_{j=1}^n (p'_j-p_j)(\mu_j - \bm{c}_j^\top\bm{\lambda})_+ < \varepsilon/4.$$
Then, we look at the variable $\bm{b}'$. Because $\mathcal{I}(\bm{b},\bm{p})$ and $\bar{\mathcal{I}}(\bm{b},\bm{p})$ are finite, there exist $L$ such that whenever $||\bm{b}' - \bm{b}||_{\infty} \leq L$, we have
$$\max_{\bm{\lambda}\in\mathcal{I}(\bm{b},\bm{p})}\left\{ \bm{b}'^\top\bm{\lambda} + \sum_{j=1}^n p_j(\mu_j - \bm{c}_j^\top\bm{\lambda})_+ \right\} < OPT(\bm{b}) + \frac{\varepsilon}{4},$$
$$\min_{\bm{\lambda}\in\bar{\mathcal{I}}(\bm{b},\bm{p})}\left\{ \bm{b}'^\top\bm{\lambda} + \sum_{j=1}^n p_j(\mu_j - \bm{c}_j^\top\bm{\lambda})_+ \right\} > OPT(\bm{b}) + \frac{3\varepsilon}{4}.$$
Combining with the fact that $\sum_{j=1}^n (p'_j-p_j)(\mu_j - \bm{c}_j^\top\bm{\lambda})_+ < \varepsilon/4$, we can arrive at the conclusion that for any $||\bm{p}' - \bm{p}|| \leq \eta$ and $||\bm{b}' - \bm{b}|| \leq L$, we have
$$\max_{\bm{\lambda}\in\mathcal{I}(\bm{b},\bm{p})}\left\{ \bm{b}'^\top\bm{\lambda} + \sum_{j=1}^n p_j'(\mu_j - \bm{c}_j^\top\bm{\lambda})_+ \right\} < OPT(\bm{b}) + \frac{\varepsilon}{2},$$
$$\min_{\bm{\lambda}\in\bar{\mathcal{I}}(\bm{b},\bm{p})}\left\{ \bm{b}'^\top\bm{\lambda} + \sum_{j=1}^n p_j'(\mu_j - \bm{c}_j^\top\bm{\lambda})_+ \right\} > OPT(\bm{b}) + \frac{\varepsilon}{2}.$$
Therefore, we can have that $\mathcal{I}(\bm{b}',\bm{p}')\subseteq\mathcal{I}(\bm{b},\bm{p})$ whenever $||\bm{p}' - \bm{p}|| \leq \eta$ and $||\bm{b}' - \bm{b}|| \leq L$. The next step is to show the same relationship for $||\bm{p}' - \bm{p}|| \leq \eta$ and
$$\bm{b}' \in \left(\bigotimes_{i\in\mathcal{B}} \left[b_{i}-L, b_{i}+L\right]\right)\bigotimes \left(\bigotimes_{i\in\mathcal{N}} \left[b_{i}-L, +\infty\right)\right).$$
To show this argument, we continue with the definition of projection $h$ such that
\begin{equation*}
    \begin{aligned}
     h(\bm{b}')_i &=
        \begin{cases}
        b_i-L,\,\,\, i\in\mathcal{N}\\
        b_i',\,\,\, i\in\mathcal{B}
        \end{cases}\\
    \end{aligned}
\end{equation*}
From the definition, we know that $||h(\bm{b}') - \bm{b}||_{\infty} \leq L$, therefore from previous argument we know $\mathcal{I}(h(\bm{b}'), \bm{p}') \subseteq \mathcal{I}(\bm{b},\bm{p})$ for $||\bm{p}'-\bm{p}||_{\infty} \leq \eta$. We aim to show $\mathcal{I}(\bm{b}',\bm{p}')  \subseteq \mathcal{I}(h(\bm{b}'),\bm{p}')$.

For any $\bm{\lambda}_1 \in \mathcal{I}(h(\bm{b}'),\bm{p}')$ and $\bm{\lambda}_2 \in \bar{\mathcal{I}}(h(\bm{b}'),\bm{p}')$, we know that 
$$h(\bm{b}')^\top\bm{\lambda}_1 + \sum_{j=1}^n p'_j(\mu_j - \bm{c}_j^\top\bm{\lambda}_1)_+ < h(\bm{b}')^\top\bm{\lambda}_2 + \sum_{j=1}^n p'_j(\mu_j - \bm{c}_j^\top\bm{\lambda}_2)_+.$$

From Assumption \ref{assp_dual_binding} and the previous conclusion $\mathcal{I}(h(\bm{b}'), \bm{p}') \subseteq \mathcal{I}(\bm{b},\bm{p})$, we know that $\bm{\lambda}_1$ must have non-binding entry being zero. Since $\bm{b}' - h(\bm{b}')$ is positive only for the non-binding constraint, we have that 
$$\bm{b}'^\top\bm{\lambda}_1 + \sum_{j=1}^n p_j'(\mu_j - \bm{c}_j^\top\bm{\lambda}_1)_+ = h(\bm{b}')^\top\bm{\lambda}_1 + \sum_{j=1}^n p_j'(\mu_j - \bm{c}_j^\top\bm{\lambda}_1)_+.$$
Moreover, since $\bm{b}' - h(\bm{b}') \geq 0$, we have 
$$h(\bm{b}')^\top\bm{\lambda}_2 + \sum_{j=1}^n p_j'(\mu_j - \bm{c}_j^\top\bm{\lambda}_2)_+ < \bm{b}'^\top\bm{\lambda}_2 + \sum_{j=1}^n p_j'(\mu_j - \bm{c}_j^\top\bm{\lambda}_2)_+,$$
and arrive at the inequality that
$$\bm{b}'^\top\bm{\lambda}_1 + \sum_{j=1}^n p_j'(\mu_j - \bm{c}_j^\top\bm{\lambda}_1)_+ < \bm{b}'^\top\bm{\lambda}_2 + \sum_{j=1}^n p_j'(\mu_j - \bm{c}_j^\top\bm{\lambda}_2)_+.$$
Since $\bm{\lambda}_1 \in \mathcal{I}(h(\bm{b}'), \bm{p}')$, the above inequality ensures that we must have $\mathcal{I}(\bm{b}', \bm{p}') \subseteq \mathcal{I}(h(\bm{b}'), \bm{p}')$.

Finally, we have shown that for $||\bm{p}' - \bm{p}|| \leq \eta$ and
$$\bm{b}' \in \left(\bigotimes_{i\in\mathcal{B}} \left[b_{i}-L, b_{i}+L\right]\right)\bigotimes \left(\bigotimes_{i\in\mathcal{N}} \left[b_{i}-L, +\infty\right)\right),$$
we have $\mathcal{I}(\bm{b}', \bm{p}') \subseteq \mathcal{I}(\bm{b}, \bm{p})$. By adjusting $L$ to be the minimum of the original $L$ and $\eta$, we arrive at result that if $(\bm{b}_t, \hat{\bm{p}}_t) \in \mathcal{K}$, the dual optimal solution set of the sampled LP \eqref{eqn:adpt_lp} is a subset of $\mathcal{D}^*$.

\subsection{Proof of Corollary \ref{cor_bindingness}}\label{sec_ap_cor2}
The proof of the first statement trivially follows the fact that the interior dual solution is a subset of $\mathcal{D}^*$, the condition of complementary slackness, and the Assumption \ref{assp_dual_binding}. Therefore we focus on the second statement. Let us first focus on solving the sampled LP with fixed $\hat{\bm{p}}$ such that $||\hat{\bm{p}} - \bm{p}|| \leq L$ but varying $\bm{b}'$ that belongs to
\begin{equation}\label{eqn:pf_region_stab}
    \begin{aligned}
     \hat{b}_i &\in
        \begin{cases}
        \{b_i\},\,\,\, i\in\mathcal{N}\\
        [b_i - L, b_i + L],\,\,\, i\in\mathcal{B}.
        \end{cases}\\
    \end{aligned}
\end{equation}
Because the perturbation of $\bm{b}'$ and $\hat{\bm{p}}$ is within the stable region, and the region \eqref{eqn:pf_region_stab} and $||\hat{\bm{p}} - \bm{p}|| \leq L$ are closed intervals, there exists $\delta$ such that $$\left(\sum_{j=1}^n \hat{p}_{j,t}\bm{c}_jy_{j}^*\right)_{\mathcal{N}} < \bm{b}_{\mathcal{N}} - \delta\cdot\bm{1}$$
for all the solution $\bm{y}^*$ from the DLP with input in \eqref{eqn:pf_region_stab}. Having shown such property, by restricting the $\hat{\bm{p}}$ such that $||\hat{\bm{p}} - \bm{p}||_{\infty} \leq \delta / 2$, we can have the result that for the deviation 
\begin{equation*}
    \begin{aligned}
     \hat{b}_i &\in
        \begin{cases}
        \{b_i\},\,\,\, i\in\mathcal{N}\\
        [b_i - L, b_i + L],\,\,\, i\in\mathcal{B}
        \end{cases}\\
     \hat{p}_{j} &\in [p_j - \delta/2, p_j + \delta/2],\,\,\, j = [n],
    \end{aligned}
\end{equation*}
we have 
$$\left(\sum_{j=1}^n p_{j}\bm{c}_jy_{j}^*\right)_{\mathcal{N}} < \left(\sum_{j=1}^n \hat{p}_{j,t}\bm{c}_jy_{j}^*\right)_{\mathcal{N}} + \delta/2\cdot\bm{1} < \bm{b}_{\mathcal{N}} - \delta/2\cdot\bm{1}.$$
Therefore, by setting $L$ to be $\min\{L, \delta/2\}$ we finish proving this corollary.

\subsection{Proof of Proposition \ref{prop_stoppingtime}}\label{sec_ap_stptime}
By the definition of $\mathcal{E}_t$. When $t\leq \kappa T$, we know that $\mathcal{E}_t$ happens with probability $1$. When $\tilde{\tau} >  t > \kappa T$ we know that $(\bm{b}_t, \hat{\bm{p}}_t)$ is in the stable region and from Corollary \ref{cor_bindingness} we know $\mathcal{G}_t$ defined by
$$\mathcal{G}_t \coloneqq \left\{ b_{i,t} = (\bm{C}_t\bm{y}_t^*)_i, \forall i \in \mathcal{B}\right\}\cap\left\{ b_{i,t} > (\bm{C}_t\bm{y}_t^*)_i, \forall i \in \mathcal{N}\right\}$$
is true. $\mathcal{G}_t$ states that the binding/nonbinding structure of the sampled LP is the same as the DLP, and this implies that Algorithm \ref{alg:DAA_fair} is the same as Algorithm \ref{alg:DAA} with modified $\bm{b}_t'$ on the right binding/nonbinding index, and an interior-point solver. Therefore, by recalling that 
\begin{equation*}
    \begin{aligned}
    \prob\left(\tau_S \leq t\right)
    &\le \prob\left(\tilde{\bm{b}}_s\notin\Omega \text{ for some }s\le t \right) + \sum_{s=1}^t \prob(\bar{\mathcal{E}}_s),
    \end{aligned}
\end{equation*}
we can bound $\prob\left(\tilde{\bm{b}}_s\notin\Omega \text{ for some }s\le t \right)$ by the similar approach in Appendix C of \cite{chen2021linear}. We denote that the modification $\bm{b}_t'$ in Algorithm \ref{alg:DAA_fair} will change the dynamic of average remaining inventory $\bm{b}_t$ compared with the $\bm{b}_t$ in \cite{chen2021linear}. However, under the event $\mathcal{G}_t$, we will show that the inventory process $\bm{b}_t$ with modified dynamic will still comply with the characterization of ``good'' event $\mathcal{E}_t$. Therefore from the similar analysis, we can state the lemma below, and its proof is left in \ref{sec_ap_mtg}.

\begin{lemma}\label{lem_mtg}
For $T\ge T_1$ and $t\le T-2$
\begin{equation*}
    \begin{aligned}
    \prob\left(\tilde{\bm{b}}_s\notin\Omega \text{ for some }s\le t \right)  \leq 2me^{-\frac{L^2(T-t)}{8}}
    \end{aligned},
\end{equation*}
where the constant $T_1$ is defined as the minimal integer such that 
\begin{equation*}
    \begin{aligned}
    T_1 \ge \left(\frac{1}{\exp{\left(\frac{L}{8}\right) - 1}}  + 2\right) \vee \left(\frac{1}{\exp{\left(\frac{L}{8(1+L-\underline{b})}\right) - 1}}  + 2\right)
    \end{aligned}
\end{equation*}
and 
$\frac{\log T_1}{T_1^{1/4}} \le \frac{\kappa^{1/4} L}{4}$, and $\kappa$ is set by $\kappa = \left(1-\exp(-\frac{L}{8})\right) \wedge \left(1-\exp(-\frac{L}{8(1+L-\underline{b})})\right)$. 
\end{lemma}
In terms of $\sum_{s=1}^t \prob(\bar{\mathcal{E}}_s)$, from the fact that $\mathcal{E}_t$ happens with probability $1$ for $t \leq \kappa T$, we know that 
\begin{equation*}
    \begin{aligned}
    \sum_{s=1}^t \prob(\bar{\mathcal{E}}_s) &= \sum_{s=\kappa T + 1}^{t} \prob(\bar{\mathcal{E}}_s).\\
    \end{aligned}
\end{equation*}
Next, by defining 
\begin{equation}\label{eqn:event_AB}
    \begin{aligned}
    \mathcal{A}_t^{(j)} \coloneqq \left\{ \left|\hat{p}_{j,t} - p_j\right| \leq L\right\},\,\,\,\,\,\mathcal{B}_t^{(j)} \coloneqq \left\{ \left\vert \hat{p}_{j,t}-p_j\right\vert \leq \frac{1}{n(t-1)^{1/4}} \right\},
    \end{aligned}
\end{equation}
we want to show $\cap_{j=1}^n\left(\mathcal{A}_t^{(j)}\cap \mathcal{B}_t^{(j)}\right) \subseteq \mathcal{E}_t$. 

To see this, for $t \geq 2$, let $\bm{y}^*_{t}=(y_{1,t}^*,...,y_{n,t}^*)^\top$ be the optimal solution of sampled LP \eqref{eqn:adpt_lp} with $\bm{b}_{t}=\bm{b}'$ for some $\bm{b}'\in \Omega$. By the algorithm, we have the expected resource consumption at time $t$
$$\E[\bm{a}_{t}x_{t}(\bm{b}')|\mathcal{H}_{t-1},\bm{b}_{t}=\bm{b}'] = \sum_{j=1}^n p_j\bm{c}_j\cdot y_{j,t}^*.$$
Moreover, we know that given the event $\cap_{j=1}^n \mathcal{A}_{t}^{(j)}$ and $\bm{b}' \in \Omega$, we are in the stable region $\mathcal{K}$. Therefore, from Corollary \ref{cor_bindingness} we have
$$\bm{b}'_{\mathcal{B}} = \left(\sum_{j=1}^n \hat{p}_{j,t} \bm{c}_j \cdot y_{j,t}^*\right)_{\mathcal{B}}.$$
Then, taking the difference, 
\begin{equation}\label{eqn_axminusd}
    \begin{aligned}
    \E[\bm{a}_{\mathcal{B}, t}x_{t}(\bm{b}')|\mathcal{H}_{t-1},\bm{b}_{t}=\bm{b}']-\bm{b}'_{\mathcal{B}} = \left(\sum_{j=1}^n \left(p_j - \hat{p}_{j,t}\right)\bm{c}_j\cdot y_{j,t}^*\right)_{\mathcal{B}}.
    \end{aligned}
\end{equation}
Next, given the event $\cap_{j=1}^n\mathcal{B}_{t}^{(j)},$ we have
\begin{align*}
 \left\|\E[\bm{a}_{\mathcal{B}, t}x_{t}(\bm{b}')|\mathcal{H}_{t-1},\bm{b}_{t}=\bm{b}']-\bm{b}'_{\mathcal{B}}\right\|_\infty & = \left\|\left(\sum_{j=1}^n \left(p_j - \hat{p}_{j,t}\right)\bm{c}_j y_{j,t}^*\right)_{\mathcal{B}}\right\|_\infty\\
 & \le \sum_{j=1}^n \left \vert p_j - \hat{p}_{j,t}\right\vert \\
 & \le \min\left\{(t-1)^{-\frac{1}{4}}, 1\right\},
\end{align*}
where we use the fact that $\|\bm{c}_j\|_\infty\le 1$ from Assumption \ref{assp}. This meets the definition of the event $\mathcal{E}_t$ and we have shown that $\cap_{j=1}^n\left(\mathcal{A}_t^{(j)}\cap \mathcal{B}_t^{(j)}\right) \subseteq \mathcal{E}_t$.

Next, by applying Hoeffding's inequality for $\mathcal{A}_t^{(j)}$ and $\mathcal{B}_t^{(j)}$ we have
\begin{align*}
\prob(\bar{\mathcal{E}}_t) \leq \prob\left(\left(\cup_{j=1}^n \bar{\mathcal{A}}_{t}^{(j)}\right)\cup \left(\cup_{j=1}^n \bar{\mathcal{B}}_{t}^{(j)}\right)\right) & \leq 2n \exp\left(-2L^2(t-1)\right) +2n\exp\left(-\frac{2(t-1)^{1/2}}{n^2}\right),
\end{align*}
\begin{equation*}
    \begin{aligned}
    \sum_{s=\kappa T + 1}^{t} \prob(\bar{\mathcal{E}}_t) &\leq \cfrac{n}{L^2} \exp\left(-2L^2 \kappa T\right) + 4n^3T^{1/2}\exp\left({-\frac{\kappa ^{1/2}}{n^2}\cdot T^{1/2}}\right),
    \end{aligned}
\end{equation*}
and
\begin{equation*}
    \begin{aligned}
    \sum_{t=1}^T\sum_{s=\kappa T + 1}^{t} \prob(\bar{\mathcal{E}}_t) &\leq \cfrac{nT}{L^2} \exp\left(-2L^2 \kappa T\right) + 4n^3T^{3/2}\exp\left({-\frac{\kappa ^{1/2}}{n^2}\cdot T^{1/2}}\right).
    \end{aligned}
\end{equation*}
Again recalling the formula we have
\begin{equation*}
    \begin{aligned}
    \prob\left(\tau_S \leq t\right)
    &\le \prob\left(\tilde{\bm{b}}_s\notin\Omega \text{ for some }s\le t \right) + \sum_{s=1}^t \prob(\bar{\mathcal{E}}_s)\\
    &\leq 2m\exp\left(-\frac{L^2(T-t)}{8}\right) + \frac{n}{L^2}\exp\left(-2L^2\kappa T\right) + 4n^3T^{1/2}\exp\left(-\frac{\kappa^{1/2}}{n^2}T^{1/2}\right).
    \end{aligned}
\end{equation*}
Lastly, 
\begin{equation*}
    \begin{aligned}
    \mathbb{E}[\tau_S] &\geq \sum_{t=1}^T (1 - \mathbb{P}(\tau_S \leq t))\\
    &\geq T - 2 - \sum_{t=1}^{T-2} \prob\left(\tilde{\bm{b}}_s\notin\Omega \text{ for some }s \leq t \right) - \sum_{t=1}^T \sum_{s=\kappa T + 1}^t \mathbb{P}(\bar{\mathcal{E}}_s)\\
    &\ge T - 2 - 2m \frac{1-e^{-(T-2)L^2/8}}{1-e^{-L^2/8}} - \cfrac{nT}{L^2} \exp\left(-2L^2 \kappa T\right) - 4n^3T^{3/2}\exp\left({-\frac{\kappa ^{1/2}}{n^2}\cdot T^{1/2}}\right)\\
    &\ge T - 2 - \frac{16m}{L^2} - \cfrac{nT}{L^2} \exp\left(-2L^2 \kappa T\right) - 4n^3T^{3/2}\exp\left({-\frac{\kappa ^{1/2}}{n^2}\cdot T^{1/2}}\right).
    \end{aligned}
\end{equation*}

\subsection{Proof of Lemma \ref{lem_bounding_acceptance}}\label{ap:pf_acceptance}
To bound $\sum_{t=1}^{\tau_S}\E\left[(r_t - \bm{a}_t^\top\bm{\lambda}_t)_+(1-x_t)\right] + \sum_{t=1}^{\tau_S}\E\left[(\bm{a}_t^\top\bm{\lambda}_t - r_t )_+x_t\right]$, we first observe the following facts.
\begin{itemize}
    \item At time $t$, we already can choose $\bm{\lambda}_{t-1} \in \mathcal{H}_{t-1}$, and the result for any $\bm{\lambda}_{t-1} \in \mathcal{H}_{t-1}$ could be used as an upper bound.
    \item Only one order type arrives, hence it can only be one of the three types : $r_t - \bm{a}_t^\top\bm{\lambda}_t > 0$, $r_t - \bm{a}_t^\top\bm{\lambda}_t = 0$, and $r_t - \bm{a}_t^\top\bm{\lambda}_t < 0$.
    \item For the bound of \eqref{eqn:cor_regret_line2}, we can ignore the type $r_t - \bm{a}_t^\top\bm{\lambda}_t = 0$.
    \item $(r_t - \bm{a}_t^\top\bm{\lambda}_t)_+(1-x_t) + (\bm{a}_t^\top\bm{\lambda}_t - r_t )_+x_t > 0$ if and only if we reject/accept the wrong type.
\end{itemize}
From the observation above, we choose $\bm{\lambda}^*_t \in \mathcal{H}_{t-1}$, the analytic center of the sampled LP \eqref{eqn:adpt_lp}, and define the event of ``making mistake at time $t$'' to be
\begin{equation*}
    \begin{aligned}
     \mathcal{E}_t :&= \{(r_t - \bm{a}_t^\top\bm{\lambda}_t)_+(1-x_t) + (\bm{a}_t^\top\bm{\lambda}_t - r_t )_+x_t > 0\}
    \end{aligned}
\end{equation*}
Now, define event $\mathcal{A}_t^{(j)}$ such that 
\begin{equation*}
    \begin{aligned}
    \mathcal{A}_t^{(j)} = \left\{\left|\hat{p}_j - p_j\right| \leq L\right\},
    \end{aligned}
\end{equation*}
and define event $\mathcal{Q}_t$ such that
\begin{equation*}
    \begin{aligned}
    \mathcal{Q}_t = \{\bm{b}_t \in \Omega\}\cap\left\{\cap_{j=1}^n\mathcal{A}_t^{(j)}\right\}.
    \end{aligned}
\end{equation*}
We then remark some important observations. First, $\mathcal{Q}_t$ is $\mathcal{H}_{t-1}$-measurable and we can already know if $\mathcal{Q}_t$ is true at time $t$. Second, At time $t$ we also have already knew the information at $\eqref{eqn:adpt_lp}$, which is $\mathcal{H}_{t-1}$-measurable, and we can calculate $\bm{y}_t^*$ and $\bm{\lambda}_t^*$ (we don't have to exactly calculate $\bm{\lambda}_t^*$ since it is only for bounding the expectation for theoretical purpose). Third, $\mathcal{Q}_t \subseteq \{(\bm{b}_t, \hat{\bm{p}}_t) \in \mathcal{K}\}$, and from Lemma \ref{lem_region_subset}, we know that if $\mathcal{Q}_t$ happens, $\mathcal{D}_t^*$, the set of optimal dual solution for the sampled LP $\eqref{eqn:adpt_lp}$, is a subset of $\mathcal{D}^*$. 

Based on the observations above, we then show that at time $t$, by setting $\bm{\lambda}_t = \bm{\lambda}_t^*$, and using the interior solution $\bm{y}_t^*$ of the smapled LP \eqref{eqn:adpt_lp}, we have

\begin{align}\label{eqn:zero_exp}
    \E\left[(r_t - \bm{a}_t^\top\bm{\lambda}_t^*)_+(1-x_t)\right] + \E\left[(\bm{a}_t^\top\bm{\lambda}_t^* - r_t )_+x_t\right] = 0
\end{align}

To show \eqref{eqn:zero_exp}, we have to validate our choice of $\bm{y}_t^*$ and $\bm{\lambda}_t^*$. From the property of primal-dual central path, we know that the interior primal solution $\bar{\bm{y}}^* = [\bm{y}^*, \bm{s}^*, \bm{z}^*]$ of \eqref{eqn:PSDDLP} and interior dual solution $\bm{\lambda}^*$ satisfy the strict complementary condition such that 
\begin{equation*}
    \begin{aligned}
    s_i^*\cdot\lambda_i^* &= (b_i - (\bm{C}\bm{y}^*)_i)\cdot\lambda_i^* = 0,\,\,\, i \in [m]\\
    z_j^*\cdot(\mu_j - \bm{c}_j^\top\bm{\lambda}^*)_+ &= (1 - y_j^*)\cdot(\mu_j - \bm{c}_j^\top\bm{\lambda}^*)_+ = 0,\,\,\,  j \in [n]\\
    y_j^*\cdot(\mu_j - \bm{c}_j^\top\bm{\lambda}^* - (\mu_j - \bm{c}_j^\top\bm{\lambda}^*)_+) &= 0,\,\,\, j \in [n],\\
    \end{aligned}
\end{equation*}
and this condition implies the following lemma
\begin{lemma}\label{lem_dual_correspondence}
 For the interior primal solution $\bm{y}^*_t$ for the DLP $\eqref{eqn:DLP}$ and the interior dual solution $\bm{\lambda}^*$ for the dual $\eqref{eqn:ADDLP}$, we know $y_i^* = 1$ if and only if $\bm{c}_i^\top\bm{\lambda}^* < \mu_i$, $y_i^* = 0$ if and only if $\bm{c}_i^\top\bm{\lambda}^* > \mu_i$, and $y_i^* \in (0, 1)$ if and only if $\bm{c}_i^\top\bm{\lambda}^* = \mu_i$. Moreover, the same holds for the interior primal and dual solutions for the perturbed LP $\eqref{eqn:adpt_lp}$ and its dual.
\end{lemma}

From Lemma \ref{lem_dual_correspondence} we know that the interior primal solution is consistent with the accept/reject rule given by the interior dual solution. More specifically, if at time $t$, we have $(r_t, \bm{a}_t) = (\mu_j, \bm{c}_j)$ for category $j$, Lemma \ref{lem_dual_correspondence} implies that we must have $y_{i,t}^* = 0$ if $r_t - \bm{a}_t^\top\bm{\lambda}_t^* < 0$ and $y_{i,t}^* = 1$ if $r_t - \bm{a}_t^\top\bm{\lambda}_t^* > 0$. In other words,
\begin{align*}
    \E\left[(r_t - \bm{a}_t^\top\bm{\lambda}_t^*)_+(1-x_t)\right] + \E\left[(\bm{a}_t^\top\bm{\lambda}_t^* - r_t )_+x_t\right] = 0.
\end{align*}
However, this is not enough to show the bound \eqref{eqn:cor_regret_line2} is small. This is because the sample LP might have different characterization of the accept/reject type from the DLP. The condition that $(\bm{b}_t, \hat{\bm{p}}_t) \in \mathcal{K}$ (it suffices to ensure that $\mathcal{Q}_t$ is true) makes sure sample LP and DLP has the same characterization of the accept/reject type. If we can show $(\bm{b}_t, \hat{\bm{p}}_t) \in \mathcal{K}$ (which is equivalent to $\bm{\lambda}_t^* \in \mathcal{D}^*$) with high probability, then \eqref{eqn:cor_regret_line2} is small.

For $\{\bm{\lambda}_t\}_{t=1}^T$ such that $\bm{\lambda}_t$ is $\mathcal{H}_{t-1}$-measurable and $\bm{\lambda}_t \in \mathcal{D}^*$,
\begin{equation*}
    \begin{aligned}
    &\hspace{5mm}\E\left[\sum_{t=1}^{\tau_S}\left((r_t - \bm{a}_t^\top\bm{\lambda}_t)_+(1-x_t)+ (\bm{a}_t^\top\bm{\lambda}_t - r_t )_+x_t\right)\right]\\
    &=\E\left[\sum_{t=1}^{\tau_S}\left(\left(\max_{j\in[n],\bm{\lambda}\in\mathcal{D}^*}|\mu_j - \bm{c}_j^\top\bm{\lambda}|\bm{1}_{\{\bm{\lambda}_t^* \notin \mathcal{D}^* \}}\right) + \left((r_t - \bm{a}_t^\top\bm{\lambda}_t^*)_+(1-x_t)+ (\bm{a}_t^\top\bm{\lambda}_t^* - r_t )_+x_t\right)\bm{1}_{\{\bm{\lambda}_t^* \in \mathcal{D}^* \}}\right)\right]\\
    &=\E\left[\sum_{t=1}^{\tau_S}\left(\max_{j\in[n],\bm{\lambda}\in\mathcal{D}^*}|\mu_j - \bm{c}_j^\top\bm{\lambda}|\bm{1}_{\{\bm{\lambda}_t^* \notin \mathcal{D}^* \}}\right)\right],
    \end{aligned}
\end{equation*}
where the last equality is because  

\begin{equation*}
    \begin{aligned}
    &\hspace{5mm}\E\left[\left((r_t - \bm{a}_t^\top\bm{\lambda}_t^*)_+(1-x_t)+ (\bm{a}_t^\top\bm{\lambda}_t^* - r_t )_+x_t\right)\bm{1}_{\{\bm{\lambda}_t^* \in \mathcal{D}^* \}}\right]\\
    &= \E\left[\bm{1}_{\{\bm{\lambda}_t^* \in \mathcal{D}^* \}}\E\left[\left((r_t - \bm{a}_t^\top\bm{\lambda}_t^*)_+(1-x_t)+ (\bm{a}_t^\top\bm{\lambda}_t^* - r_t )_+x_t\right)|\mathcal{H}_{t-1}\right]\right] = 0.
    \end{aligned}
\end{equation*}
Then, since $\{t<\tau_S\} \cap \{\bm{\lambda}_t^* \notin \mathcal{D}^*\} \subseteq \{t<\tau_S\} \cap \{(\bm{b}_t, \hat{\bm{p}}_t) \notin \mathcal{K}\} \subseteq \{\bm{b}_t\in\Omega\} \cap \{\bar{\mathcal{Q}}_t\} = \cup_{j=1}^n \bar{\mathcal{A}}_t^{(j)}$
\begin{equation*}
    \begin{aligned}
    &\hspace{5mm}\E\left[\sum_{t=1}^{\tau_S}\left(\max_{j\in[n],\bm{\lambda}\in\mathcal{D}^*}|\mu_j - \bm{c}_j^\top\bm{\lambda}|\bm{1}_{\{\bm{\lambda}_t^* \notin \mathcal{D}^* \}}\right)\right]\\
    &= \E\left[\sum_{t=1}^{T}\mathbf{1}_{\{t<\tau_S\}}\left(\max_{j\in[n],\bm{\lambda}\in\mathcal{D}^*}|\mu_j - \bm{c}_j^\top\bm{\lambda}|\bm{1}_{\{\bm{\lambda}_t^* \notin \mathcal{D}^* \}}\right)\right]\\
    &\leq \max_{j\in[n],\bm{\lambda}\in\mathcal{D}^*}|\mu_j - \bm{c}_j^\top\bm{\lambda}|\sum_{t=1}^{T}\sum_{j=1}^{n}P\left(\bar{\mathcal{A}}_t^{(j)}\right)\\
    &\leq \max_{j\in[n],\bm{\lambda}\in\mathcal{D}^*}|\mu_j - \bm{c}_j^\top\bm{\lambda}|\left(n + \sum_{t=1}^{T-1}2n\exp(-2L^2t)\right)\\
    &\leq \frac{2n\cdot\max_{j\in[n],\bm{\lambda}\in\mathcal{D}^*}|\mu_j - \bm{c}_j^\top\bm{\lambda}|}{1-\exp(-2L^2)}.
    \end{aligned}
\end{equation*}

\subsection{Proof of Lemma \ref{lem_bt}}\label{sec_ap_lem_bt}
In the proof we analyze the binding resource and non-binding resource separately. Without loss of generalization we can permute the order of the resource such that $\bm{b} = [\bm{b}_{\mathcal{B}}, \bm{b}_{\mathcal{N}}]$, we will work with $\mathbb{E}\left[\sum_{t=1}^T||\bm{b}_{\mathcal{B}, t} - \bm{b}_{\mathcal{B}}||_2\right]$ first. The reason we work with $\bm{b}_t$ instead of $\bm{b}_t'$ is that by definition, $||\bm{b}_t - \bm{b}||_2$ is an upper bound of $||\bm{b}_t' - \bm{b}||_2$. For the convenience of notation, for now let's denote $\bm{b} = \bm{b}_{\mathcal{B}}$.

From the update rule we have
\begin{align*}
    \bm{b}_{t+1} - \bm{b} = \bm{b}_t - \bm{b} - \frac{1}{T-t}(\bm{a}_tx_t - \bm{b}_t).
\end{align*}
We also have the geometry propoerty that
\begin{equation*}
    \begin{aligned}
     (\bm{b}_{t+1} - \bm{b})^\top(\bm{b}_{t+1} - \bm{b}) &=  (\bm{b}_{t} - \bm{b})^\top(\bm{b}_{t} - \bm{b}) -  \frac{2}{T-t}(\bm{b}_{t} - \bm{b})^\top(\bm{a}_tx_t - \bm{b}_t)\\ 
     &\hspace{15mm} + \frac{1}{(T-t)^2}(\bm{a}_tx_t - \bm{b}_t)^\top(\bm{a}_tx_t - \bm{b}_t).
    \end{aligned}
\end{equation*}
Because the analysis of each $i \in \mathcal{B}$ will be the same, we use $b_t, a_t$ to denote $b_{1,t}, a_{1,t}$. Observe that

\begin{equation*}
    \begin{aligned}
     (b_{t+1} - b)^2 &=  (b_{t} - b)^2 -  \frac{2}{T-t}(b_{t} - b)(a_tx_t - b_t) + \frac{1}{(T-t)^2}(a_tx_t - b_t)^2.
    \end{aligned}
\end{equation*}
Since we are dealing with binding dimensions, from $||\bm{c}_j||_{\infty} \leq 1$ we have $b_t \leq 2$ before the jump time $\tau_S$. This is because for binding resource we must have $b \leq 1$, and by setting $L \leq 1$ we know $b_t \leq 2$ before the stopping time. Since $T - \mathbb{E}[\tau_S] = O(1)$, for bounding the difference we can just treat $\tau_S = T$ and we know the difference will also be $O(1)$. Next, we have that
\begin{equation}\label{ineq_lemb_recursive}
    \begin{aligned}
     \mathbb{E}[(b_{t+1} - b)^2] &\leq  \mathbb{E}[(b_{t} - b)^2] +  \mathbb{E}\left[\frac{2}{T-t}(b_{t} - b)(a_tx_t - b_t)\right] + \frac{4}{(T-t)^2}.
    \end{aligned}
\end{equation}
We want the above inequality to be in a recursive form of $\mathbb{E}[(b_{t} - b)^2]$, therefore we need to bound $\mathbb{E}\left[\frac{2}{T-t}(b_{t} - b)(a_tx_t - b_t)\right]$. The corresponding result is summarized from the following lemma:
\begin{lemma}\label{lem_sqrt_term}
Under the binding case, there exists positive constants $K, A_1, A_2, B_1, B_2, C_1$ and $C_2$ such that
\begin{equation*}
    2\mathbb{E}\left[(b_{t} - b)(a_tx_t - b_t)\right] \leq \sqrt{\mathbb{E}\left[(b_t - b)^2\right]}\left(\frac{\sqrt{K}}{\sqrt{t-1}\wedge 1} + \sqrt{A_1e^{-A_2(T-t+1)} + B_1e^{-B_2t} + C_1T^{1/2}e^{-C_2T^{1/2}}} \right).
\end{equation*}
\end{lemma}

Next, by denoting $z_t = \mathbb{E}[(b_t - b)^2]$, form $\eqref{ineq_lemb_recursive}$ and the above lemma we know that
\begin{equation}\label{ineq_zt_sequence}
    \begin{aligned}
     z_{t+1} \leq z_t + \frac{1}{T-t}\sqrt{z_t}\left(\frac{\sqrt{K}}{{\sqrt{t-1}\wedge 1}} + \sqrt{A_1e^{-A_2(T-t+1)} + B_1e^{-B_2t} + C_1T^{1/2}e^{-C_2T^{1/2}}} \right) + \frac{4}{(T-t)^2}.
    \end{aligned}
\end{equation}
Then we can finish the proof from the following lemma
\begin{lemma}\label{lem_zt}
    For $\{z_t\}_{t=1}^{+\infty}$ satisfying $\eqref{ineq_zt_sequence}$, we have $\sum_{t=1}^T z_t = O(\bar{N}\log(T))$, where $\bar{N} = 4(\bar{M}^2 \vee 16^2)$ and $\bar{M} = 4\sqrt{K} + 2\sqrt{\frac{2A_1}{eA_2} + \frac{B_1}{eB_2} + \frac{27C_1}{e^3C_2^3}}$.
\end{lemma}
Lastly, let us identify the order for $\bar{N}$. By comparing the parameter in \eqref{ineq_zt_sequence} and \eqref{eqn:para_acbd} we find that $K$ is of the order $n^3$ and $C_1/C_2^3$ is of the order $n^9$. Therefore we have the bound $\sum_{t=1}^T\mathbb{E}[(b_{i,t} - b_i)^2] = O(n^9\log(T))$ holds for every $i \in \mathcal{B}$, and performing the same analysis for all the binding resources yields $\mathbb{E}\left[\sum_{t=1}^T ||\bm{b}_t' - \bm{b}||_2^2\right] \leq \mathbb{E}\left[\sum_{t=1}^T ||\bm{b}_t - \bm{b}||_2^2\right] \leq O(mn^9\log(T))$.

Next, let's denote $\bm{b}_t' = \bm{b}_{\mathcal{N}, t}'$ and analyze the non-binding case. Again because each dimension $i \in \mathcal{N}$ can be treated the same way, we denote $b_t' = b_{i, t}'$. Like in Lemma \ref{lem_sqrt_term}, we define event 
\begin{equation*}
    \begin{aligned}
     \mathcal{A}_t^{(j)} = \left\{\left|\hat{p}_{j,t} - p_j\right| \leq L\right\},
    \end{aligned}
\end{equation*}
and analyze the good event that $\{\bm{b}_t \in \Omega\} \cap \{\cap_{j=1}^n\mathcal{A}_t^{(j)}\}$. Under $\{\bm{b}_t \in \Omega\} \cap \{\cap_{j=1}^n\mathcal{A}_t^{(j)}\}$, we know that the index $i \in \mathcal{N}$ is non-binding hence Algorithm \ref{alg:DAA_fair} will set $b_{t}' = b$. Therefore, it suffices to bound the probability of bad events. We have

\begin{equation*}
    \begin{aligned}
     \sum_{t=1}^T\mathbb{E}[(b_{t}' - b)^2] &= \sum_{t=1}^T\mathbb{E}[\mathbf{1}_{\{\tau_S > t\} \cap \{\cap_{j=1}^n\mathcal{A}_t^{(j)}\}}(b_{t}' - b)^2] + \sum_{t=1}^T\mathbb{E}[\mathbf{1}_{\{\tau_S \leq t\} \cup \{\cup_{j=1}^n\bar{\mathcal{A}}_t^{(j)}\}}(b_{t}' - b)^2]\\
     &\leq 0 + \bar{b}^2 \sum_{t=1}^T\left(P\left(\tau_S \leq t\right) + \sum_{j=1}^nP(\bar{\mathcal{A}}_t^{(j)})\right).\\
    \end{aligned}
\end{equation*}
From Proposition \ref{prop_stoppingtime} and calculation we know that 
\begin{equation*}
    \begin{aligned}
     \sum_{t=1}^TP(\tau_S \leq t) &\leq \frac{16m}{L^2} + \frac{nT}{L^2}\exp\left(-2L^2\kappa T\right) + 4n^3T^{3/2}\exp\left(-\frac{\kappa^{1/2}}{n^2}T^{1/2}\right)\\
     \sum_{t=1}^T\sum_{j=1}^nP(\bar{\mathcal{A}}_t^{(j)}) &\leq  n +  \sum_{t=1}^{T-1}2n\exp\left(-2tL^2\right) \leq \frac{2n}{1-\exp(-2L^2)}.
    \end{aligned}
\end{equation*}
Therefore, for $i \in \mathcal{N}$ we know $\sum_{t=1}^T\mathbb{E}[(b_{t}' - b)^2] = O(m)$. To summarize, we have
$$||\bm{b}_t' - \bm{b}||_2 = ||\bm{b}_{\mathcal{B}, t}' - \bm{b}_{\mathcal{B}}||_2 + ||\bm{b}_{\mathcal{N}, t}' - \bm{b}_{\mathcal{N}}||_2 \leq O(mn^9\log T) + O(m^2).$$

\subsection{Proof of Lemma \ref{lem_lipchitz}}\label{sec_ap_lem_lipchitz}
We prove by contradiction. If instead
\begin{align}
        ||\bar{\bm{y}}^*(\theta, \Delta\bar{\bm{b}}) - \bar{\bm{y}}^*(0, \Delta\bar{\bm{b}})||_2 > \chi_{\bm{\bar{C}}}||\Delta\bar{\bm{b}}||_2|\theta|,
    \end{align}
From the fact that $||\bar{\bm{y}}^*(\theta, \Delta\bar{\bm{b}}) - \bar{\bm{y}}^*(0, \Delta\bar{\bm{b}})||_2 = ||\bar{\bm{y}}^*(\theta, \Delta\bar{\bm{b}}) - \bar{\bm{y}}^*(\theta/2, \Delta\bar{\bm{b}})||_2 + ||\bar{\bm{y}}^*(\theta/2, \Delta\bar{\bm{b}}) - \bar{\bm{y}}^*(0, \Delta\bar{\bm{b}})||_2$. It must be the case that either $$||\bar{\bm{y}}^*(\theta/2, \Delta\bar{\bm{b}}) - \bar{\bm{y}}^*(0, \Delta\bar{\bm{b}})||_2 > \chi_{\bm{\bar{C}}}||\Delta\bar{\bm{b}}||_2|\theta|/2$$ or $$||\bar{\bm{y}}^*(\theta, \Delta\bar{\bm{b}}) - \bar{\bm{y}}^*(\theta/2, \Delta\bar{\bm{b}})||_2 > \chi_{\bm{\bar{C}}}||\Delta\bar{\bm{b}}||_2|\theta|/2.$$ 
Repeating such argument will generate a sequence $\{\theta_i\}_{i=1}^{+\infty}$ such that $2|\theta_{i+1} - \theta_{i}| = |\theta_{i} - \theta_{i-1}|$ and $||\bar{\bm{y}}^*(\theta_{i+1}, \Delta\bar{\bm{b}}) - \bar{\bm{y}}^*(\theta_i, \Delta\bar{\bm{b}})||_2 > \chi_{\bm{\bar{C}}}||\Delta\bar{\bm{b}}||_2|\theta_{i+1} - \theta_i|
$. This implies that there exists $\theta^* = \lim_{i\to+\infty}\theta_i$ and a vector $\bar{\bm{b}}^* = \bar{\bm{b}} + \theta^*\Delta\bar{\bm{b}}$ such that the left or right derivative on direction $\Delta\bar{\bm{b}}$ is greater than $\chi_{\bar{\bm{C}}}||\Delta\bar{\bm{b}}||_2$, which contradicts Theorem 3.4 of \cite{holder2001marginal}.

\subsection{Proof of Lemma \ref{lem_sqrt_term}}
We define event 
\begin{equation*}
    \begin{aligned}
     \mathcal{A}_t^{(j)} = \left\{\left|\hat{p}_{j,t} - p_j\right| \leq L\right\},
    \end{aligned}
\end{equation*}
and analyze the good event that $ \mathcal{C}_{t-1} = \{\bm{b}_t \in \Omega\} \cap \{\cap_{j=1}^n\mathcal{A}_t^{(j)}\}$. Notice that we use $\mathcal{C}_{t-1}$ because we want to stress that $\mathcal{C}_{t-1}$ is $\mathcal{H}_{t-1}$-measurable. From the definition of $\Omega$, $\mathcal{A}_t^{(j)}$, the convention $\bm{b} = \bm{b}_{\mathcal{B}}$, and Algorithm \ref{alg:DAA_fair} we know that $\bm{b}_t = \sum_{j=1}^n \hat{p}_{j,t}\bm{c}_jy_{j,t}^*$ and $\mathbb{E}\left[\bm{a}_tx_t |\mathcal{C}_{t-1} \right] = \sum_{j=1}^n p_j\bm{c}_jy_{j,t}^*$. Then
\begin{equation*}
    \begin{aligned}
     \mathbb{E}\left[\bm{a}_tx_t - \bm{b}_t|\mathcal{C}_{t-1  } \right] = \mathbb{E}\left[\bm{a}_tx_t |\mathcal{C}_{t-1} \right] - \bm{b}_t = \sum_{j=1}^n\left(p_j - \hat{p}_{j,t}\right)\bm{c}_jy_{j,t}^*.
    \end{aligned}
\end{equation*}
Therefore, by taking the one-dimensional convention ($a = \bm{a}_1$ and the same for others) we have that
\begin{equation}\label{lem_pf_sqrt_term_two_terms}
    \begin{aligned}
     \mathbb{E}\left[(b_{t} - b)(a_tx_t - b_t)\right] &= \mathbb{E}\left[(b_{t} - b)(a_tx_t - b_t)1_{\{\mathcal{C}_{t-1}\}}\right]+ \mathbb{E}\left[(b_{t} - b)(a_tx_t - b_t)1_{\{\bar{\mathcal{C}}_{t-1}\}}\right]\\
    \end{aligned}
\end{equation}
For the first term
\begin{equation*}
    \begin{aligned}
     \mathbb{E}\left[(b_{t} - b)(a_tx_t - b_t)1_{\{\mathcal{C}_{t-1}\}}\right] &= \mathbb{E}\left[\mathbb{E}\left[(b_{t} - b)(a_tx_t - b_t)1_{\{\mathcal{C}_{t-1}\}}|\mathcal{H}_{t-1}\right]\right]\\
     &= \mathbb{E}\left[(b_{t} - b)1_{\{\mathcal{C}_{t-1}\}}\mathbb{E}\left[(a_tx_t - b_t)|\mathcal{H}_{t-1}\right]\right]\\
     &= \mathbb{E}\left[(b_{t} - b)1_{\{\mathcal{C}_{t-1}\}}\sum_{j=1}^n\left(p_j - \hat{p}_{j,t}\right)c_jy_{j,t}^*\right]\\
     &\leq \mathbb{E}\left[|b_{t} - b|\left(\sum_{j=1}^n\left|p_j - \hat{p}_{j,t}\right|c_jy_{j,t}^*\right)\right]\\
     &\leq \sqrt{\mathbb{E}\left[(b_{t} - b)^2\right]}\sqrt{\mathbb{E}\left[\left(\sum_{j=1}^n\left|p_j - \hat{p}_{j,t}\right|\right)^2\right]}
    \end{aligned}
\end{equation*}
From Hoeffding's inequality we know that 
\begin{equation*}
    \begin{aligned}
     \mathbb{E}\left[\left(\sum_{j=1}^n\left|p_j - \hat{p}_{j,t}\right|\right)^2\right] &= \int_0^{+\infty}\cfrac{y}{2}P\left(\sum_{j=1}^n\left|p_j - \hat{p}_{j,t}\right| > y\right)dy\\
     &\leq \int_0^{+\infty}\frac{y}{2}\sum_{j=1}^nP\left(|p_j - \hat{p}_{j,t}|>y/n\right)dy\\
     &\leq \int_0^{+\infty}y\sum_{j=1}^n\exp\left(-\frac{2(t-1)y^2}{n^2}\right)dy\\
     &= \frac{n^3}{4(t-1)}.
    \end{aligned}
\end{equation*}
Therefore, for the first term in $\eqref{lem_pf_sqrt_term_two_terms}$ we have that
\begin{equation*}
    \begin{aligned}
     \mathbb{E}\left[(b_{t} - b)(a_tx_t - b_t)1_{\{\mathcal{C}_{t-1}\}}\right] \leq \frac{\sqrt{n^3/4}}{\sqrt{t-1}}\sqrt{\mathbb{E}\left[(b_{t} - b)^2\right]}.
    \end{aligned}
\end{equation*}
Next, for the second term in $\eqref{lem_pf_sqrt_term_two_terms}$, we have that
\begin{equation*}
    \begin{aligned}
     \mathbb{E}\left[(b_{t} - b)(a_tx_t - b_t)1_{\{\bar{\mathcal{C}}_{t-1}\}}\right] &\leq \mathbb{E}\left[|b_{t} - b||a_tx_t - b_t|1_{\{\bar{\mathcal{C}}_{t-1}\}}\right]\\
     &\leq 2\mathbb{E}\left[|b_{t} - b|1_{\{\bar{\mathcal{C}}_{t-1}\}}\right]\\
     &\leq 2\sqrt{\mathbb{E}\left[(b_{t} - b)^2\right]}\sqrt{P(\bar{\mathcal{C}}_{t-1})}.\\
    \end{aligned}
\end{equation*}
Lastly, it suffices to give a bound for $P(\bar{\mathcal{C}}_{t-1})$. Notice that $\bar{\mathcal{C}}_{t-1} \subseteq \{\tau_S \leq t-1\}\cup\left\{\cup_{j=1}^n\bar{\mathcal{A}}_{t}^{(j)}\right\}$, therefore from Proposition \ref{prop_stoppingtime} and Hoeffding's inequality we have 
\begin{equation*}
    \begin{aligned}
     P(\bar{\mathcal{C}}_{t-1}) &\leq P(\tau_S \leq t-1) + \sum_{j=1}^nP\left(\bar{\mathcal{A}}_{t}^{(j)}\right)\\
     &\leq 2m\exp\left(-\frac{L^2(T-t+1)}{8}\right) + \frac{n}{L^2}\exp\left(-2L^2\kappa T\right) + 4n^3T^{1/2}\exp\left(-\frac{\kappa^{1/2}}{n^2}T^{1/2}\right)\\
     &\hspace{20mm}+2n\exp\left(-2L^2(t-1)\right),      
    \end{aligned}
\end{equation*}
and the last term can be dominated by the previous exponential terms. Therefore, by combining the results above we have 
\begin{equation}\label{eqn:para_acbd}
    \begin{aligned}
     2\mathbb{E}\left[(b_{t} - b)(a_tx_t - b_t)\right] &\leq \sqrt{\mathbb{E}\left[(b_t - b)^2\right]}\left(\frac{\sqrt{n^3}}{\sqrt{t-1}\wedge 1}  \right.\\
     &\hspace{5mm}+\left. \sqrt{16m\exp\left(-\frac{L^2(T-t+1)}{8}\right) + \frac{8n}{L^2}\exp\left(-2L^2\kappa T\right) + 32n^3T^{1/2}\exp\left(-\frac{\kappa^{1/2}}{n^2}T^{1/2}\right)} \right). 
    \end{aligned}
\end{equation}

\subsection{Proof of Lemma \ref{lem_zt}}
For
\begin{align*}
 z_{t+1} \leq z_t + \frac{2}{T-t}\sqrt{z_t}\left(\frac{\sqrt{K}}{{\sqrt{t-1}\wedge 1}} + \sqrt{A_1e^{-A_2(T-t+1)} + B_1e^{-B_2t} + C_1T^{1/2}e^{-C_2T^{1/2}}} \right) + \frac{4^2}{(T-t)^2}.
\end{align*}
From the fact that $tc_1e^{-c_2t} \leq \frac{c_1}{c_2e}$ and $t^3c_1e^{-c_2t} \leq \frac{27c_1}{c_2^3e^3}$ for any $t, c_1, c_2 > 0$, we have that for all $t \leq T/2$
\begin{equation*}
    \begin{aligned}
     &\hspace{5mm}\frac{\sqrt{K}}{{\sqrt{t-1}\wedge 1}} + \sqrt{A_1e^{-A_2(T-t+1)} + B_1e^{-B_2t} + C_1T^{1/2}e^{-C_2T^{1/2}}}\\
     &\leq \frac{2\sqrt{K}}{\sqrt{t}} + \frac{\sqrt{\left(\frac{2A_1}{eA_2} + \frac{B_1}{eB_2} + \frac{27C_1}{e^3C_2^3}\right)}}{\sqrt{t}}\\
     &= \frac{1}{\sqrt{t}}\left(2\sqrt{K} + \sqrt{\frac{2A_1}{eA_2} + \frac{B_1}{eB_2} + \frac{27C_1}{e^3C_2^3}}\right).\\ 
    \end{aligned}
\end{equation*}
If we denote $\bar{M} = 4\sqrt{K} + 2\sqrt{\frac{2A_1}{eA_2} + \frac{B_1}{eB_2} + \frac{27C_1}{e^3C_2^3}}$, for $t \leq T/2$ we have
$z_{t+1} \leq z_t + \frac{\bar{M}}{(T-t)\sqrt{t}}\sqrt{z_t} + \frac{4^2}{(T-t)^2}$. Denote $\bar{N} = 4(\bar{M}^2 \vee 16^2)$, and suppose that we have $z_t \leq \frac{\bar{N}}{(T-t)^2}t$. We know the base case is true since $z_1 = 0$ and the increment is at the order $\frac{4^2}{(T-t)^2}$. Then,
$$z_{t+1} \leq \frac{\bar{N}t}{(T-t)^2} + \frac{\bar{M}\sqrt{\bar{N}}}{(T-t)^2}+\frac{4^2}{(T-t)^2}\leq \frac{\bar{N}(t+1)}{(T-t-1)^2}.$$
Therefore, we have that $z_t \leq \frac{\bar{N}}{(T-t)^2}t$ for all $t \leq T/2$. 

Such bound will be insufficient to generate a $\log (T)$ regret when $t$ gets large. So for $t > T/2$, we need a different bound and choose to test if $z_t \leq \frac{\bar{N}}{T-t}$. The base case is true from the result on $t \leq T/2$, and for larger $t$ we have
\begin{equation*}
    \begin{aligned}
     z_{t+1} \leq \frac{\bar{N}}{T-t} + \frac{\bar{M}\sqrt{\bar{N}}}{(T-t)^{3/2}\sqrt{t}} + \frac{4^2}{(T-t)^2} \leq \frac{\bar{N}}{T-t-1}.
    \end{aligned}
\end{equation*}
Therefore, combining it together yields
\begin{equation*}
    \begin{aligned}
     \sum_{t=1}^T z_t &= \sum_{t=1}^{T/2} z_t +\sum_{t=T/2 + 1}^{T} z_t \leq \sum_{t=1}^{T/2} \frac{\bar{N}t}{(T-t)^2} +\sum_{t=T/2 + 1}^{T} \frac{\bar{N}}{T-t-1} = O(\bar{N}\log(T)).
    \end{aligned}
\end{equation*}

\subsection{Proof of Lemma \ref{lem_mtg}} \label{sec_ap_mtg}

We first analyze the process $\tilde{b}_{i,t}$, where $i \in \mathcal{B}$. Then we analyze the non-binding process. Lastly we do a union bound and conclude the result. Throughout the proof, we have to use the following theorem.

\begin{theorem}{(Hoeffding's inequality for dependent data \citep{van2002hoeffding})} \label{thm_dehling}
Consider a sequence of random variables $\{X_t\}_{t=1}^T$ adapted to the filtration $\mathcal{F}_t$'s and 
$$\E[X_t | \mathcal{F}_{t-1}] = 0 \text{ \ for \ } t=1,...,T$$
where $\mathcal{F}_0=\emptyset.$
Suppose $L_t, U_t$ are $\mathcal{F}_{t-1}$-measurable random variables such that $L_t\le X_t\le U_t$ almost surely for $t=1,...,T$. Let $S_t =\sum_{s=1}^t X_t$ and $V_t=\sum_{s=1}^t(U_s-L_s)^2$. Then, the following inequality holds for any $b>0, c>0$ and $T \in \mathbb{N}_+,$
\begin{align*}
\prob(|S_t|\ge b, V_t \le c^2 \text{ for some } t\in\{1,...,T\}) \leq 2e^{-\frac{2b^2}{c^2}}.
\end{align*}
\label{Hf_dependent}
\end{theorem}

Now we begin the proof. For $i \in \mathcal{B}$, let $$Y_t\coloneqq\tilde{b}_{i,t+1}-\tilde{b}_{i,t}$$ for $t\ge1$ and $$X_t \coloneqq Y_t - \mathbb{E}[Y_t|\mathcal{H}_{t-1}].$$
In this way, to analyze the process $\tilde{b}_{i,t}$, we can equivalently analyze the summation $\sum_{s=1}^{t-1} Y_t$. From the definition of the process $\tilde{\bm{b}}_t$, we know that when $t> \tilde{\tau} - 1$, we have $$\tilde{b}_{i,t+1}=\tilde{b}_{i,t},$$
and when $1\le t<\tilde{\tau},$ we have
$$\tilde{b}_{i,t+1}= \tilde{b}_{i,t} +\frac{1}{T-t}(\tilde{b}_{i,t}-{a}_{i,t}x_{t}).$$
From the fact that $\tilde{b}_{i,t}$ is $\mathcal{H}_{t-1}$-measurable, we can bound the absolute value of $|x_t|$ such that
\begin{equation*}
    \begin{aligned}
        |X_t| &= \left|\frac{1}{T-t}(\tilde{b}_{i,t}-{a}_{i,t}x_{t}) - \mathbb{E}\left[\left. \frac{1}{T-t}(\tilde{b}_{i,t} - {a}_{i,t}x_{t})\right|\mathcal{H}_{t-1}\right]\right|\\
        &=\frac{1}{T-t}\left|\mathbb{E}\left[{a}_{i,t}x_t|\mathcal{H}_{t-1}\right] - {a}_{i,t}x_{t}\right| \leq \frac{1}{T-t}
    \end{aligned}
\end{equation*}
for each $t\le T-1.$
So we can define $L_t$ and $U_t$ as 
\begin{align*}
    L_t & \coloneqq -\frac{1}{T-t},\\
    U_t & \coloneqq \frac{1}{T-t},
\end{align*}
and the conditions of Theorem \ref{thm_dehling} are met for the process $X_t, L_t$ and $U_t$. Then as in Theorem \ref{thm_dehling}, 
$$V_t= \sum_{s=1}^t (U_s-L_s)^2 = \sum_{s=1}^t \frac{4}{(T-s)^2} \le \frac{4}{T-t-1}$$
for $t=1,...,T-2.$
From Theorem \ref{Hf_dependent}, we know that
\begin{equation}
  \prob\left(\left\vert\sum_{j=1}^s X_j \right\vert \ge L \text{ for some } s\le t\right)\le 2e^{-\frac{L^2(T-t-1)}{2}} \label{sum_X}
\end{equation}
holds for all $L>0$ and $t\le T-2.$ 

With this bound on the summation of $X_t$, we return to analyze the summation of $Y_t$ by bounding the difference between these two sequences. By the definition, we have
  \begin{align}
    |X_t-Y_t| & = |\mathbb{E}[Y_t|\mathcal{H}_{t-1}]| \nonumber \\
    & =|\mathbb{E}[\tilde{b}_{i,t+1}-\tilde{b}_{i,t}|\mathcal{H}_{t-1}]|\nonumber \\
    & = \left\vert\frac{1}{T-t}I(t<\tilde{\tau})\mathbb{E}[({a}_{i,t}x_{t}-b_{i,t})|\mathcal{H}_{t-1}] \right\vert \nonumber \\& \le \frac{\epsilon_t^*}{T-t} = \frac{1}{T-t}I(t\leq \kappa T) + \frac{1}{(T-t)t^{1/4}}I(t >\kappa T)
    \label{difference_X_Y}
  \end{align}
for $1\le t\le T-1.$ The second line comes from the definition of $Y_t$. The third line comes from the definition of $\mathcal{E}_t$ and the definition of $\tilde{\tau}$. For the last line, we denote that since $t < \tilde{\tau}$, for $t \leq \kappa T$ we have that $\left|\mathbb{E}[({a}_{i,t}x_{t}-b_{i,t})|\mathcal{H}_{t-1}]\right| \leq 1$ no matter if we put $\bm{b}_t$ \eqref{eqn:LP_algorithm_beforefair} or $\bm{b}_t'$ \eqref{eqn:LP_algorithm_afterfair} in the right-hand-side of the sampled LP; when $t > \kappa T$, since $t < \tilde{\tau}$, we know that $\mathcal{G}_t$ is true such that we are putting $b_{i,t}'=b_{i,t}$ (because $i\in\mathcal{B}$) in the right-hand-side of the sampled LP, and by definition of $\mathcal{E}_t$ we know that $\mathbb{E}[\bm{a}_{\mathcal{B}, t}x_{t}(\bm{b}_t')|\mathcal{H}_{t-1}]-\bm{b}_{\mathcal{B},t} \leq \epsilon_{t}^*$.
  
By taking summation of (\ref{difference_X_Y}), we have for $s\leq T-2$,
  $$\left\vert\sum_{j=1}^s X_j - \sum_{j=1}^s Y_j \right\vert \le  \sum_{j=1}^{\kappa T} \frac{1}{T-j} + \sum_{j=\kappa T +1 }^{s} \frac{1}{(T-j)j^{1/4}}.$$
The next is to find a proper value of $\kappa$ such that the equation above is bounded by $\frac{L}{2}$. For the first part, we have
 \begin{equation*}
    \begin{aligned}
    \sum_{j=1}^{\kappa T} \frac{1}{T-j} &\leq  \int_{T - \kappa T - 1}^{T-1} \frac{1}{x}dx = \log\left(\frac{T-1}{T -\kappa T - 1}\right)\\
    &\leq\log\left(\frac{T-1}{T - \kappa T- 2 - 2\kappa}\right) = \log\left(\frac{T-1}{T-2}\right) - \log\left(1-\kappa\right).
    \end{aligned}
\end{equation*}
For the second part,
\begin{equation*}
    \begin{aligned}
    \sum_{j=\kappa T+ 1}^{T-1} \frac{1}{(T-j)j^{1/4}} 
    &\leq \frac{1}{(\kappa T)^{1/4}} \sum_{j=\kappa T + 1}^{T-1} \frac{1}{T-j} \leq \frac{\log T}{(\kappa T)^{1/4}}.
    \end{aligned}
\end{equation*}
Henceforth, if we set $\kappa = 1-\exp(-\frac{L}{8})$ and define $T_1$ as the minimal integer such that $T_1 \ge \frac{1}{\exp{\left(\frac{L}{8}\right) - 1}}  + 2$ and 
$\frac{\log T_1}{T_1^{1/4}} \le \frac{\kappa^{1/4}L}{4}$, then the following inequality holds for $T\ge T_1$
\begin{align*}
\sum_{j=1}^{\kappa T} \frac{1}{T-j} + \sum_{j=\kappa T +1 }^{T-1} \frac{1}{(T-j)j^{1/4}} \leq \frac{L}{4} + \frac{L}{4} = \frac{L}{2}.
\end{align*}
With the choice of $\kappa$ and $T\ge T_1$, we have  $$\left\vert\sum_{j=1}^s X_j - \sum_{j=1}^s Y_j \right\vert \leq \frac{L}{2}$$
holds almost surely. Consequently,
\begin{equation*}
    \begin{aligned}
    \left\{|\tilde{b}_{i,s}-b_i| > L \text{ for some } s \leq t\right\} &= \left\{\left\vert\sum_{j=1}^{s-1} Y_j \right\vert > L \text{ for some } s \leq t\right\}\\
    &= \left\{\left\vert\sum_{j=1}^{s} Y_j \right\vert > L \text{ for some } s \leq t-1\right\}\\
    &\subseteq \left\{\left\vert\sum_{j=1}^{s} X_j \right\vert > L/2 \text{ for some } s \leq t-1\right\}.\\
    \end{aligned}
\end{equation*}
Therefore, if we apply union bound with respect to binding constraint index, we have
\begin{equation*}
    \begin{aligned}
    \prob\left(\tilde{\bm{b}}_{\mathcal{B}, s}\notin \bigotimes_{i\in\mathcal{B}} \left[b_{i}-L, b_{i}+L\right] \text{ for some }s\le t\right)\le 2|\mathcal{B}|e^{-\frac{L^2(T-t)}{8}}
    \end{aligned}
\end{equation*}
for $t\le T-2$ and $T\ge T_1$.

Next, for non-binding index $i\in \mathcal{N}$, we use the same definition on $Y_t, X_t$ such that
$$Y_t\coloneqq\tilde{b}_{i,t+1}-\tilde{b}_{i,t},\,\,\,\,\,\,X_t \coloneqq Y_t - \mathbb{E}[Y_t|\mathcal{H}_{t-1}],$$
and want to show
$$\sum_{j=1}^s X_j - \sum_{j=1}^s Y_j \leq \frac{L}{2}.$$
Since in Algorithm \ref{alg:DAA_fair} use $\bm{b}_t'$ instead for solving the sampled LP, we have to make sure the following is true for $t < \tilde{\tau}$
\begin{equation}\label{eqn:mtg_to_show}
    \begin{aligned}
     \mathbb{E}[b_{i,t}  - a_{i,t}x_t(\bm{b}_t')|\mathcal{H}_{t-1}] \geq 
\begin{cases}
\underline{b} - L - 1 & t \leq \kappa T, \\
0 & t >  \kappa T.
\end{cases}
    \end{aligned}
\end{equation}
To show the property \eqref{eqn:mtg_to_show}, for $t \leq \kappa T$, we have
$$\mathbb{E}[b_{i,t}  - a_{i,t}x_t(\bm{b}_t')|\mathcal{H}_{t-1}] \geq \underline{b} - L - 1$$
because for $t < \tilde{\tau}$ we know $b_{i,t} \geq \underline{b} - L $ and $||\bm{a}_t||_{\infty} \leq 1$ from Assumption \ref{assp}. Next, for $t > \kappa T$, since $t < \tilde{\tau}$, from definition of $\mathcal{E}_t$, we know $(\bm{b}_t, \hat{\bm{p}}_t) \in \mathcal{K}$, and $\mathcal{G}_t$ is true; From the second statement of Corollary \ref{cor_bindingness}, we have we that 
$$\mathbb{E}[b_{i,t}  - a_{i,t}x_t(\bm{b}_t')|\mathcal{H}_{t-1}] \geq b_{i,t} + L - b_i \geq 0.$$
From the same analysis we know 
\begin{equation*}
    \begin{aligned}
     X_t - Y_t &= -\mathbb{E}\left[Y_t|\mathcal{H}_{t-1}\right]= -\mathbb{E}\left[\left.\frac{1}{T-t}(b_{i,t}  - a_{i,t}x_t(\bm{b}_t')) I(\tilde{\tau} > t)\right|\mathcal{H}_{t-1}\right]\\
     &\leq \frac{1+L-\underline{b}}{T-t}I(t \leq \kappa T) + 0\leq \frac{1+L-\underline{b}}{T-t}I(t \leq \kappa T) + \frac{1}{t^{1/4}(T-t)}I(t > \kappa T).
    \end{aligned}
\end{equation*}
Then we apply the same approach. By defining $\kappa \leq 1-\exp(-\frac{L}{8(1+L-\underline{b})})$, and $T_1$ to be the minimal integer such that $T_1 \ge \frac{1}{\exp{\left(\frac{L}{8(1+L-\underline{b})}\right) - 1}}  + 2$ and $\frac{\log T_1}{T_1^{1/4}} \leq \frac{\kappa^{1/4} L}{4}$, we know that for $T > T_1$,
\begin{equation*}
    \begin{aligned}
    \sum_{j=1}^{\kappa T} \frac{1+L-\underline{b}}{T-j} &\leq  \left(1+L-\underline{b}\right)\left(\log\left(\frac{T-1}{T-2}\right) - \log\left(1-\kappa\right)\right)\leq \frac{L}{4},\\
    \sum_{j=\kappa T +1}^{ T-1}\frac{1}{(T-j)j^{1/4}}
    &\leq \frac{1}{(\kappa T)^{1/4}} \sum_{j=\kappa T + 1}^{T-1} \frac{1}{T-j} \leq \frac{\log T}{(\kappa T)^{1/4}} \leq \frac{L}{4}.
    \end{aligned}
\end{equation*}
Therefore, with the choice of $\kappa$ and $T\ge T_1$, we have
$$\sum_{j=1}^s X_j - \sum_{j=1}^s Y_j =  \sum_{j=1}^{\kappa T} \frac{1+L-\underline{b}}{T-j} + \sum_{j=\kappa T + 1}^{s} \frac{1}{(T-j)j^{1/4}} \leq \frac{L}{2}.$$
Next, for $t\le T-2.$ we have that  
\begin{equation*}
  \prob\left(\sum_{j=1}^s X_j  \le -\frac{L}{2} \text{ for some } s\le t\right)\le e^{-\frac{L^2(T-t-1)}{8}}.
\end{equation*}
Then, we have for $i \in \mathcal{N}$
\begin{equation*}
    \begin{aligned}
    \left\{\tilde{b}_{i,s}-b_i \leq -L \text{ for some } s \leq t\right\} &= \left\{\sum_{j=1}^{s-1} Y_j \leq -L \text{ for some } s \leq t\right\}\\
    &= \left\{\sum_{j=1}^{s} Y_j \leq -L \text{ for some } s \leq t-1\right\}\\
    &\subseteq \left\{\sum_{j=1}^{s} X_j \leq -\frac{L}{2} \text{ for some } s \leq t-1\right\}.\\
    \end{aligned}
\end{equation*}
Therefore, for $i \in \mathcal{N}$ we have that
\begin{equation*}
    \begin{aligned}
    \prob\left(\tilde{b}_{i,s} \notin \Omega \text{ for some } s\le t\right)\le 2|\mathcal{N}|e^{-\frac{L^2(T-t)}{8}}.
    \end{aligned}
\end{equation*}
Summing up and taking the union bound, we know that for all $T\ge T_1$ and $t\le T-2$, we have
\begin{equation*}
    \begin{aligned}
    \prob\left(\tilde{\bm{b}}_s\notin\Omega \text{ for some }s\le t\right)\le 2me^{-\frac{L^2(T-t)}{8}}.
    \end{aligned}
\end{equation*}

\end{document}